\documentclass[a4paper,twocolumn,floatfix,superscriptaddress,longbibliography,notitlepage, nofootinbib]{quantumarticle}
\pdfoutput=1

\usepackage[utf8]{inputenc}
\usepackage{braket}
\usepackage{amsmath}
\DeclareMathOperator{\Tr}{Tr}
\usepackage{dcolumn}   
\usepackage{bm}        
\usepackage{amssymb}
\usepackage{mathtools}
\usepackage{tikz}
\usepackage{qcircuit}
\usepackage{appendix}
\usepackage{hyperref}
\usepackage{subcaption}
\usepackage{amsmath,amsfonts,amssymb}
\usepackage{mathtools}
\usepackage{thmtools,thm-restate}
\usepackage{algorithm}
\usepackage{mathrsfs}
\usepackage{tikz}
\usepackage{subcaption}
\usepackage{pgfplots}
\usetikzlibrary{3d}
\usepackage{tkz-euclide}
\usetikzlibrary{arrows.meta, decorations.pathreplacing}
\usepackage{tikz-3dplot}
\usetikzlibrary{shapes.geometric}
\usetikzlibrary{3d, calc}
\usepackage{soul}

\usepackage{algpseudocode}

\newtheorem{definition}{Definition}
\newtheorem{theorem}{Theorem}
\newtheorem{lemma}{Lemma}

\newcommand{\Ibb}{\mathbb{I}}

\newcommand{\Rbb}{\mathbb{R}}

\newcommand{\revise}[1]{\textcolor{black}{#1}}
\usetikzlibrary{arrows.meta}

\def\BibTeX{{\rm B\kern-.05em{\sc i\kern-.025em b}\kern-.08em
    T\kern-.1667em\lower.7ex\hbox{E}\kern-.125emX}}

\begin{document}
\title{Quantum Algorithm for Estimating Betti Numbers Using Cohomology Approach}
\author{Nhat A. Nghiem}
\affiliation{QuEra Computing Inc., Boston, Massachusetts 02135, USA}
\affiliation{Department of Physics and Astronomy, State University of New York at Stony Brook, Stony Brook, NY 11794-3800, USA}
\affiliation{C. N. Yang Institute for Theoretical Physics, State University of New York at Stony Brook, Stony Brook, NY 11794-3840, USA}
\author{Xianfeng David Gu}
\affiliation{Department of Computer Science, State University of New York at Stony Brook, Stony Brook, NY 11794, USA}
\affiliation{Department of Applied Mathematics \& Statistics, State University of New York at Stony Brook, Stony Brook, NY 11794, USA }
\author{Tzu-Chieh Wei}
\affiliation{Department of Physics and Astronomy, State University of New York at Stony Brook, Stony Brook, NY 11794-3800, USA}
\affiliation{C. N. Yang Institute for Theoretical Physics, State University of New York at Stony Brook, Stony Brook, NY 11794-3840, USA}

\begin{abstract}
 Topological data analysis has emerged as a powerful tool for analyzing large-scale data. An abstract simplicial complex, in principle, can be built from data points, and by using tools from homology, topological features could be identified. Given a simplex, an important feature is called the Betti numbers, which roughly count the number of `holes' in different dimensions. Calculating Betti numbers exactly can be  $\#$P-hard, and approximating them can be NP-hard, which rules out the possibility of any generic efficient algorithms and unconditional exponential quantum speedup. Here, we explore the specific setting of a triangulated manifold instead of the general TDA. In contrast to most known quantum algorithms to estimate (normalized) Betti numbers, which rely on homology, we exploit the `dual' approach, namely, cohomology, combining the insight of the Hodge theory and de Rham cohomology.  By first randomizing the so-called $r$-forms as vectors, we leverage several recipes from the block-encoding framework to deform them to their corresponding \textit{hamornic forms}. Then the $r$-th Betti number can be found by examining the number of linearly independent harmonic forms. A detailed analysis is provided, showing that our cohomology framework can perform exponentially faster than previous homology methods in several regimes. In particular, our method can be effective when $\beta_r \ll |S_r^K|$, which can offer more flexibility and practicability than existing quantum algorithms that achieve the best performance in the regime $\beta_r \approx |S_r^K|$. 
   
\end{abstract}
\maketitle

\section{Introduction}
\begin{figure}[htbp]
    \begin{flushleft}
      (a)
\end{flushleft} \vspace{-0.9cm}\begin{subfigure}[b]{0.7\linewidth}
        \includegraphics[width = \linewidth]{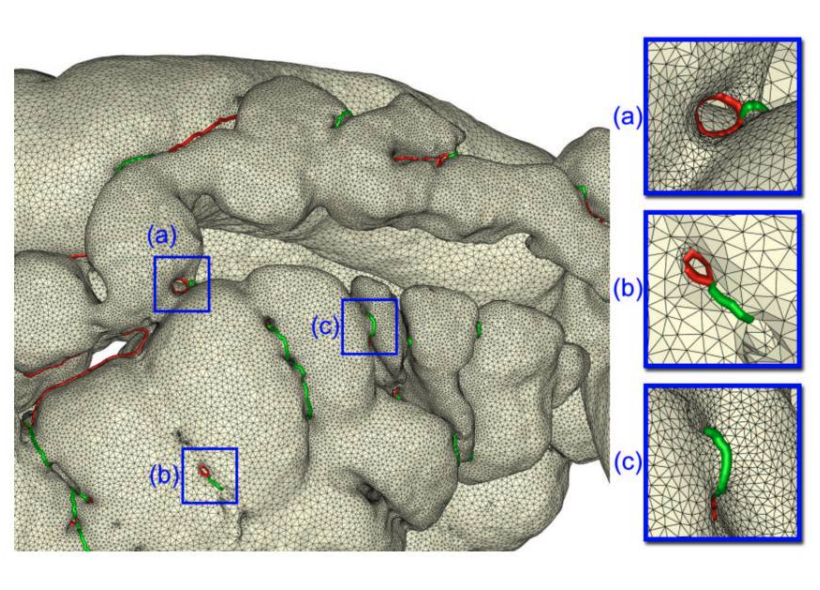}
        \label{fig: tdamedical}
    \end{subfigure}
    \begin{flushleft}
      (b)
\end{flushleft} \vspace{-2cm}
\begin{subfigure}[b]{\linewidth}
    \centering
\includegraphics[width = 0.88\linewidth]{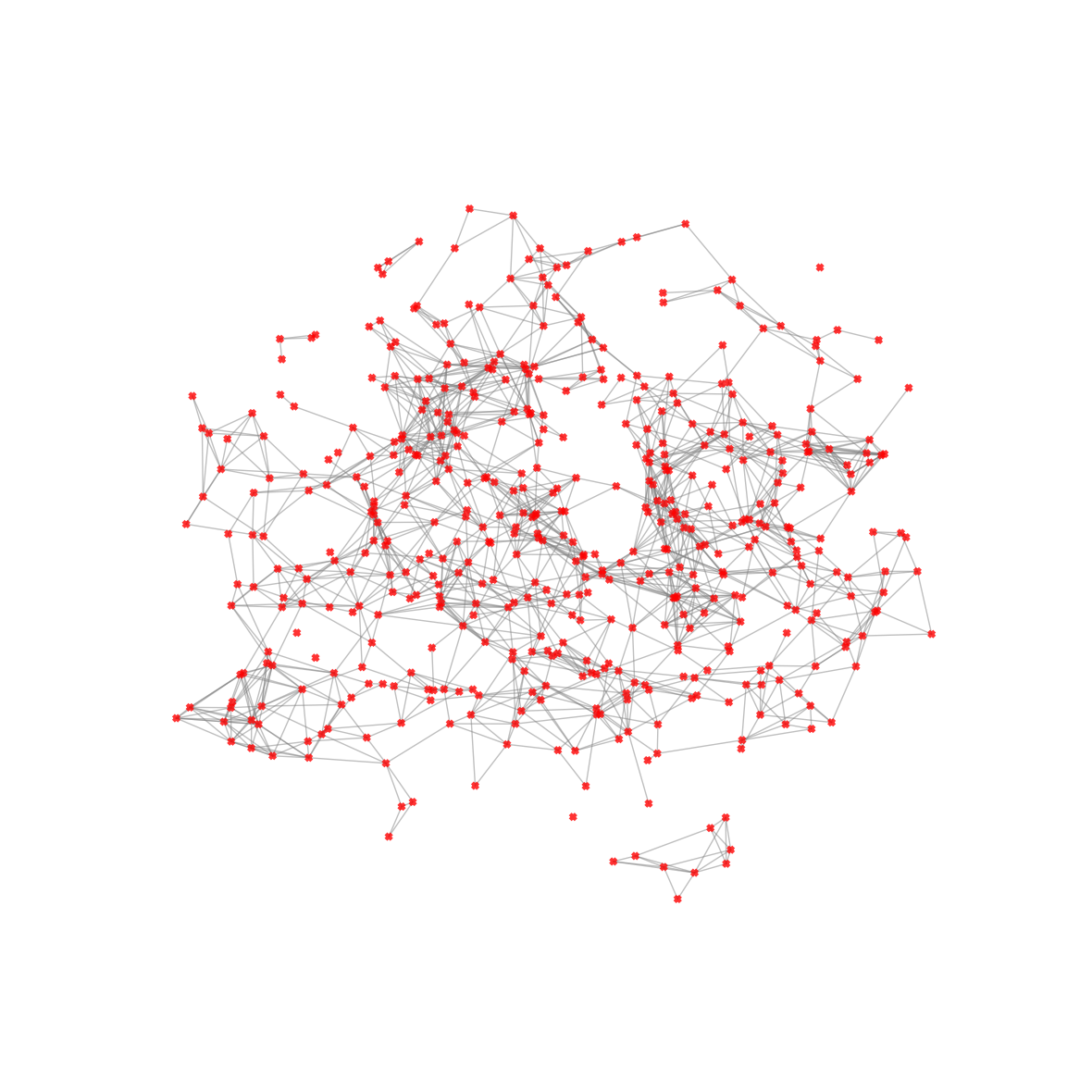}
        \label{fig: scomplex}
    \end{subfigure}
    \vspace{-1.2cm}
    \caption{(a) Application of TDA in medical imaging: performing topological denoise.  (b) An example of point cloud. Each point might be a vector in a very high dimension. Upon an appropriate metric, one can define distance between two points, and set a threshold for connectivity, e.g., two points are connected if their distance is less than some value. Roughly speaking, a point is called 0-simplex, an edge connecting two points is called 1-simplex, a triangle having 3 points mutually connected is called 2-simplex. Higher dimensional simplex is generalized in a straightforward manner. Then the resultant configuration forms a so-called simplicial complex. The goal of TDA is to analyze the underlying structure of given data points by using tools from algebraic topology, more concretely, homology theory. }
    \label{fig: TDA}
\end{figure}

Topology and geometry are lasting and vibrant areas in mathematics. Despite being abstract, topology has laid the groundwork for many important tools that have been widely applied in science and engineering~\cite{wasserman2016topological, bubenik2015statistical, gu2003genus, gu2008computational, gu2020computational}. Among them, topological data analysis (TDA) is gaining much attention due to its utility in revealing important features of datasets, which, in reality, can be sensitive to noise or sampling errors. This capability gives rise to useful applications, such as in computer vision and medical imaging; see Figs.~\ref{fig: TDA}a and~\ref{fig: TDA}b for illustration. A popular technique within TDA is persistent homology, which is built upon algebraic topology, showcasing a powerful method that can probe the underlying structure of a given dataset.  {Although intrinsically high-dimensional data, such as feature vectors, impose a large amount of computational hurdles for machine learning methods, intriguingly, TDA does not directly involve entries of these vectors in the computation.} Instead, by defining a suitable metric in the space of the given data points, one can associate connectivity between the data points (point cloud), forming a configuration called a simplicial complex (see Fig.~\ref{fig: TDA}), which is formally defined in Appendix~\ref{sec: crashcoursehomology}. 

Given a simplicial complex, one is interested in its Betti numbers, which reveal the number of connected components, loops, holes, etc., of such a configuration. {These numbers capture the structure of the dataset, which is topological by nature. Thus, TDA features a distinct way to analyze a large-scale dataset, which is fundamentally different from other popular methods such as supervised learning, unsupervised learning, support vector machine, neural network, etc. } The Abelian property and linearity of homology allow the formulation of problems in a linear algebra framework, which can be more convenient to work with practically. However, the massive dataset volume and its exponential growth in dimensionality induce a very large computational cost, and hence pose a severe obstacle in practical execution. 

{At the same time, by utilizing intrinsic properties of quantum mechanics as resources for processing information, quantum computers hold a great promise in solving challenging computational problems. From early proposals~\cite{manin1980computable, benioff1980computer, feynman2018simulating}, tremendous progress has been made in realizing the potential advantage of quantum computers relative to their classical counterparts. For example, Refs.~\cite{deutsch1985quantum,deutsch1992rapid} showed that quantum computers can probe black-box's property exponentially more efficiently than classical computers, Shor's algorithm~\cite{shor1999polynomial} showed that quantum computers can factorize a large integer with superpolynomial speed-up,  while Grover's algorithm~\cite{grover1996fast} showed that quantum searching can be quadratically faster than classical ones. In addition, quantum computers also have a great ability to simulate quantum systems, which is naturally difficult for classical computers~\cite{aharonov2006polynomial, berry2007efficient, berry2012black, berry2014high, berry2015hamiltonian, berry2015simulating,berry2017quantum, low2017optimal,low2019hamiltonian,childs2010relationship,childs2018toward, childs2019nearly}. Recent progress involves variational quantum algorithms and their interaction with machine learning as well as optimization~\cite{mitarai2018quantum, schuld2014quest,schuld2018supervised,schuld2019evaluating,schuld2019machine,schuld2019quantum,schuld2020circuit,schuld2020effect, farhi2018classification, zhou2020quantum}. }

Lloyd,  Garnerone, and Zanardi considered the problem of computing Betti numbers in the quantum setting, and their proposed algorithm (the so-called LGZ algorithm) claimed to yield exponential speedup compared to classical methods~\cite{lloyd2016quantum}. Their underlying approach essentially relies on (simplicial) homology: Given a simplicial complex $K$, the $r$-th Betti number is the rank of the $r$-th homology group $H^K_r$.   The rank of such a group (or, equivalently, the dimension of such a space) is revealed by analyzing the spectrum of the so-called boundary map $\partial$, which is a linear map between chain spaces (as will be formally defined subsequently). In Ref.~\cite{lloyd2016quantum}, this analysis is then done by applying quantum techniques, such as the quantum phase estimation~\cite{kitaev1995quantum} and sparse-matrix simulation~\cite{berry2007efficient}, leading to their LGZ quantum algorithm, which we review in Sec.~\ref{sec: existingquantumalgorithmsforTDA}. Following Ref.~\cite{lloyd2016quantum}, a series of works~\cite{ubaru2021quantum, mcardle2022streamlined, gunn2019review, ameneyro2022quantum} have substantially improved the running time of the original LGZ algorithm~\cite{lloyd2016quantum}. Most recently, a striking result from Ref.~\cite{schmidhuber2022complexity} clarified some assumptions and the performance of the LGZ-like algorithms~\cite{lloyd2016quantum}. In particular,  Schmidhuber and Lloyd~\cite{schmidhuber2022complexity} showed that computing Betti numbers is a \#P-hard in the general case and that approximating them up to a multiplicative error is NP-hard, ruling out the generic efficient estimation of Betti numbers. The potentially exponential advantage of the quantum algorithm can only be obtained if the input is \textit{a specified complex} instead of a list of vertices and pairwise connectivity. 

Inspired by the exciting development of quantum applications in computational topology-related problems, and, in particular, the complexity hardness result of estimating Betti numbers in~\cite{schmidhuber2022complexity}, we focus on a different scenario from the TDA, where simplicial complex $S$ is a triangulation of some closed manifold. For example, a triangulation can be built via standard methods such as Poisson reconstruction and Delaunay triangulation. Such methods are widely used in the fields of medical imaging, computer graphics, computer-aided design, and computational conformal geometry. In practice, many manifolds admit uniform or nearly uniform triangulation (e.g., can be abstract/topological triangulation). Although the above-mentioned \#P and NP-hardness of estimating Betti numbers is for general complexes, and it is not known whether the complexity reduces in triangulated manifolds, our aim is to provide a quantum algorithm that can efficiently compute them.  

Here, we provide a `dual' approach to~\cite{lloyd2016quantum} in computing Betti numbers on manifolds via cohomology theory. In this simplicial complex model, we can use the powerful de Rham cohomology and Hodge theory, as they provide a direct link between the $k$-th homology or cohomology group and a special group called the harmonic group. Roughly speaking, the building blocks of homology theory are simplexes. Instead of simplexes, within the cohomology framework, a real number is assigned to each simplex via a functional called \textit{form}. Thus, this cohomology approach is the `dual' to homology, as the collection of different functionals (i.e., forms) results in a (linear) vector space, which is the `dual' to the space of simplexes within the homology framework. Given an arbitrary form, one can deform it into the so-called harmonic form via the Hodge decomposition theorem. \textit{As a fundamental result in the field of differential geometry, the dimension of the space composed of harmonic forms is equal to the corresponding Betti number}. Hence, our strategy is to track the dimension of such a space. Other enabling elements for our algorithm come from recent advances in quantum algorithmic techniques~\cite{harrow2009quantum, lloyd2014quantum, childs2017quantum, nghiem2023improved, low2017optimal, low2019hamiltonian, gilyen2019quantum} that allow the handling of large matrix operations, such as multiplication, inversion, and polynomial transformation. A high-level overview of our quantum algorithm will be described in the next section. 

This article is organized as follows. First, in Section~\ref{sec: overview}, we provide an overview of our problem setup (Section~\ref{sec: datainputmodel}), as well as introducing the main cohomology theory underlying our work and statement of key results (in Section~\ref{sec: highleveloverview}). Section~\ref{sec: existingquantumalgorithmsforTDA} contains a summary of existing progress in quantum topological data analysis. In Section~\ref{sec: quantumalgorithm}, we outline a detailed procedure of our main quantum algorithm for estimating Betti numbers. Further analysis and related discussion on the advantage of our cohomology approach compared to the homology one and the classical approach will be provided in Section~\ref{sec: furtheranalysisanddiscussion}. In Appendix~\ref{sec: QSVT}, we summarize the necessary definitions and related arithmetic techniques used throughout our work, drawn mainly from the block encoding framework~\cite{gilyen2019quantum}. Appendices~\ref{sec: crashcoursehomology} and \ref{sec: cohomologicalframework} present a brief overview of homology and cohomology theory, which are designed for interested readers. Appendices~\ref{sec: elaboration} and~\ref{sec: generalhodge} provide more details about some aspects of our method outlined in Section~\ref {sec: quantumalgorithm}.    

\section{Overview}
\label{sec: overview}

\revise{The building blocks of TDA are simplexes and simplicial complex. } As illustrated in Fig.~\ref{fig: TDA}, a simplicial complex is composed of simplexes, e.g., 0-simplex is a point, 1-simplex is an edge, 2-simplex is a triangle, etc. In Appendix~\ref{sec: crashcoursehomology}, we provide a detailed review on this and describe the standard homology theory that leads to an important result of the $r$-th Betti number being the rank of the corresonding $r$-th homology group ($\beta_r={\rm rank} H_r^K$). 

\begin{figure}[htbp]
    \centering
    \includegraphics[width=0.6\linewidth]{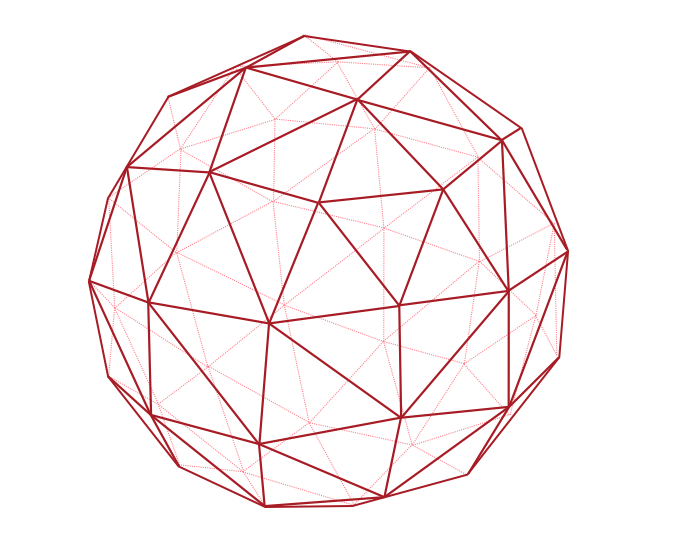}
    \caption{A triangulated sphere}
    \label{fig:triangulatedsphere}
\end{figure}

The complex model that we consider in our work is called the triangulation of a manifold in some dimension. As an example, Figure~\ref{fig:triangulatedsphere} illustrates a 2-d triangulated sphere, which consists of triangles (2-simplexes) being glued together. 
Using the standard terminology, reviewed in Appendix~\ref{sec: crashcoursehomology}, if we treat the configuration as a complex, then we have a set of $|S_0|$ points (0-simplexes), $|S_1|$ edges (1-simplexes), $|S_2|$ triangles (2-simplexes), and so on for higher-dimensional objects. 
\begin{figure}[H]
    \centering
    \begin{tikzpicture}[scale=2]
\begin{scope}[rotate=-45]
  \coordinate (v1) at (0,0);
  \coordinate (v2) at (1,1.732); 
  \coordinate (v3) at (2,0);
  \coordinate (v4) at (1,-1.732);

  \fill (v1) circle (1pt) node[below left] {$v_1$};
  \fill (v2) circle (1pt) node[above] {$v_2$};
  \fill (v3) circle (1pt) node[below right] {$v_3$};
  \fill (v4) circle (1pt) node[below] {$v_4$};

  \draw (v1) -- (v2) node[midway, left, blue] {$e_1$};
  \draw (v2) -- (v3) node[midway, right, blue] {$e_2$};
  \draw (v3) -- (v1) node[midway, below, blue] {$e_3$};

  \draw (v1) -- (v4) node[midway, below left, blue] {$e_4$};
  \draw (v4) -- (v3) node[midway, below right, blue] {$e_5$};

  \node[red] at (1,0.5) {$f_1$};
  \node[red] at (1,-0.9) {$f_2$};
  \end{scope}
\end{tikzpicture}
    \caption{A part of a 2-D triangulated manifold, which contains two triangles (or 2-simplex) glued together. From a graph perspective, we have a set of vertices $v_1,v_2,v_3,v_4$, edges (1-simplex) $e_1,e_2,e_3,e_4$, and faces (2-simplex) $f_1,f_2$. The above figure could be specified by some functions that describe the local connectivity between points/edges/faces as well as the inclusion relation between point-edge or edge-face.  }
    \label{fig:celltriangle}
\end{figure}

In the following, we first describe the setup for our computational input and then a high-level overview of the cohomology approach that we shall quantize into a quantum algorithm.
\subsection{Setup of the Input Model}
\label{sec: datainputmodel}
The computational input model, which is also the classical description of a complex that we rely on, is the mutual relation between simplexes within our complex. More specifically, for arbitrary $r$-simplex $\sigma_{r_i}$ and $r'$-simplex $\sigma_{r'_j}$, we are provided with a computable function $f$ indicating their adjacency (meaning that if $r=r'$, $\sigma_{r_i}$ and $\sigma_{r_j}$ share a common face), or inclusion (meaning that $\sigma_{r_i} \subset \sigma_{r'_j}$ if $r < r'$ or otherwise). In general, these conditions amount to the following:
\begin{align}
    f ( \sigma_r, \sigma_{r'}) = \begin{cases}
        1 \text{\ if $\big( \sigma_r \cap \sigma_{r'}\big) \neq \emptyset$}, \\
        0 \text{\ if $\big( \sigma_r \cap \sigma_{r'}\big) = \emptyset$}.
    \end{cases}
\end{align}
It is straightforward to see that $f$ encodes the classical knowledge of the complex. Alternatively, for each $r$, this classical description can be provided via two matrices. \revise{The first one $F_r$ is of size $|S_r| \times |S_r|$ with entries defined as following: $F_r(i,j) = 1$ if $f(\sigma_{r_i}, \sigma_{r_j}) = 1$ and 0 otherwise. The second one $G_r$ is of size $|S_r| \times |S_{r-1}|$, with entries $G_r(i,j) = 1$ if $f(\sigma_{r_i}, \sigma_{(r-1)_j}) =1$ and 0 otherwise. } One can see that for all $r$, the matrices $F_r$ and $G_r$ encode the locally mutual relation of building components of a complex, in a very analogous way to the adjacency matrix encoding pairwise connectivity within the graph theory context. Since each $r$-th simplex has $r+1$ faces, it is only adjacent to another $r$-th simplex via one of its faces, we can see that the matrix $F_r$ is ($r+1$)-sparse symmetric, and the matrix $G_r$ is row-wise ($r+1$)-sparse, column-wise 2-sparse, but not symmetric. As an example, shown in Figure~\ref{fig:celltriangle}, which features a portion of a 2-d triangulated manifold, we have that $F_2(1,2) = 1; F_1(1,2) = 1; F_1(1,3) = 0; G_2(1,1) = 1; G_2(3,1)=0$. 


The above description highlights \textit{the main difference between the computational input model of our work and previous works.} To emphasize, the setting in our work is more restrictive and informative. While prior work relies solely on the pairwise connectivity between vertices, our work requires the classical knowledge of the mutual relation between simplexes as input. Additionally, the simplexes in~\cite{lloyd2016quantum} are labeled combinatorially, e.g., using a bitstring, while we label/order simplexes by integers. We emphasize that our input model is inspired by what was concluded in~\cite{schmidhuber2022complexity}, that estimating Betti numbers is NP-hard in the general case, and that \textit{one needs to specify the complex in a different manner to potentially gain more speedup}. In our setting, a triangulated manifold model is the specification. Another motivation is to extend the reach of cohomology theory and, particularly, (discrete) differential geometry to the exciting area of data analysis.  

One can ask in what context does this model arise? Or equivalently, what can be a good description of the model, besides the set of vertices and pairwise connectivity? We recall from the above that there are two matrices $F_r$, which encodes the mutual relation between two $r$-simplexes, and $G_r$, which encodes the mutual relation between an $r$-simplex and an $(r-1)$-simplex. Because an $r$-simplex has $r+1$ faces, and it can only be adjacent (sharing a face) with at most one other $r$-simplex, any column of $F_r$ has $r+1$ non-zero entries having value $1$. As two $r$-simplexes can share at most one adjacent $r$-simplex, any two columns of $F_r$ have only one non-zero entry at the same location. 
For the matrix $G_r$, which is of size $|S_r^K| \times |S_{r-1}^K|$, the $i$-th row contains the information about which $(r-1)$-simplex belongs to the $i$-th $r$-simplex. Therefore, any row of $G_r$ has $r+1$ non-zero entries. Any arbitrary two $(r-1)$-simplexes share at most one $r$-simplex, and thus, any two columns of $G_r$ have at most one non-zero entry having the same row index.

The above description can be straightforwardly extended to lower-order simplexes. Essentially, these matrices $\{ G_r, F_r\}$ contain the local information of the triangulated manifold. The prescription of the triangulation model is very popular in the context of computational geometric processing~\cite{gu2003genus, gu2008computational, gu2020computational}, where a mesh typically represents a geometric object. Equivalently, we can frame our problem as follows: \\


\noindent
\textit{\textbf{Input Model:} Given matrices $\{ G_r, F_r \} $ having the properties as mentioned above, one treats the row indices of $G_r, F_r$ as $r$-simplexes, and column indices of $G_r$ as $(r-1)$-simplexes.} \\

\noindent
 \textit{\textbf{Objective:} What are the (normalized) Betti numbers $\beta_r$'s of the resulting configuration?   } \\

Apparently, the homology approach can be used to solve the above problem, as the locally mutual information between $r$-simplexes and $(r-1)$-simplexes can be used to define the boundary map. By analyzing the spectrum of this map, we can determine the Betti numbers. However, as we will see below, cohomology theory allows for an alternative solution with significant speed-up. 

We note that, as will be shown in Appendix~\ref{sec: elaboration}, another crucial input information we need is the angle (within a 2-simplex/triangle), area (of a 2-simplex/triangle), volume (of a 3-simplex), etc. For uniform triangulation, this information can be trivially obtained, e.g., the area of an equilateral triangle with an edge of length $a$ is $\frac{\sqrt{3}}{4}a^2$. For non-uniform triangulation, information about the non-uniform part needs to be supplied in addition to the matrices $G_r, F_r$ above. For example, we know the indexes of these non-uniform simplexes, and their corresponding area/volume, etc. However, as we mentioned, in practice, most common manifolds admit uniform or nearly uniform triangulation, e.g., via combinatorial/abstract triangulations (in this regard, we are having a topological triangulation, instead of strict geometrical triangulation). Exceptions include non-compact manifold, or manifold with singularities or exotic topology \cite{cheeger1984curvature, sharp2021geometry, attie2003surgery}. Throughout this work, for convenience, we assume uniformity/near uniformity. A discussion on how to extend our soon-to-be-outlined method to deal with nonuniformity will be provided in the appendix \ref{sec: elaboration}. In fact, the topological structure of underlying manifold does not depend on the uniformity, thus, this assumption is without loss of generality. Rather, it is aimed at simplifying the representation of certain matrices, as we will see shortly.

\subsection{Cohomology Approach and Our Key Result}
\label{sec: highleveloverview}
As mentioned above, the cohomology framework relies on a functional called a form, which assigns a real number to the corresponding simplexes. In Figs.~\ref{fig: 1form} and ~\ref{fig: 2form}, two examples are given to exemplify the concept of 1-form and 2-form. While a detailed review of cohomology theory is not possible and a brief one is in Appendix~\ref{sec: cohomologicalframework}, we point out a few key ideas in the following.  A 1-form $\omega^1$ maps an edge, or 1-simplex, to a real number, a 2-form $\omega^2$ maps a triangle, or 2-simplex, to a real number, and a higher-order form, say, $r$-form $\omega^r$, maps a higher-order $r$-simplex to some real number. The collection of $r$-forms constitutes a vector space $\Omega^r$ (which is dual to the chain group/space formed by $r$-simplexes, reviewed in Appendix~\ref{sec: crashcoursehomology}). As a fundamental result within the cohomology theory, or more precisely, differential geometry, this space can be decomposed into a direct sum of three subspaces, one of which is called the harmonic (sub)space $H^r_{\Delta}$. An element $\omega^r \in \Omega^r$ can therefore be expressed as 
$$\omega^r = \delta \Omega + d\eta + h^r,$$ 
where $h^r \in H^r_{\Delta}$ is called a \textit{harmonic form}. An important result underlying our method is that the dimension of this harmonic space $H^r_{\Delta}$ is equal to the $r$-th Betti number of the given triangulated manifold.  This gives rise to an alternative classical algorithm based on cohomology; see Algorithm~\ref{alg: bettinumber}. 
\begin{figure*}[htbp]
    \centering
    \begin{subfigure}[b]{0.45\textwidth}
    \centering
        \begin{tikzpicture}[scale = 1.5]
\draw (0,0.75) -- (1,-1) -- (-1,-1) -- cycle;
\node[below left] at (-1,-1) {$p_1$};
\node[below right] at (1,-1) {$p_2$};
\node[above] at (0,0.75) {$p_0$};
\draw (0,0.75) -- (1,-1) -- (1.7, 0.5) -- cycle;
\node[right] at (1.7,0.45) {$p_3$};
\draw (0,0.75) -- (-1,-1) -- (-1.7, 0.45) -- cycle;
\node[left] at (-1.7, 0.5) {$p_4$}; 
\draw (1,-1) -- (-1,-1) -- (0,-2.6) -- cycle;
\node[below] at (0,-2.6) {$p_5$};
\node at (-0.6, 0) { \textcolor{blue}{0.2}}; 
\node at (0.6, 0) { \textcolor{blue}{0.1}}; 
\node at (0, -1.2) { \textcolor{blue}{0.15}};
\node at (0.8, 0.7) { \textcolor{blue}{0.1}};
\node at (1.6, 0.0) {\textcolor{blue}{0.4}};
\node at (-0.8, 0.7) {\textcolor{blue}{0.05}};
\node at (-1.6, 0.0) {\textcolor{blue}{0.5}};
\node at (0.7, -1.8) {\textcolor{blue}{0.1} };
\node at (-0.7, -1.8) {\textcolor{blue}{0.05}};
\end{tikzpicture}
\caption{Illustration of 1-form: $\omega^1$ essentially assigns each edge, or 1-simplex, to a real number}
    \label{fig: 1form}
    \end{subfigure}
    \hfill
    \begin{subfigure}[b]{0.45\textwidth}
    \centering
        \begin{tikzpicture}[scale = 1.5]
\draw (0,0.75) -- (1,-1) -- (-1,-1) -- cycle;
\node[below left] at (-1,-1) {$p_1$};
\node[below right] at (1,-1) {$p_2$};
\node[above] at (0,0.75) {$p_0$};
\draw (0,0.75) -- (1,-1) -- (1.7, 0.5) -- cycle;
\node[right] at (1.7,0.45) {$p_3$};
\draw (0,0.75) -- (-1,-1) -- (-1.7, 0.45) -- cycle;
\node[left] at (-1.7, 0.5) {$p_4$}; 
\draw (1,-1) -- (-1,-1) -- (0,-2.6) -- cycle;
\node[below] at (0,-2.6) {$p_5$};
\node at (0,-0.2) {\textcolor{purple}{0.2}};
\node at (1.0,-0.0) {\textcolor{purple}{0.25}};
\node at (-1.0,-0.0) {\textcolor{purple}{0.1}};
\node at (0,-1.6) {\textcolor{purple}{0.15}};
\end{tikzpicture}
\caption{Illustration of 2-form: $\omega^2$ essentially assigns each triangle, or 2-simplex, to a real number}
    \label{fig: 2form}
    \end{subfigure}
    \caption{Illustration of the key difference between homology and cohomology. In homology, the main working objects are simplexes, as will be formally defined in Sec.~\ref{sec: crashcoursehomology}. A central recipe within homology theory is the chain group/space, for which the simplexes serve as the basis. The cohomology theory is based on the dual of such a chain group/space, called \textit{form space}, with the main working objects being linear functionals, or \textit{forms}, that map simplexes to real numbers. As illustrated above, a 1-form maps an edge, or 1-simplex, to a real value; a 2-form maps a triangle, or 2-simplex, to a real value. Higher-dimensional forms are further extended and defined for higher-dimensional objects in a similar manner.   }
\end{figure*}

Our quantum algorithm, as summarized in Algorithm~\ref{alg: quantumbettinumber}, is based on this link, and we aim to find the dimension of $\mathcal{H}^r$, which directly yields the $r$-th Betti number. Another crucial recipe is that, for a given $r$-form $\omega^r$ and a specification of triangulated manifold (via matrices $G$'s and $F$'s), there is a systematic procedure for finding $h^r$, e.g., $h^r = \omega^r - \delta \Omega - d\eta$. More specifically, to find $\Omega$ and $\eta$ for a given $\omega^r $, we need to solve a linear equation for each of them, respectively. This enables us to leverage recent advances in quantum algorithmic frameworks, such as the quantum linear solver~\cite{childs2017quantum, harrow2009quantum} and the quantum singular value transformation/block-encoding technique~\cite{gilyen2019quantum, low2017optimal, low2019hamiltonian}. We note that the differential operator $d$ and the codifferential operator $\delta$ can be built from the input of the above matrices $G$'s and $F$'s, supplemented with additional information on the areas/volumes of simplexes constituting the triangulation (will be shown in Appendix~\ref{sec: elaboration}). Therefore, we can employ the QSVT/block-encoding technique to obtain $\delta \Omega, d\eta$, and eventually $h^r = \omega^r - \delta \Omega - d\eta$. Once we succeed in drawing many elements $\in H^r_{\Delta}$, we can track the number of linearly independent terms, and then the dimension of $H^r_{\Delta}$ can be revealed. 
\begin{algorithm}[H]
    \begin{algorithmic}[1]
\Require A simplicial complex $K = \{ S_r^K \}$. 
\State Generate the randomized $|S_r^K|$ $r $-forms $\{\omega^r_i\}_{i=1}^{|S_r^K|} \in \Omega^r$\\
For all $i$, define $\overrightarrow{\omega_i^r}$ as a $|S_r^K|$-dimensional vector containing the action of $\omega_i^r$ on all r-simplexes $\sigma_{r_j}$, i.e., $\omega_i^r (\sigma_{r_j}) \in \mathbb{R}$.   For each $i =1,2,.., F$, do the following steps
\State Compute the coexact term $\overrightarrow{ \delta \Omega_i}$ --  which is a vector containing the action of $\delta \Omega_i$ on all $r$-simplexes
\State Compute the exact term $\overrightarrow{d\eta_i}$ -- which is a vector containing the action of $ d \eta_i$ on all $r$-simplexes
\State Find the harmonic form $\overrightarrow{h^r_i}  = \overrightarrow{\omega^r_i} - \overrightarrow{d\eta_i} - \overrightarrow{\delta\Omega_i } $ 
\State Arrange $\{ \overrightarrow{h^r_i} \}$ as a matrix $\mathbb{H}$. Find the number $\mathcal{K}$ of linearly independent harmonic forms among $\Gamma$ columns of $\mathbb{H}$, where $\Gamma$ is some integer (of our choice) and $ \Gamma \leq |S_r^K|$
\Ensure The $r $-th Betti number is $\mathcal{K}$.
    \end{algorithmic} 
\caption{Algorithm for Finding the $r$-th Betti Number}
\label{alg: bettinumber}
\end{algorithm}

Figure~\ref{fig: comparison} features a high-level comparison between our cohomology approach and others' homology approach for estimating Betti numbers. Our quantum algorithm, as summarized in Algo.~\ref{alg: quantumbettinumber}, is a direct translation of the above procedure into the quantum setting. For demonstration, we summarize our results for estimating the normalized $r$-th Betti number and the $r$-th Betti number as follows.
\begin{theorem}\label{theorem:rthBetti}
    Let $K$ be a simplicial complex with $n$ points, corresponding to the triangulation of an $n$-manifold. The r-th normalized Betti number $\beta_r/|S_r^K|$ can be estimated to additive accuracy $\epsilon$ with success probability $1- \xi$, by choosing $\Gamma = |S_r^K|$ in the Algo.~\ref{alg: bettinumber}, in time complexity
\begin{align*}
    \mathcal{O}\Big( \log\big( \frac{1}{\xi } \big) \frac{\log (|S_r^K| |S_{r+1}^K|)}{\sqrt{\epsilon}} r \log \frac{1}{\epsilon}   \Big).
\end{align*}
The $r$-th Betti number $\beta_r$ can be estimated to a multiplicative accuracy $\epsilon'$, with success probability $1-\xi$, by setting $\epsilon = \epsilon' \frac{\beta_r}{|\Gamma|}$, thus the complexity is 
\begin{align*}
    \mathcal{O}\Big( \log\big( \frac{1}{\xi}\big)\frac{ \log (|S_r^K| |S_{r+1}^K|)}{\sqrt{\epsilon'}}  \sqrt{\frac{ \Gamma}{\beta_r} } r \log \frac{1}{\epsilon}   \Big).
\end{align*}
\end{theorem}
\begin{figure*}[htbp]
    \centering
    \begin{tikzpicture}[scale=2, line join=round]
\coordinate (P0) at (-1.2,1);
\node[above] at (P0) {$p_0$};
\coordinate (P1) at (-2.0,0);
\node[left] at (P1) {$p_1$};
\coordinate (P2) at (-0.8,-0.2);
\node[below right] at (P2) {$p_2$};
\coordinate (P3) at (-0.8,0.4);
\node[right] at (P3) {$p_3$};
\coordinate (P4) at (-2.2, 1.0);
\node[above] at (P4) {$p_4$};
\coordinate (P5) at (-1.6, -0.5);
\node[below] at (P5) {$p_5$};
\coordinate (P6) at (-0.3, 0.8) ;
\node[right] at (P6) {$p_6$}; 
\coordinate (P7) at (-1.7, 0.8); 
\node at (P7) {$p_7$};
\draw[dashed, thick] (P0) -- (P1);
\draw[thick] (P2) -- (P0);
\draw[thick] (P1) -- (P2);
\draw[thick] (P0) -- (P3) -- (P2);
\draw[thick] (P1) -- (P5);
\draw[thick] (P2) -- (P5);
\draw[dashed, thick] (P5)--(P3);
\draw[dashed, thick] (P1) -- (P3);
\draw[thick] (P4)--(P1); 
\draw[thick] (P4)--(P0);
\draw[thick] (P4) -- (P2);
\draw[thick] (P6) -- (P0);
\draw[thick] (P6) -- (P2);
\draw[thick] (P6) -- (P3); 
\draw[dashed] (P7) -- (P0);
\draw[dashed] (P7) -- (P1);
\draw[dashed] (P7) -- (P3);
\node (A) at (1.3,1.2) {Encoding simplexes};
\node at (1.3, 1.0) {$\{\sigma_{r_i}\}_{i=1}^{|S_r^K|} \longrightarrow \{ \ket{\sigma_{r_i}} \in \mathbb{C}^{2^n} \}_{i=1}^{|S_r^K|}$};
\node at (1.3, 0.6) {Build a vector space};
\node at (1.3,0.4) {$ C_r^K =\big\{ \rm span \{ \ket{\sigma_{r_i}} \}_{i=1}^{|S_r^K|} \big\} $};
\node at (1.3, 0.0) {Define boundary maps $ \partial_r$};
\node at (1.3, -0.2) {and combinatorial Laplacian $\Delta_r $ };
\node at (1.3, -0.6) {r-th Betti number};
\node at (1.3,-0.8) { $\beta_r = \dim \rm Ker(\Delta_r)$ };
\node (B) at (-4.3,1.2) {Randomize $r$-forms};
\node at (-4.3, 1.0) { $\omega_1,\omega_2,..., \omega_{|S_r^K|}$}; 
\node at (-4.3, 0.6) {Build a dual vector space};
\node at (-4.3,0.4) {$\Omega^r =\big\{ \rm span \{ \omega_i \}_{i=1}^{|S_r^K|} \big\} $};
\node at (-4.3, 0.0) {Find a so-called harmonic space};
\node at (-4.3, -0.2) {$H^r_\Delta$ which is a subspace of $\Omega^r$  };
\node at (-4.3, -0.6) {r-th Betti number};
\node at (-4.3,-0.8) { $\beta_r = \dim \rm (H^r_\Delta)$ };
\node (C) at (-1.2, 1.5) {Simplicial complex K};
\draw[->] (-1.0,1.7) to[out = 30, in = 150 ] (A);
\draw[->] (-1.4,1.7) to[out = 150, in = 30 ] (B);
\node at (0.5, 1.9) {Homology };
\node at (-3.5, 1.9) {Cohomology};
\end{tikzpicture}
    \caption{A diagram capturing the essential steps and main flow of quantum algorithms for estimating Betti numbers using the cohomology and homology frameworks, respectively.}
    \label{fig: comparison}
\end{figure*}
In order to execute our quantum algorithm, we need a few recipes from block-encoding and the quantum singular value transformation framework~\cite{gilyen2019quantum}, for which we list a few relevant results in Appendix~\ref{sec: QSVT}. In the next sections, we will proceed to describe the LGZ quantum algorithm in greater detail and contrast it with our setting, before describing our quantum algorithm.  

A detailed overview of previous quantum algorithms for estimating Betti numbers, which are based on homology, is provided in Section~\ref{sec: existingquantumalgorithmsforTDA}. A detailed analysis of the best-performance regime and comparison between our cohomology approach versus the homology as well as the classical approach is provided in Section~\ref{sec: furtheranalysisanddiscussion}. Here, we provide two tables (i.e., Tables~\ref{tab: comparisonhomology} and~\ref{tab: comparisoncohomology}) summarizing the complexities (stated in terms of $n$ -- the number of data points), in the corresponding regimes, of the homology approach (previous works), cohomology approach (our work), thus demonstrating the advantage of our cohomology framework. 
\begin{table}[H]
    \centering
    \resizebox{0.5\textwidth}{!}{
    \begin{tabular}{|c|c|c|}
    \hline
      $\beta_r$   & $|S_r^K| \ll \binom{n}{r+1} $ & $|S_r^K| \approx \binom{n}{r+1}$  \\
         \hline
     $ \ll |S_r^K|$ & $\mathcal{O}\left( \rm poly(n) \right)$, c.b. $\mathcal{O}\left( \exp(n)\right) $   & $\mathcal{O}\left( \rm poly(n) \right)$, c.b. $\mathcal{O}\left( \exp(n)\right) $ \\
     \hline
     $ \approx |S_r^K|$  & $\mathcal{O}\left( \rm poly(n) \right)$, c.b. $\mathcal{O}\left( \exp(n)\right) $  & $\mathcal{O}(n^2)$\\
     \hline
    \end{tabular}}
    \caption{Summary of the best-known complexity of homology approach, e.g., LGZ algorithm and improved versions~\cite{lloyd2016quantum, schmidhuber2022complexity}. ``c.b.'' is the abbreviation of ``can be''. More specifically, when $r$ scales with $n$, the value of $ \binom{n}{r+1}$ can be $\mathcal{O}(\exp n)$. Otherwise, it is $\mathcal{O}\left(\rm poly (n) \right)$.  }
    \label{tab: comparisonhomology}
\end{table}
\begin{table}[H]
    \centering
    \begin{tabular}{|c|c|c|}
    \hline
      $\beta_r$   & $|S_r^K| \ll \binom{n}{r+1} $ & $|S_r^K| \approx \binom{n}{r+1}$  \\
         \hline
     $ \ll |S_r^K|$ & $ \mathcal{O}\left( \log^2 n \right) $   & $\mathcal{O}\left( \log^2 n \right)$, c.b. $\mathcal{O}\left( n\right) $ \\
     \hline
     $ \approx |S_r^K|$  & $\mathcal{O}( \log^2 n)$  &  $\mathcal{O}(n)$\\
     \hline
    \end{tabular}
    \caption{ Summary of the complexity of our cohomology approach, in appropriate regimes.  }
    \label{tab: comparisoncohomology}
\end{table}
From the above tables, we can see that in most cases our cohomology approach for estimating Betti numbers offers superpolynomial speed-up, except the case where $ \beta_r \approx |S_r^K|$ and $ |S_r^K| \approx \binom{n}{r+1}$ (clique-dense regime), where only a quadratic speed-up w.r.t. the homology approach is obtained. 

\section{Existing Quantum Algorithms for Betti Numbers and TDA}
\label{sec: existingquantumalgorithmsforTDA}

 We shall define the problem of estimating Betti numbers in the context of topological data analysis, followed by a discussion on related works regarding classical and quantum algorithms for estimating Betti numbers, including the LGZ algorithm. We then proceed to describe the intuition underlying our work and explicitly discuss the input model with which we will be working, with emphasis on the difference between our model and existing prior literature. 

In a typical setting of topological data analysis, we are provided with $n$ data points (which may be vectors in a high-dimensional space) and a suitable metric that defines the distance between them. Two points are connected if their pairwise distance is less than, say, a threshold $\bar{\epsilon}$. From such connections, a simplicial complex is then defined accordingly, by treating fully connected subgraphs as appropriate simplexes. For example, we treat a point as a 0-simplex, a subgraph consisting of two connected points as a 1-simplex, a subgraph of three points pairwise connected as a 2-simplex, and so on. Once the simplicial complex is constructed, a central goal is to estimate its Betti numbers. As mentioned in the previous section, these numbers define the topological structure of the given dataset, and thus of both fundamental and practical importance. Another aspect of TDA is tracking the topological feature across different scales. For example, for different thresholds $\bar{\epsilon_1},\bar{\epsilon_2}, ..., \bar{\epsilon_m} $, the connectivity can change, and the resulting complex might not be the same. Some ``holes'', as indicated by Betti numbers, might appear and disappear, suggesting some subtle structure of given data points. \\

\subsection{The Lloyd-Garnerone-Zanardi (LGZ) Quantum Algorithm}
\label{sec: relatedworks}

A quantum algorithm for estimating Betti numbers was first introduced in~\cite{lloyd2016quantum}, also referred to as the LGZ algorithm. Their core ideas consist of the following parts. 

\smallskip\noindent
\textbf{Step 1.}  First, encode a simplex into a quantum state. Let the given simplicial complex $K$ have dimension $n$, and $S_r^K$ denote the set of $ r$-simplexes in $K$. A simplex is then represented by an $n$-qubit string, where the value of bit `1' indicates the vertex at the corresponding position, as illustrated below.
\begin{center}
    \begin{tikzpicture}[scale=2, line join=round]
\coordinate (P0) at (-1.2,1);
\coordinate (P1) at (-2.0,0);
\coordinate (P2) at (-0.8,-0.2);
\coordinate (P3) at (-0.8,0.4);
\draw[thick] (P0) -- (P1) -- (P2) -- (P0);
\draw[thick] (P0) -- (P3) -- (P2);
\draw[dashed, thick] (P1) -- (P3);
\node[above] at (P0) {$p_0$};
\node[left] at (P1) {$p_1$};
\node[below right] at (P2) {$p_2$};
\node[right] at (P3) {$p_3$};
\node at (1.3,0.7) {$\ket{1111}$};
\node at (0.2,0.7) {$[p_0,p_1,p_2,p_3]$};
\draw[->] (0.7,0.7) -- (1.0, 0.7);
\node at (1.3,0.5) {$\ket{1110}$};
\node at (0.2,0.5) {$[p_0,p_1,p_2]$};
\draw[->] (0.7,0.5) -- (1.0, 0.5);
\node at (1.3,0.3) {$\ket{1101}$};
\node at (0.2,0.3) {$[p_0,p_1,p_3]$};
\draw[->] (0.7,0.3) -- (1.0, 0.3);
\node at (1.3,0.1) {$\ket{0111}$};
\node at (0.2,0.1) {$[p_1,p_2,p_3]$};
\draw[->] (0.7,0.1) -- (1.0, 0.1);
\node at (1.3, -0.1) {$\ket{1100}$};
\node at (0.2,-0.1) {$[p_0,p_1]$};
\draw[->] (0.7,-0.1) -- (1.0, -0.1);
\end{tikzpicture}
\end{center}

Under this encoding, the set of $r$-simplexes $S_r^K \equiv \{ \sigma_{r_i} \}$ is mapped to computational basis states $\{ \ket{\sigma_{r_i}}\}$ of the Hilbert space of $n$ qubits, i.e., $\mathbb{C}^{2^n}$.  The chain groups are then associated with a vector space spanned by corresponding simplexes, e.g., by a slight abuse of notation,  $C_r^K \equiv \rm span \{ \ket{\sigma_{r_i}} \}_{i=1}^{|S_r^K|} $. The boundary map $\partial_r$ is then a linear map from $C_r^K \longrightarrow C_{r-1}^K$. 

\smallskip\noindent
\textbf{Step 2.} The next step is to prepare a superposition of those $ \ket{\sigma_{r_i}}$ in which $\sigma_{r_i} \in K$, i.e.,
\begin{align}
    \ket{\sigma}_r =  \frac{1}{ \sqrt{|S_r^K|}} \sum_{i=1}^{|S_r^K|} \ket{\sigma_{r_i}}.
\end{align}
According to~\cite{lloyd2016quantum}, the connectivity (or distance) between points can be stored in a quantum random access memory (QRAM), and one can construct the following oracle 
\begin{align}
    O_r^K \ket{ \sigma_{r_i}} \ket{0} = \begin{cases}
        \ket{\sigma_{r_i}} \ket{1} \text{\ if $\sigma_{r_i} \in K$}, \\
         \ket{\sigma_{r_i}} \ket{0}  \text{\ otherwise},
    \end{cases}
    \label{eqn: oracle}
\end{align}
by using these QRAM-stored connectivity and performing a pairwise checking for all vertices of the candidate simplex $\sigma_{r_i}$. Thus, it results in a complexity $\mathcal{O}(r^2)$ -- where the complexity refers to the number of QRAM calls, and one can equivalently understand it as the circuit depth required. Given the above oracle, the state $\ket{\sigma}_r$ can be constructed by using a multi-solution version of Grover's algorithm.  

Next, it is desired to prepare the following mixed state:
\begin{align}
   \rho_r = \frac{1}{ |S_r^K|} \sum_{i=1}^{|S_r^K|} \ket{\sigma_{r_i}} \bra{\sigma_{r_i}}, 
\end{align}
which can be done by appending ancillas and use CNOT gates to transform $\ket{\sigma}_r \ket{000..0} \longrightarrow  \frac{1}{ \sqrt{|S_r^K|}} \sum_{i=1}^{|S_r^K|} \ket{\sigma_{r_i}}\ket{\sigma_{r_i}} $. Tracing out either the first or second register yields the desired mixed state. 

\smallskip\noindent
\textbf{Step 3.} As the next step, a so-called Dirac operator $\mathscr{D}_r$ is defined as:
\begin{align}
    \mathscr{D}_r = \begin{pmatrix}
        0 & \partial_{r-1} & 0 \\
        \partial^\dagger_{r-1} & 0 & \partial_r\\
        0 & \partial^\dagger_r & 0 
    \end{pmatrix},
\end{align}
and a fundamental result in homology theory states that $\dim \rm Ker(\mathscr{D}_r) = \beta_r$. Therefore, to find the Betti numbers, it is sufficient to find the dimension of the kernel space of $\mathscr{D}_r$. As the second part of the algorithm in~\cite{lloyd2016quantum}, the authors used the fact that the boundary operator $\partial_r$, and hence, the Dirac operator $\mathscr{D}_r$ is row/column computable, and, therefore, it is possible to apply the Hamiltonian simulation techniques~\cite{berry2007efficient, berry2012black,berry2015hamiltonian,berry2015simulating} to construct $\exp(-i \mathscr{D}_r t)$. 

\smallskip\noindent
\textbf{Step 4.} As the final step, the algorithm in~\cite{lloyd2016quantum} executes the quantum phase estimation algorithm with unitary $\exp(-i \mathscr{D}_rt)$ and input state $ \rho_r$. Then one measures the phase register, with the probability of measuring zeros (i.e., the kernel) given by $p_0= \frac{\beta_r}{|S_r^K|}$ -- a quantity often called the \textit{normalized r-th Betti number}. Such a probability $p_0$ can be estimated by either amplitude estimation or just direct sampling. 

\smallskip The above algorithm has motivated a few subsequent developments that seek to improve, as well as address some missing ingredients of, the LGZ algorithm~\cite{hayakawa2022quantum, ubaru2021quantum, berry2024analyzing, schmidhuber2022complexity, crichigno2024clique, gyurik2022towards}. In the following, we briefly summarize some major points and refer the interested readers to the original works for more details.
\subsection{Other related works}

\smallskip\noindent
$\bullet$ \revise{(Near-term friendliness)} The work in~\cite{ubaru2021quantum} aims to produce a near-term friendly quantum algorithm for TDA. They replace the use of Grover's algorithm within \textbf{Step 1} of the LGZ algorithm above by a technique called rejection sampling, which lowers the circuit depth. In addition, they show that the Dirac operator used in \textbf{Step 3} of the LGZ algorithm can be expressed as $\sum_i a_i + a_i^\dagger$, where $a_i,a_i^\dagger$ are fermionic annihilation and creation operators, which in turn can be represented by (a sum of) Pauli operators. Lastly, they replace the phase estimation algorithm in \textbf{Step 4} by the stochastic rank estimation technique~\cite{ubaru2016fast, ubaru2017fast}. 

Recently, the work of~\cite{berry2024analyzing} features another improved quantum algorithm for estimating Betti numbers. They introduce a new way to construct the oracle as described earlier in the LGZ algorithm, based on Toffoli gates, and prepare the state $\rho_r$ in \textbf{Step 2} of the LGZ algorithm above, based on an improved method for constructing a, namely, Dicke states. Then introduce a new kernel space's dimension estimation algorithm based on Chebyshev polynomials, which bypasses the quantum phase estimation algorithm. In particular, they explicitly provide a few instances of graphs where superpolynomial quantum advantage is possible. 

In a parallel attempt, the work \cite{mcardle2022streamlined} introduces a new strategy to encode simplexes into quantum state, that reduces the total number of qubits compared to the LGZ algorithm. For a given point cloud having $N$ data points, while the LGZ algorithm requires $N$ qubits for encoding purpose, the method in \cite{mcardle2022streamlined} requires $\log N$ qubits, which is exponentially more efficient.

\smallskip
$\bullet$ \revise{(Persistence)} The work in~\cite{hayakawa2022quantum,ameneyro2022quantum} explores the so-called persistent Betti numbers, which are more general than the usual Betti numbers defined above. We recall that previously, we have mentioned another aspect of TDA where the scale, or threshold $\bar{\epsilon}$, can be varying, and as a result, the complex configuration can be changed, which further induces a change in Betti numbers. A meaningful task, then, is to have a means to track this change, and a useful quantity for such a purpose is the persistent Betti numbers, which are those Betti numbers that remain the same upon the change of threshold. More formally, let $K$ and $L$ denotes two simplicial pairs at two thresholds $\bar{\epsilon_1} < \bar{\epsilon_2}$ (it is apparent that $K \subset L$), the $r$-th persistent homology group is defined as:
\begin{align}
    H_r^{K,L} = \frac{\rm Ker (\partial_r^K)}{ \rm Im (\partial_{r+1}^L) \cap \rm Ker (\partial_r^K)}.
\end{align}
The algorithm in~\cite{hayakawa2022quantum} assumes two oracles $O_r^K$ and $O_r^L$ for simplicial complexes $K$ and $L$, respectively, working in a similar manner to what appeared in Eq.~(\ref{eqn: oracle}). From these oracles, the author in~\cite{hayakawa2022quantum} showed that they can be used to construct the (block encoding of) the so-called \textit{persistent combinatorial Laplacian} $\Delta_r^{K,L}$. The dimension of its kernel space is equal to the $r$-th persistent Betti number, and finding such a dimension can be done using a few techniques from the block encoding framework~\cite{gilyen2019quantum}. 

\smallskip
$\bullet$ \revise{(Complexity-theoretic limitation)} The work of~\cite{schmidhuber2022complexity} explores the complexity-theoretic limit of topological data analysis. In particular, they show that, in a generic setting, where only data points and connectivity are provided, estimating Betti numbers exactly is $\#$P-hard, while determining whether $\beta_r \geq 0$ is NP-hard. These results have ruled out the possibility of exponential speedup in general. As discussed in~\cite{schmidhuber2022complexity}, the only room for large speedup is where the simplex needs to be directly specified, instead of constructing it from a mere description of the graph (as in \textbf{Step 1} and $\textbf{Step 2}$ of the LGZ algorithm above). 

At the same time, the works of~\cite{crichigno2024clique} and~\cite{gyurik2022towards} provide more complexity-theoretic results regarding quantum algorithm for TDA. More specifically, the result in~\cite{crichigno2024clique} reveals that deciding whether the combinatorial Laplacian $\Delta_r$ has a trivial or non-trivial kernel, or equivalently, determining whether $\beta_r \geq 0$ is QMA-hard. At the same time, the result of~\cite{gyurik2022towards} shows that estimating the normalized Betti numbers, or more generally, a so-called low-lying spectral density, is a DQC1-hard instance.

\subsection{Our perspective}
The above results have featured an exciting progress of quantum TDA, suggesting that the main problem itself is non-trivial and, in general, exponential quantum speedup in estimating Betti numbers is not possible. In order to gain speedup, as emphasized in~\cite{schmidhuber2022complexity} and~\cite{berry2024analyzing}, one needs to have \textit{appropriate input, e.g., a specified simplicial complex} instead of a graph with mere vertices and edges connecting them. Another fundamental aspect of the aforementioned works is that they all build on the homology theory. In the following, we take another route, based on the cohomology theory instead.  We shall see that, upon appropriate conditions, our approach can provide significant improvement over previous homological methods. To emphasize how the cohomology approach is advantageous (in terms of qubits required, and subsequently, running time) compared to the homology one, we recall \textbf{Step 1} of the LGZ algorithm above, in which a simplex is combinatorially encoded into the computational basis state. To encode $r$-simplexes, the required qubit number is at least $r$. On the other hand, within the cohomology framework, instead of working on the vector space spanned by simplexes, we work on the dual linear functional space, i.e., the space of mappings $\omega: C_r^K \longrightarrow \mathbb{R}$. 

As illustrated below,  we see that to encode the given simplex via the homology, it requires 4 qubits. 
\begin{center}
    \begin{tikzpicture}[scale=2, line join=round]
\coordinate (P0) at (-1.2,1);
\coordinate (P1) at (-2.0,0);
\coordinate (P2) at (-0.8,-0.2);
\coordinate (P3) at (-0.8,0.4);
\draw[thick] (P0) -- (P1) -- (P2) -- (P0);
\draw[thick] (P0) -- (P3) -- (P2);
\draw[dashed, thick] (P1) -- (P3);
\node[above] at (P0) {$p_0$};
\node[left] at (P1) {$p_1$};
\node[below right] at (P2) {$p_2$};
\node[right] at (P3) {$p_3$};
\node at (1.3,1.0) {$\ket{1110}$};
\node at (0.2,1.0) {$[p_0,p_1,p_2]$};
\draw[->] (0.7,1.0) -- (1.0, 1.0);
\node at (1.3,0.8) {$\ket{1101}$}; 
\node at (0.2,0.8) {$[p_0,p_1,p_3]$};
\draw[->] (0.7,0.8) -- (1.0, 0.8);
\node at (1.3,0.6) {$\ket{0111}$};
\node at (0.2,0.6) {$[p_1,p_2,p_3]$};
\draw[->] (0.7,0.6) -- (1.0, 0.6);
\node at (1.3, 1.2) {$\ket{1011}$};
\node at (0.2, 1.2) {$[p_0,p_2,p_3]$};
\draw[->] (0.7,1.2) -- (1.0, 1.2);
\node at (0.3,1.4) {Homology encoding};
\node at (0.2, 0.2) {$[p_0,p_2,p_3]$};
\node at (0.2,0.0) {$[p_0,p_1,p_2]$};
\node at (0.2, -0.2) {$[p_0,p_1,p_3]$};
\node at (0.2, -0.4) {$[p_1,p_2,p_3]$};
\node at (1.4, 0.2) {$\omega([p_0,p_2,p_3]) $};
\node at (0.4, 0.4) {Cohomology encoding}; 
\draw[->] (0.6,0.2) -- (0.85, 0.2);
\draw[->] (0.6,0.0) -- (0.85, 0.0);
\draw[->] (0.6,-0.2) -- (0.85, -0.2);
\draw[->] (0.6, -0.4) -- (0.85, -0.4);
\node at (1.4, 0.0) {$\omega([p_0,p_1,p_2]) $}; 
\node at (1.4, -0.2) {$\omega([p_0,p_1,p_3]) $}; 
\node at (1.4, -0.4) {$\omega([p_1,p_2,p_3]) $}; 
\end{tikzpicture}
\end{center}
For cohomology, if we organize these scalar values $\omega([p_0,p_2,p_3]) $, $\omega([p_0,p_1,p_2]) $, $\omega([p_0,p_1,p_3]) $, and $\omega([p_1,p_2,p_3]) $ inside a column vector, then its dimension is equal to 4, which can be stored using $2$ qubits. More generally, for a complex $K$ having $|S_r^K|$ number of $r$-simplexes, the homology method requires at least $|S_r^K| $ qubits, and the cohomology method requires $\log_2 |S_r^K|$ number of qubits, as it is sufficient to store a vector of dimension $|S_r^K|$. Figure~\ref{fig: comparison} illustrates schematically the main flow of our cohomology algorithm for estimating Betti numbers, compared to the previous works using homology.

\section{Quantum Algorithm for Estimating Betti Numbers Using Cohomology}
\label{sec: quantumalgorithm}
The classical algorithm~\ref{alg: bettinumber}, on which our quantum algorithm is based, can, in principle, be used to find all the Betti numbers. To explain our quantum algorithm, for simplicity, we first consider the triangulation of a 2-dimensional manifold (which means that it is composed of many 2-simplexes glued together; see Fig.~\ref{fig:triangulatedsphere}), and specifically describe how to find the first Betti number. Afterwards, we describe how to generalize to higher Betti numbers and higher-dimensional manifolds. As we shall show, \textit{the only difference with higher Betti numbers is the specific entries or coefficients of the corresponding linear equations}. In the following, we first give the definition of the key quantum framework of block encoding and QSVT. Then we proceed to quantize each step of the Algorithm~\ref{alg: bettinumber} one by one. 

\subsection{Block Encoding and Quantum Singular-Value Transformation Techniques}
\label{sec: summaryofnecessarytechniques}
Here, we define what the block encoding is. Once a matrix $A$ is block-encoded, one can apply various QSVT techniques, some of which are relevant to this work and are summarized in Appendix~\ref{sec: QSVT}.  

\begin{definition}[Block Encoding Unitary]~\cite{low2017optimal, low2019hamiltonian, gilyen2019quantum}
\label{def: blockencode} 
Let $A$ be some Hermitian matrix of size $N \times N$ whose matrix norm $|A| < 1$. Let a unitary $U$ have the following form:
\begin{align*}
    U = \begin{pmatrix}
       A & \cdot \\
       \cdot & \cdot \\
    \end{pmatrix}.
\end{align*}
Then $U$ is said to be an exact block encoding of matrix $A$.{The circuit complexity, e.g., depth, of $U$ is called the complexity of block encoding A}. Equivalently, we can write $U = \ket{ \bf{0}}\bra{ \bf{0}} \otimes A + (\cdots)$, where $\ket{\bf 0}$ refers to the ancillary system required for the block encoding purpose. In the case where the $U$ has the form $ U  =  \ket{ \bf{0}}\bra{ \bf{0}} \otimes \Tilde{A} + (\cdots) $, where $|| \Tilde{A} - A || \leq \epsilon$ (with $||.||$ being the matrix norm), then $U$ is said to be an $\epsilon$-approximated block encoding of $A$. Furthermore, the action of $U$ on some quantum state $\ket{\bf 0}\ket{\phi}$ is:
\begin{align}
    \label{eqn: action}
    U \ket{\bf 0}\ket{\phi} = \ket{\bf 0} A\ket{\phi} +  \ket{\rm Garbage},
\end{align}
where $\ket{\rm Garbage }$ is a redundant state that is orthogonal to $\ket{\bf 0} A\ket{\phi}$. 
\end{definition}
The above definition has multiple natural \textbf{corollaries}: 
\begin{itemize}
    \item First, an arbitrary unitary $U$ block encodes itself.
    \item Second, suppose that $A$ is block encoded by some matrix $U$, then $A$ can be block encoded in a larger matrix by simply adding any ancilla (supposed to have dimension $m$), then note that \,
 $\Ibb_m \otimes U$ contains $A$ in the top-left corner, which is the block encoding of $A$ again by definition. 
    \item Third, it is almost trivial to block encode the identity matrix of any dimension. For instance, we consider $\sigma_z \otimes \Ibb_m$ (for any $m$), which contains $\Ibb_m$ in the top-left corner. 
\end{itemize}
Given some block-encoded operators, there is a variety of arithmetic operations that we can perform on them, and we provide a few of them that we will use subsequently in Appendix~\ref{sec: QSVT}.

\subsection{Computing Coexact Terms $\delta \Omega$}
We now discuss details of key steps in Algorithm~\ref{alg: quantumbettinumber}, beginning with the first step. Let us recall the notation from Appendix~\ref{sec: crashcoursehomology} that the set of 0-simplexes, 1-simplexes and 2-simplexes as $S_0^K = \{ \sigma_{0_i} \}, S_1^K = \{ \sigma_{1_i}\}$, and $S_2^K = \{\sigma_{2_i}  \} $, respectively. Let $\omega^1 \in \Omega^1$ be some random 1-form, which means that $\omega^1 ( \sigma_{1_i}) \in \mathbb{R} $ for all $i= 1,2,..., |S_1^K|$.  
The first step in  Algorithm~\ref{alg: bettinumber} is to compute the coexact term $\delta \Omega$. From equation~(\ref{eqn: hodge}) we have
\begin{align}
    d\omega^1 (\sigma_{2_i} ) = d \delta \Omega (\sigma_{2_i} ),
    \label{eqn: 17}
\end{align}
where $\sigma_{2_i} $ is some 2-simplex. Roughly speaking, the action of $d\omega^1$ on the $i$-th 2-simplex $\sigma_{2_i}$ is $d\omega^1 (\sigma_{2_i}) = \sum_{j, \sigma_{1_j} \subseteq \sigma_{2_i}} \omega^1(\sigma_{1_j}) $. The action of $d \delta \Omega $ on $\sigma_{2_i}$ is a linear combination $d \delta \Omega (\sigma_{2_i} ) = \sum_{j, \sigma_{2,j} \cap \sigma_{2,i} \neq \emptyset }  A_{ij} \Omega (\sigma_{2_j} )$, where $A_{ij}$ is classically computable. A more detailed elaboration on operators $d, \delta$ can be found in Appendix~\ref{sec: elaboration} and \ref{sec: generalhodge}. Here we point out that as for each 2-simplex $\sigma_{2_i}$ we have: 
\begin{align}
     \sum_{j, \sigma_{2,j} \cap \sigma_{2,i} \neq \emptyset }  A_{ij} \Omega (\sigma_{2_j} ) =\sum_{j, \sigma_{1_j} \subseteq \sigma_{2_i}} \omega^1(\sigma_{1_j}). 
\end{align}
So by considering all 2-simplexes $\{ \sigma_{2_i} \}$, we obtain a large linear equation: 
\begin{align}
    A\cdot \Vec{\Omega} =   C\cdot \vec{\omega^1},
    \label{eqn: coexactequation}
\end{align}
where $A$ is a sparse matrix  $\in R^{|S_2^K| \times |S_2^K|}$ which encodes the linear action of $d \delta$, $\Vec{\Omega}$ is a column vector containing all $\{\Omega(\sigma_{2_i})\}_{i=1}^{|S_2^K|}$ and $C \in R^{|S_2^K| \times |S_1^K|}$, which encodes explicitly the operation $d$, and $\overrightarrow{\omega^1}$ is a vector of size $|S_1^K| \times 1$ whose entries contain the values of 1-form $\omega^1(\sigma_{1_i}) $ on all 1-simplexes. In order to find $\Vec{\Omega}$, in principle, we can use the quantum linear solver~\cite{harrow2009quantum}, e.g., $\Vec{\Omega} = A^{-1} C \ \overrightarrow{\omega^1}$. 

We remark that we need to find the coexact term $\overrightarrow{\delta \Omega}$, whose entries are the action of $\delta \Omega$ on 1-simplexes. As details will be provided in Sec.~\ref{sec: elaboration} (see Lemma \ref{lemma: discreteform} and its generalization, Lemma \ref{lemma: generalcodiffoperator}), we mention here that for $i$-th 1-simplex $\sigma_{1_i}$, we have $ \delta \Omega ( \sigma_{1_i}) = \sum_{j, \sigma_{1_i} \subseteq \sigma_{2_j} } P_{ij} \Omega(\sigma_{2_j})$, where $P_{ij}$ is classically computable. Therefore, the vector $\overrightarrow{\delta \Omega}$ can be given as $\overrightarrow{\delta \Omega} = P \cdot \Vec{\Omega}$. So we end up having the following:  
\begin{align}
\label{eqn: coexactdeltaomega}
    \overrightarrow{\delta \Omega} = P A^{-1} C \ \vec{\omega^1}.
\end{align}
As mentioned, in Appendix~\ref{sec: elaboration}, we explicitly show that all matrices $P$, $A$, and $C$ have sparsity $\mathcal{O}(1)$, and are classically entry-computable, meaning that for each row, we know the location of non-zero entries and their corresponding values. More specifically, we have the following lemma regarding the structure of $A,P,C$, with the derivation provided in \ref{sec: elaboration}.
\begin{lemma}[Appendix \ref{sec: elaboration}]
\label{lemma: ACP}
The matrices $A,C, P$ have the following structure, respectively:
\begin{itemize}
\item The matrix $A$ of Eqn.~\ref{eqn: coexactdeltaomega} has size $|S_2^K| \times |S_2^K|$ with $-6$ on the diagonals. For an $i$-th row, the off-diagonal elements of $A$ are 2 and their column indexes are those faces, or 2-simplexes that are adjacent to $\sigma_{2_i}$.  
    \item  The matrix $C$ in Eqn.~\ref{eqn: coexactdeltaomega} has size $|S_2^K|\times |S_1^K|$. For a given row $i$-th, the $C$ has 3 non-zero elements (which are 1), and their column positions are those edges, or 1-simplexes that are boundary of $\sigma_{2_i}$. 
    \item  The matrix P of Eqn.~\ref{eqn: coexactdeltaomega} has size $|S_1^K| \times |S_2^K|$. For a given row $i$-th, $P$ has 2 non-zero entries element, which are $2$ and $-2$. Their column indexes are those 2-simplexes, or triangles that include $\sigma_{1_i}$ as a face/ or a part of the boundary.
\end{itemize}
\end{lemma}
Therefore, the matrices $F_1,F_2,G_1,G_2$ described earlier, in Section \ref{sec: datainputmodel}, which encodes the classical description of the simplex, sufficiently provide classical description/knowledge of $A,C,P$. From here, there are two ways to block-encode them. First, by virtue of Lemma \ref{lemma: entrycomputablematrix} (second version), $P/||P||_F, A/||A||_F, C/||C||_F$ can be efficiently block-encoded. The Frobenius norm of $A$ is $||A||_F = \sqrt{40}|S_2^K| $. The Frobenius norm is $ \sqrt{3}| S_2^K|$.  The Frobenius norm $||P||_F$ is $ \sqrt{2} |S_1^K|$. Then Lemma \ref{lemma: scale} can be used to remove the factor $||P||_F, ||A||_F,||C||_F $ respectively, resulting in the block-encoding of $\frac{P}{\Lambda}, \frac{A}{\Lambda}, \frac{C}{\Lambda}$ where $\Lambda$ is the upper bound of the eigenvalues of $P,A,C$, which is assumed to be some constant. In fact, it can be seen from the lemma above that in each row or column of $A,C,P$, there are at most $\mathcal{O}(1)$ elements with values $\mathcal{O}(1)$, and thus by Gershgorin circle theorem, their maximum eigenvalues are $\Lambda = \mathcal{O}(1)$. Second, if we are provided with the oracle/black-box access (in a similar manner to sparse-access model in Hamiltonian simulation \cite{berry2007efficient,berry2012black,berry2015hamiltonian, harrow2009quantum})  to entries of $ A,C,P$, then Lemma \ref{lemma: As} can be used to block-encode $\frac{P}{s_p \Lambda}, \frac{A}{s_A\Lambda}, \frac{C}{s_C \Lambda}$ where $s_p,s_A,s_C$ are the sparsity of $P,A,C$. Then Lemma \ref{lemma: scale} can be applied to remove these sparsity parameters, resulting in the block-encoding of $ \frac{P}{\Lambda}, \frac{A}{\Lambda}, \frac{C}{\Lambda}$. We note that as also pointed out in Lemma \ref{lemma: ACP}, the sparsity of $A,C,P$ are all $\mathcal{O}(1)$. 

In the following, we proceed with the second model, which is the oracle/black-box access. The analysis for the first model can be done in a straightforward manner (the role of sparsity parameters $s_p,s_A,s_C$ will interchange with the Frobenius norm $||P||_F, ||A||_F, ||C||_F$). Recall that in the first step of Algorithm~\ref{alg: quantumbettinumber}, we need to randomize $|S_1^K|$ different 1-forms $\{\omega^1_i \}_{i=1}^{|S_1^K|}$. Since it can be randomized, we can pick a random $|S_1^K| \times |S_1^K|$ matrix $W$, and treat each column as a randomized 1-form $\omega^1_i$. For example, we can pick a random unitary of size $|S_1^K| \times |S_1^K| $, or of size $\mathscr{C} \times \mathscr{C}$ (where $\mathscr{C} \geq |S_1^K|$) and pay attention to its top-left block of size $ |S_1^K|\times |S_1^K|$ W, i.e., W is block-encoded. As $\frac{P}{\Lambda}, \frac{A}{\Lambda}, \frac{C}{\Lambda} $ can be block-encoded, one can first apply the Lemma~\ref{lemma: matrixinversion} to obtain the block encoding of $ \frac{A^{-1}}{\kappa_A}$ (where $\kappa_A$ is the condition number of $A$), and then Lemma~\ref{lemma: product} to construct the block encoding of 
$$\frac{1}{ \Lambda^2 \kappa_A } P A^{-1} C W \equiv \frac{1}{\mathscr{N}} P A^{-1} C W, $$ 
where $\kappa_A$ is the condition number of $A$ and we have defined $ \mathscr{N} \equiv  \Lambda^2 \kappa_A$ for brevity.  


The matrix $P A^{-1} C W$ contains $\overrightarrow{\delta \Omega}_i $ in its $i$-th column, where $ \overrightarrow{\delta \Omega}_i$ is a vector whose entries are the action of $\delta \Omega_i$ on all 1-simplexes. This completes the step 3 of Algorithm~\ref{alg: quantumbettinumber}.

\subsection{Computing exact term $d \eta$}
Now we proceed to the computation of the exact term $d\eta$. This step is very similar to what we had for $\delta \Omega$. From equation~(\ref{eqn: hodge}), we have: 

\begin{align}
    \delta \omega ( \sigma_{0_i}) = \delta d\eta (  \sigma_{0_i} ),
\end{align}
where $ \sigma_{0_i}$ is some 0-simplex, e.g., a vertex. As shown in Appendix~\ref{sec: elaboration}, the action $\delta \omega ( \sigma_{0_i})$ results in $\sum_{ j, \sigma_{0,i} \subseteq \sigma_{1,j}} D_{ij} \omega^1 (\sigma_{1_j})$  and $\delta d\eta (  \sigma_{1_i} ) = \sum_{j, \sigma_{0_j} \subseteq \sigma_{1_i}} K_{ij} \eta(\sigma_{0_j})  $. So we have
\begin{align}
    \sum_{j, \sigma_{0_j} \subseteq \sigma_{1_i}} K_{ij} \eta(\sigma_{0_j})  = \sum_{ j, \sigma_{0,i} \subseteq \sigma_{1,j}}D_{ij} \omega^1 (\sigma_{1_j}).
\end{align}
Once applied to all 0-simplexes, the above yields the following linear equation:
\begin{align}
    K\cdot \Vec{\eta} = D\cdot \vec{\omega^1}.
    \label{eqn: exact}
\end{align}

As will be shown in Appendix \ref{sec: elaboration}, the action of the exact term $d \eta$ on the $i$-th 1-simplex $\sigma_{1_i}$ is given by $d\eta (\sigma_{1_i}) = \sum_{j,  \sigma_{0_j}\subseteq \sigma_{1_i} } \eta( \sigma_{0_j} ) $. Therefore, the vector $\overrightarrow{d \eta}$ which contains the action of $d\eta$ on all 1-simplexes can be written as $\overrightarrow{ d \eta} = Q \overrightarrow{\eta}$ (thereby defining the matrix $Q$), which further gives 
\begin{align}
\label{eqn: exacttermdeta}
    \overrightarrow{d \eta} = Q \cdot K^{-1} \cdot D \cdot \overrightarrow{\omega^1}.
\end{align}

Similar to the context of Lemma \ref{lemma: ACP}, these matrices $Q, K, D$ are classically computable, as shown in Appendix~\ref{sec: elaboration}. In fact, they share very similar structure. Therefore, they (up to a scaling by their Frobenius norm) can be block-encoded via Lemma~\ref{lemma: entrycomputablematrix}. Following the same procedure, we can obtain the block encoding of 
$$\frac{1}{ \Lambda^2  \kappa_K} Q K^{-1} D W \equiv \frac{1}{\mathscr{M}} Q K^{-1} D W,$$ 
where we define $\mathscr{M} \equiv \Lambda^2 \kappa_K$. This completes step 5 of Algorithm \ref{alg: quantumbettinumber}.

\subsection{Computing the Harmonic Form}
As in the 6th step in Algorithm~\ref{alg: bettinumber}, we need to find the harmonic forms. One observes that, for a given $i$, $W$ contains $\overrightarrow{w^1_i}$ in its $i$-th column. Therefore, as $\overrightarrow{h^1_i}  = \overrightarrow{\omega^1_i} - \overrightarrow{d\eta_i} - \overrightarrow{\delta\Omega_i } $ , we have that: 
\begin{align}
    \mathbb{H} = W -  P A^{-1} C W -  Q K^{-1} D  W.
\end{align}
In order to obtain $\mathbb{H}$, one can then first use Lemma~\ref{lemma: scale} to obtain the following block-encoded operators:
\begin{align}
  W &\longrightarrow \frac{1}{ \mathscr{M}\mathscr{N}}W, \\
  \frac{1}{\mathscr{M}} Q K^{-1} D W &\longrightarrow \frac{1}{\mathscr{M}\mathscr{N}} Q K^{-1} D W,\\
   \frac{1}{\mathscr{N}} P A^{-1} C W &\longrightarrow \frac{1}{\mathscr{M}\mathscr{N}} P A^{-1} C W.
\end{align}
 Then we can use Lemma~\ref{lemma: sumencoding} to obtain the block encoding of 
 \begin{align}
     \frac{1}{ \mathscr{M}\mathscr{N}}\Big(  W - Q K^{-1} D W - P A^{-1} C W \Big) =  \frac{1}{ \mathscr{M}\mathscr{N}} \mathbb{H}.
 \end{align}
The last step, e.g., Step 7 of Algo \ref{alg: quantumbettinumber} is finding the rank of $\mathbb{H}$,  which is also the $r$-th Betti number $\beta_r$ (with $r=1$ here). In fact, there are multiple ways to do so. As the first solution, one can use the quantum phase estimation algorithm, in a similar manner to \textbf{Step 4} of the LGZ algorithm summarized in Section~\ref{sec: relatedworks}, to find the ratio $ \frac{\dim  \rm Ker(\mathbb{H})}{|S_1^K|}$. The quantity $\frac{\dim  \rm Rank(\mathbb{H})}{|S_1^K| }$ can be estimated via $1-  \frac{\dim  \rm Ker(\mathbb{H})}{|S_1^K| }$. Another solution is to use the so-called block measurement, as provided in~\cite{hayakawa2022quantum}. Roughly speaking, this method works by applying the quantum singular value transformation technique from~\cite{gilyen2019quantum} to the block-encoded operator $\varpropto \mathbb{H}$, transforming it into $\sum_{i, \lambda_i =0} \ket{\psi_i}\bra{\psi_i}$, where $\ket{\psi_i},\lambda_i$ are the eigenvector-eigenvalue pair of $\mathbb{H}$. This transformation step essentially ``projects'' $\mathbb{H}$ into its kernel space, and then estimates the ratio $\frac{\dim  \rm Ker(\mathbb{H})}{|S_1^K| }$ by applying the block encoding of $ \sum_{i, \lambda_i \leq \delta} \ket{\psi_i}\bra{\psi_i}$ to some easy-to-prepare state, and then performs the measurement on an ancilla qubit to extract the ratio, which is also the probability of obtaining the outcome $1$ in the ancilla. A more direct approach to estimate $\frac{\dim  \rm Rank(\mathbb{H})}{|S_1^K| }$ is given in~\cite{ubaru2021quantum}, based on a technique called stochastic rank estimation~\cite{ubaru2016fast,ubaru2017fast}. In the Appendix \ref{sec: stochasticrankestimation}, we provide a review of stochastic rank estimation and describe our preferred way for rank estimation, which is summarized in the following lemma:
\begin{lemma}[Appendix \ref{sec: stochasticrankestimation}]
\label{lemma: traceestimation}
    Given a block-encoding of a matrix $A$ of size $N\times N$, with a premise that $|A| \in (\frac{1}{\kappa_A},1)$ where $|A|$ is the operator norm of $A$. Then with success probability $1- \xi$, the normalized rank $\frac{\rm rank A}{N}$ can be estimated to a precision $\epsilon$ using a quantum algorithm of complexity $\mathcal{O}\left(  T_A\kappa_A  \frac{1}{\sqrt{\epsilon}} \log (\frac{1}{\epsilon}) \log \frac{1}{\xi}\right) $ where $T_A$ is the complexity for block-encoding $A$. 
\end{lemma}


Thus, we have completed all steps in our quantum algorithm for estimating the first Betti number $\beta_1$, as summarized in Algorithm~\ref{alg: quantumbettinumber}.  
\begin{algorithm}[H]
    \begin{algorithmic}[1]
\Require A simplicial complex $K$ with classical description encoded in matrices $F$s and $G$s. 
\State Block-encode a random matrix $W$ of size $|S_1^K| \times |S_1^K|$
\State Block-encode matrices $\frac{P}{\Lambda}, \frac{A}{\Lambda}, \frac{C}{\Lambda} $. 
\State Use Lemma~\ref{lemma: product}, Lemma~\ref{lemma: matrixinversion} to construct the block encoding of $\varpropto P A^{-1} C W$
\State Block-encode matrix $Q/||Q||_F, K/||K||_F, D/||D||_F$ (defined in Eqn.~\ref{eqn: exacttermdeta}) 
\State Use Lemma~\ref{lemma: product}, Lemma~\ref{lemma: matrixinversion} to construct the block encoding of $\varpropto Q K^{-1} D W$
\State Use Lemma~\ref{lemma: sumencoding} to construct the block encoding of $\varpropto  W -  R_1 A^{-1} C W -  R_2 K^{-1} D  W \equiv \mathbb{H} $
\State Choose an integers $0 < \Gamma  \leq |S_1^K|$. Consider the sub-matrix of $\mathbb{H}$ of size $\Gamma \times \Gamma$ (top left corner), denoted as $\mathbb{H}_\Gamma$. Use Lemma \ref{lemma: traceestimation} to estimate $\rm rank(\mathbb{H}_\Gamma)/\Gamma$ to accuracy $\epsilon$ (see subsequent discussion for the role of $F$).
\Ensure The ratio $\beta_1/\Gamma$.
    \end{algorithmic} 
\caption{Quantum Algorithm for Finding the 1st Betti Number of a Triangulated Manifold}
\label{alg: quantumbettinumber}
\end{algorithm}

\subsection{Complexity analysis}
\label{sec: complexityanalysis}
\revise{
To see the complexity, we analyze the algorithm above step by step:
\begin{enumerate}
    \item The first step has $\mathcal{O}(1)$ time, as we can pick a random unitary (no need to be Haar random) from short-depth circuits. 
    \item The second step first uses Lemma~\ref{lemma: As}, followed by the Lemma \ref{lemma: amp_amp} to remove the sparsity parameters. As we pointed out in Lemma \ref{lemma: ACP}, the sparsity of $P,A,C$ is $\mathcal{O}(1)$; therefore the complexity for block-encoding each $\frac{P}{\Lambda}, \frac{A}{\Lambda}, \frac{C}{\Lambda} $ is $\mathcal{O}\big( \log |S_1^K|  \big) ,\mathcal{O}\big( \log |S_2^K|  \big), \mathcal{O}\big( \log |S_2^K|  \big) $, respectively. 
    \item The third step involves an inversion of $A/\Lambda$,  which has complexity $\mathcal{O}\big( \log(\frac{1}{\epsilon})  \kappa_A \log |S_1^K|  \big) $, followed by using Lemma \ref{lemma: product} to construct the block-encoding of products of $P,A^{-1},C ,W$. So the total complexity for this step is $\mathcal{O}\left( \log(\frac{1}{\epsilon})  \kappa_A \log (|S_1^K| |S_2^K|) \right) $.
    \item Similarly, steps 4 and 5 have similar complexity to steps 2 and 3. The complexity of step 6 is the sum of the complexity of previous steps, thus resulting in a total complexity $\mathcal{O}\Big( \log(\frac{1}{\epsilon})  \kappa_A  \log (|S_1^K| |S_2^K| \Big) $. 
    \item The last step uses the rank estimation method Lemma \ref{lemma: traceestimation}, and assuming that the minimum eigenvalue of $\mathbb{H}$ is some constant, the final total complexity can be shown to be $\mathcal{O}\big( \frac{1}{\sqrt{\epsilon}} \log(\frac{1}{\epsilon})  \kappa_A   \log (|S_1^K| |S_2^K| \big) $. 
\end{enumerate} }

To summarize, we state our main result in the following theorem.

\begin{theorem}[Estimating the 1st Betti number]
\label{theorem: mainresult}
Let $K$ be a simplicial complex with $n$ points, corresponding to the triangulation of a 2-manifold. By choosing $\Gamma= |S_1^K|$ in the above algorithm, the 1st normalized Betti number $\beta_1/|S_1^K|$ can be estimated to additive accuracy $\epsilon$, with success probability $1-\xi$, in time  
\begin{align*}
    \mathcal{O}\Big( \log\big( \frac{1}{\xi} \big) \frac{\log(|S_1^K| |S_2^K|)}{\sqrt{\epsilon}} \log(\frac{1}{\epsilon}) \Big).
\end{align*}
The 1st Betti number $\beta_1$ can be estimated to a multiplicative accuracy $\epsilon'$, with success probability $1-\xi$,  by choosing arbitrary $\Gamma$ and setting $\epsilon = \epsilon' \frac{\beta_1}{\Gamma}$, thus the complexity is 
\begin{align*}
    \mathcal{O}\Big( \log\big( \frac{1}{\xi} \big)\frac{\log(|S_1^K| |S_2^K|)}{\sqrt{\epsilon'}} \sqrt{ \frac{ \Gamma}{\beta_1}}  \log(\frac{1}{\epsilon'})  \Big).
\end{align*}
\end{theorem}

One may wonder about the role of $\Gamma$ in the estimation of the 1st Betti number $\beta_1$. If we choose $\Gamma = |S_1^K|$ as in the normalized Betti number case, then the complexity would scale as $\frac{|S_1^K|}{\beta_1}$. Thus, the algorithm is only efficient when $\beta_1 \approx |S_1^K|$, i.e., relatively large Betti number. On the other hand, if $\beta_1$ is low (e.g., low genus surfaces), then in principle, the value of $\Gamma$ does not need to be large. Intuitively, this comes from the fact that our algorithm is based on the Hodge decomposition (see Theorem~\ref{theorem: hodgedecompo}), and eventually we obtain the Betti numbers by finding the dimension of the so-called harmonic space. Our strategy begins by randomly drawing vectors, or 1-forms $\{ \omega^1_i\}$, then deforming them into the corresponding harmonic part. The dimension of the harmonic space is estimated by finding the \textit{maximal number} of linearly independent vectors drawn from such a space. If the Betti numbers are low, then the dimension of the harmonic space is low. Thus, in principle, we do not need to take too many elements from such a space to find the dimension. This property allows us to choose a smaller value of $\Gamma$ than $|S_1^K|$, thereby reducing the computational complexity. Thus, our method can be effective with respect to low Betti numbers, which is opposite to the LGZ algorithms (and related versions), as they are only efficient in the clique-dense regime where the Betti numbers $\beta_r$'s are large. 

Now we move on to generalize our framework to higher Betti-number constructions.

\subsection{Generalization to Different Betti Numbers and Higher Dimensional Manifolds}
\label{sec: generalization}
Given a triangulated 2-manifold, aside from the first, there are also the zeroth and the second Betti numbers $\beta_0, \beta_2$. The zeroth Betti number is always one here, as we already assume that the given manifold is a connected graph that is triangulated. We have presented a quantum algorithm for the first Betti number. In order to compute $\beta_2$, we first randomize a different 2-form $\omega^2$. We then deform them to the harmonic form by using the Hodge decomposition~(\ref{theorem: hodgedecompo}). The difference is that for a 2-manifold, there is no higher than 2-simplexes; therefore, there will be no co-exact term. The decomposition is thus
\begin{align}
    \omega^2= d\Omega + h^2,
\end{align}
where $\Omega$ is a 1-form, and $h^2$ is the corresponding harmonic 2-form of $\Omega$. We then follow the same procedure as we did for calculating $\beta_1$. \\

The next important and somewhat subtle point is about higher-dimensional cases. As we have emphasized from the beginning, the underlying groundwork of our method is the discrete Hodge theory. In order to apply this, for example, using Eqn.~\ref{theorem: hodgedecompo}, we need to be able to formulate the operator $d \delta$ for all forms (see Eqn.~\ref{eqn: coexactequation}). This formula depends on the dimension of the given manifold. The main reason is that, while the operator $d$ is easily formulated in terms of a matrix given the simplex, the formulation of the operation $\delta$ relies on the Poincar\'e duality, which uses the concept of \textit{dual cell}. 

\begin{figure}[htbp]
    \centering
    \includegraphics[width = 0.4\textwidth]{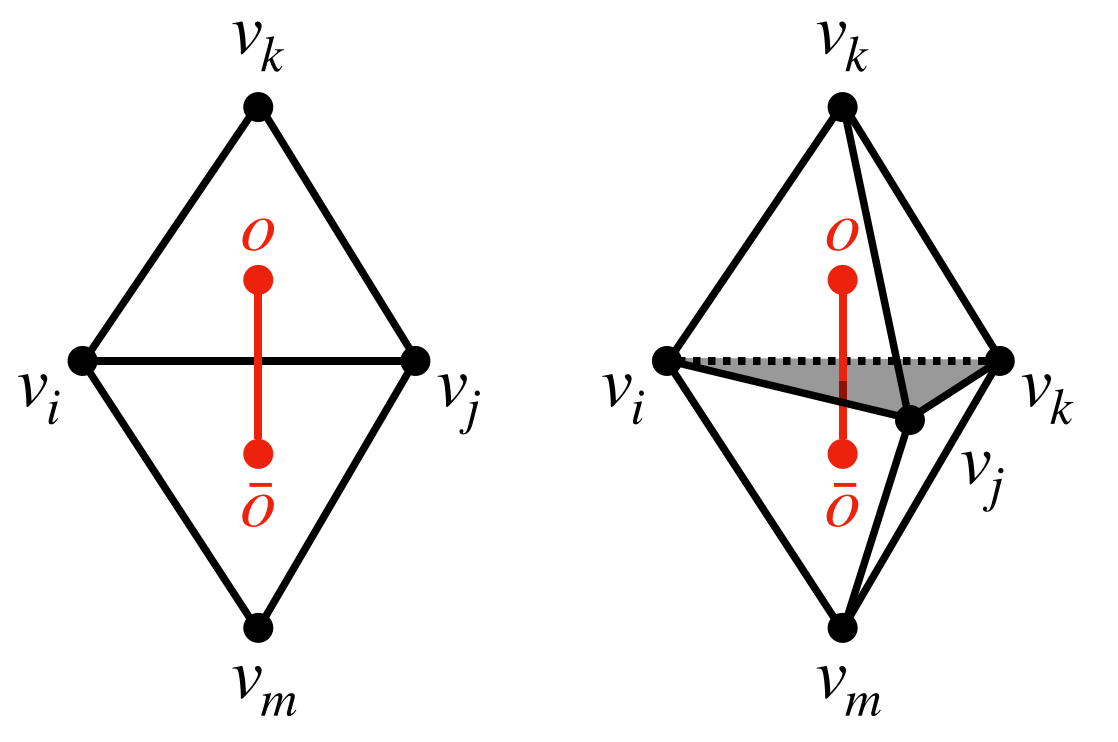}
    \caption{Dual cell complexes in 2 and 3 dimensions. Left figure: In a triangulated 2-manifold, the dual to 2-simplex $[v_i,v_j,v_k]$ is a point o. The dual to an edge $[v_i,v_j]$ is an edge $o\Bar{o}$. Right figure: In a triangulated 3-manifold, the dual to 3-simplex $[v_i,v_j,v_k,v_m]$ is a point o. The dual to a face (2-simplex) $[v_i,v_j,v_k]$ is an edge $o\Bar{o}$.   }
    \label{fig: dualcell}
\end{figure}

A formal definition of dual cell and its construction on a general triangulation of arbitrary dimension will be provided in Lemma \ref{lemma: poincaredualityconstruction} (see Appendix \ref{sec: generalhodge}, \ref{sec: poincaredualcell}). For a triangulated 2-manifold, a dual cell to an edge (1-simplex) is an edge. However, for a triangulated 3-manifold, a dual cell to an edge is no longer an edge but a face (2-simplex). It means that even for the same first Betti number $\beta_1$, the matrix coefficients (see Eqn.~\ref{eqn: coexactequation}) are different. Fortunately, the computation procedure is the same, as it still relies on deforming a given $k$-form to its harmonic part. More concretely, to estimate the $r$-th Betti number, we begin with a random $r$-form $\omega^r$. Then, according to the Hodge decomposition, we have $\omega^r = \delta\Omega + d\eta  + h^r$. To find $\Omega$ (which is a $(r+1)$-form), we have that for a $(r+1)$-simplex $\sigma_{(r+1)_i}$, $d\omega^r (\sigma_{(r+1)_i}) = d \delta \Omega (\sigma_{(r+1)_i}) $, which is analogous to Eqn.~\ref{eqn: exact}. The left-hand term turns out to be a linear combination $ \sum_{j, \sigma_{r_i} \subseteq \sigma_{(r+1)_i} } \omega^r (\sigma_{r_i})$. The right-hand term turns out to be a linear combination (with coefficients appropriately determined, e.g., see Appendix~\ref{sec: generalhodge}) of $ \{  \Omega( \sigma_{(r+1)_i}  \} $. Thus, we obtain a linear system 
\begin{align}
\label{eqn: generallinearequation}
    A^r \overrightarrow{\Omega} = C^r \overrightarrow{ \omega^r}
\end{align}
Eventually, the vector $\overrightarrow{\delta \Omega} = P^r (A^r)^{-1} C^r \overrightarrow{ \omega^r}$ is obtained in a similar manner to Eqn.~\ref{eqn: coexactequation} and Eqn.~\ref{eqn: coexactdeltaomega}. The biggest difference between this linear system and the one in Eqn.~\ref{eqn: coexactdeltaomega} is reflected in the following lemma:
\begin{lemma}
\label{lemma: generallinearequation}
The sparsity of $ A^r, C^r, P^r$ is $\mathcal{O}(r)$, and their size is $|S_r^K| \times |S_r^K|$. In addition, their Frobenius norm is $\mathcal{O}\left(  \sqrt{r} |S_r^K| \right)$    
\end{lemma}
Therefore, the complexity for block-encoding them, per Lemma~\ref{lemma: As} (plus Lemma \ref{lemma: amp_amp} afterward) is $\mathcal{O}\big( r \log (|S_r^K| |S_{r+1}^K|)  \big)$. Similarly, one proceeds to find the exact term in the decomposition $\omega^r = \delta\Omega + d\eta  + h^r $ by solving $ \delta \omega^r (\sigma_{(r-1)_i}) = \delta d \eta (\sigma_{(r-1)_i} ) $ for all $(r-1)$-simplexes $ \{ (\sigma_{(r-1)_i}  \} $, in an analogous manner to Eqn.~\ref{eqn: exact}. Eventually, the procedure results in $ \Vec{ d\eta } = Q^r \cdot (K^r)^{-1} \cdot D^r \Vec{\omega^r} $.

Then one uses the same procedure as in Algorithm~\ref{alg: quantumbettinumber} to find the $r$-th Betti numbers.  We can proceed with the computation of arbitrary Betti numbers and in an arbitrary dimensional manifold in exactly the same way as we did in the above algorithm, provided the matrix form of $\delta$ (and hence $d\delta$) is given. Despite that there is no exact expression for $\delta$, its matrix form can be efficiently calculated given the triangulation of the manifold (see Appendix~\ref{sec: generalhodge}). 
Our result for estimating $r$-th Betti number was already stated in Theorem~\ref{theorem:rthBetti}.

\section{Further Analysis and Discussion}
\label{sec: furtheranalysisanddiscussion}
 To make a comparison, we recall, summarize, and remark on some aspects of existing results, including ours, regarding the calculation of the (normalized) Betti numbers.

\smallskip
\noindent 
\textbf{Quantum Homology Approach.}  As reviewed in Sec.~\ref{sec: relatedworks}, previous efforts in quantum topological data analysis were primarily based on the homology theory. As also mentioned in~\cite{schmidhuber2022complexity}, given a simplicial complex $K$ having $n$ points and an oracle that encodes pairwise connectivity, the best running time of the improved-LGZ algorithm to estimate the $r$-th normalized Betti number to precision $\epsilon$ is
\begin{align*}
    \mathcal{O} \left( \Big(  n^2 \sqrt{ \frac{ \binom{n}{r+1}  }{|S_r^K|}} +   n\kappa  \Big) \frac{1}{\epsilon}\right),
\end{align*}
where $\kappa$ condition number of the Dirac operator. The complexity for estimating $r$-th Betti number $\beta_r$ to multiplicative accuracy $\epsilon'$ is
\begin{align*}
    \mathcal{O} \left( \Big(  n^2 \sqrt{ \frac{ \binom{n}{r+1}  }{\beta_r }} +   n\kappa \sqrt{\frac{|S_r^K|}{\beta_r} }  \Big) \frac{1}{\epsilon'}\right).
\end{align*}

\smallskip
\noindent
\textbf{Cohomology Approach.} Our cohomology method achieves the following time complexity for estimating the $r$-th normalized  Betti number to accuracy $\epsilon$
\begin{align*}
    \mathcal{O}\Big(\frac{\log(|S_r^K| |S_{r+1}^K|)}{ \sqrt{\epsilon}} r \log \frac{1}{\epsilon}   \Big).
\end{align*}
and for estimating the $r$-th Betti number to multiplicative accuracy $\epsilon'$
\begin{align*}
    \mathcal{O}\Big(\frac{\log(|S_r^K| |S_{r+1}^K|)}{\sqrt{\epsilon'}} \sqrt{\frac{ \Gamma }{\beta_r} } r \log \frac{1}{\epsilon'}    \Big),
\end{align*}
where $\Gamma \leq |S_r^K|$ is some integer of our choice. \\

\smallskip
\noindent
\textbf{Classical Homology Approach.} Additionally, we point out that the classical homology approach on which our quantum algorithm is based can (and has been) carried out on classical computers. The basic steps within our method are matrix multiplication, inversion, subtraction, as well as finding the rank of matrices of size $|S_r^K| \times |S_r^K|$. The most dominant classical time step is apparently finding the rank of the final dense matrix (which was $\mathbb{H}$ in section~\ref{sec: quantumalgorithm}). A straightforward classical method is the Gaussian elimination, which has a running time $\mathcal{O}(|S_r^K|^3)$. For 2-dimensional manifolds, there are more efficient algorithms \cite{mohar2001graphs, erickson2005greedy, chambers2009homology}, which have complexity $\mathcal{O}\left( n + |S_1^K| \right)$ in estimating the first Betti number $\beta_1$. Recently, the work \cite{apers2023simple} proposed a new approach for estimating Betti numbers.  \\

\noindent
\textbf{Comparison.} Let us now make some comparisons in respective regimes to demonstrate the advantage of our cohomology approach compared to previous quantum algorithms, which were built upon homology approach.

\smallskip\noindent
$\star$ In the clique-sparse regime, i.e., $|S_r^K| \ll \binom{n}{r+1} $ and two different limits:
\begin{itemize}
    \item Relatively large Betti numbers ($ \beta_r \approx |S_r^K|$) : the complexity of the homology approach (LGZ algorithm) is  $\mathcal{O}\Big( n^2\sqrt{\binom{n}{r+1}} + n \kappa \Big)$ for both normalized Betti number $\beta_r/|S_r^K|$ and Betti number $\beta_r$. At the same time, our cohomology approach has complexity $\mathcal{O}\big( r \log(|S_r^K| |S_{r+1}^K|) \big)$ in estimating the $r$-th normalized Betti number $\frac{\beta_r}{|S_r^K|}$. As $|S_r^K| \ll \binom{n}{r+1}$, our cohomological algorithm thus yields exponential speedup compared to that of the LGZ algorithm. For estimating Betti number $\beta_r$ up to a chosen multiplicative accuracy, we need to choose $\Gamma \approx |S_r^K|$, so the fraction $\frac{\Gamma}{\beta_r} \in \mathcal{O}(1)$, resulting in the complexity being $\mathcal{O}\left(   r \log(|S_r^K| |S_{r+1}^K|) \right)$. In this case, our cohomology approach still offers exponential speedup.
    \item Relatively small Betti numbers $(\beta_r \ll |S_r^K|$):  For relatively small $\beta_r$, e.g., $\beta_r \ll |S_r^K|$, the complexity of LGZ algorithm for estimating normalized Betti numbers and Betti numbers is (we are ignoring error tolerance dependence), respectively,
     $$ \mathcal{O}\left(  n^2 \sqrt{ \binom{n}{r+1} } + n\kappa    \right), $$
     $$ \mathcal{O}\left( n^2 \sqrt{\binom{n}{r+1}} +n \kappa \sqrt{|S_r^K|}  \right). $$
     In this case, we can choose $\Gamma$ to be considerably smaller than $|S_r^K|$ (still greater than $\beta_r$), yet the ratio $ \frac{\Gamma}{\beta_r} \in \mathcal{O}(1)$. So our complexity for estimating $r$th normalized Betti number and Betti number is the same, which is:
     $$ \mathcal{O}\left( r \log\left( |S_r^K||S_{r+1}^K| \right)  \right).$$
     As $|S_r^K|\ll \binom{n}{r+1}$, our complexity above achieves exponential speed-up compared to the homology approach, and also the classical approach. 
\end{itemize}

\smallskip\noindent
$\star$ In the clique-dense regime, i.e., $|S_r^K| \approx \binom{n}{r+1}$, and two different limits:
\begin{itemize}
    \item Relatively large Betti numbers ($ \beta_r \approx |S_r^K|$): the ratio $ \frac{\binom{n}{r+1}}{|S_r^K|}$ would approach 1, so the homology approach can estimate the $r$-th normalized Betti number and also $r$-th Betti number with complexity $\mathcal{O} (n^2 + n\kappa)$. In this case, we choose bigger value of $\Gamma$, e.g., $ \Gamma \approx |S_r^K|$, our complexity for estimating both $r$-th normalized Betti number and Betti number is $\mathcal{O}(  r \log(|S_r^K| |S_{r+1}^K|) ) \in \mathcal{O}\left(\log \binom{n}{r+1} \right)$. For a sufficiently low value of $r$, e.g., $r=1,2,3,4$, the value of $\binom{n}{r+1} $ is of ${\rm poly}($n$)$ and thus $\log \binom{n}{r+1} \in \mathcal{O}\big( \log n \big)$, and, thus, our method is still exponentially faster than the homology approach. For higher values of $r$, e.g., around the middle of the range $[1,n]$ like $\sim n/2$, $ \binom{n}{r+1} \in \mathcal{O}\big( \exp(n) \big)$, so the complexity for our cohomology approach is 
    $$\mathcal{O}\left(   r \log(|S_r^K| |S_{r+1}^K|) \right) \in \mathcal{O}\left( \log \binom{n}{r+1} \right) = \mathcal{O}\left( n \right)$$
    \revise{In this same regime, the complexity of LGZ algorithm, or the quantum homology approach, is 
    $$ \mathcal{O}\left(n^2 + n\kappa  \right)  $$}    
    Thus, the degree of speed-up \revise{of our cohomology approach} is lowered, yielding quadratic speed-up compared to the homology approach. We remark that, as also emphasized in~\cite{schmidhuber2022complexity}, this is also the regime of best performance of the LGZ algorithm. \revise{To compare with classical algorithm, we remind that the classical running time is $\mathcal{O}\left( |S_r^K|^3 \right) $. As we can always choose $\Gamma$ so that the ratio $\Gamma/\beta_r$ is bounded, this means that our quantum algorithm can achieve complexity $\mathcal{O}\left(  \log |S_r^K| \log (r \log |S_r^K|)  \right) $  in all regimes, and thus there is a superpolynomial speedup compared to the classical one.}
    
    \item Relatively small Betti numbers $(\beta_r \ll |S_r^K|)$): the complexity of homology approach in estimating $\frac{\beta_r}{|S_r^K|}$ (to additive accuracy) and $\beta_r$ (to multiplicative accuracy) is, respectively 
    $$ \mathcal{O}\left(  n^2  + n \kappa  \right), $$
    $$  \mathcal{O}\left(  n^2  + n \kappa \sqrt{|S_r^K|} \right). $$ 
    For our cohomology algorithm, in this case, we can choose $\Gamma$ (in Algorithm \ref{alg: quantumbettinumber}) to be sufficiently small, so that the ratio $ \frac{\Gamma}{\beta_r} \in \mathcal{O}(1)$. Our complexity for estimating $ \frac{\beta_r}{|S_r^K|}$ and $\beta_r$ is (asymptotically) the same:
    $$ \mathcal{O}\left( \log\left( r \log (|S_r^K| \right) \log ( |S_r^K|  ) \right) \approx  \mathcal{O}\left( \log \binom{n}{r+1} \right), $$
    because $|S_r^K| \leq \binom{n}{r+1}$. The complexity above indicates that our cohomology approach provides an exponential speed-up compared to the homology one. 
    \revise{In comparison with the classical method, which has complexity $\mathcal{O}\left( |S_r^K|^3\right)$, our cohomology method offers superpolynomial speedup. }
\end{itemize}
\revise{Further, we comment that our cohomology method requires solving a linear system. According to~\cite{harrow2009quantum}, this can be BQP-complete, which implies that simulating our quantum procedure on a classical computer will be hard. That being said, within our cohomology approach and with the same input, whether there is a clever, efficient classical solution that could bypass the need to solve linear equations to estimate Betti numbers is not known to us. }

The above details have revealed that our cohomology approach offers significant speed-up, ranging from exponential to quadratic speed-up, compared to the homology one in all regimes, which highlights the power and flexibility of the cohomology framework. However, this framework has the limitation that it can only work for a triangulated closed manifold.

Another subtle point that we would like to discuss is the assumption that we made regarding the shape of the simplicial complex. In our work, we only deal with a triangulated manifold, i.e., all the composing simplexes are (almost) similar in size, and the manifold is built by properly gluing them together. In some previous works, such as~\cite{lloyd2016quantum, mcardle2022streamlined, schmidhuber2022complexity, ubaru2021quantum}, the setting is seemingly more general where the shape can be arbitrary, as two points only get connected if their distance is smaller than a known threshold. One may wonder if our assumption would severely limit the practicality of the outlined algorithm. The answer is no, as the topology of the underlying manifold only depends on the connectivity, but not the actual distance between any two points. As an example, let us consider the 2-dimensional case. One can imagine that, given a triangle with arbitrary angles/lengths, it is deformable or topologically equivalent to an equilateral triangle. Therefore, the topological space formed by the union, or by gluing different triangles together, is topologically equivalent to the union of equilateral triangles, which form the triangulated manifold. Since they are topologically equivalent, their underlying topological properties are the same. Consequently, performing computation on the triangulated manifold is more convenient, as we discuss further in Appendix~\ref{sec: generalhodge}, where we provide an explicit formula for the codifferential operator. Finally, we remark that, in general, the Hodge theory (see Theorem~\ref{theorem: hodgedecompo}) works with closed manifolds (in both smooth and discrete settings). Therefore, the specification of a given simplicial complex requires an extra criterion. In the 2d case that we worked out earlier, each edge (1-simplex) is supposed to be adjacent to two triangles (2-simplex). The generalization to higher dimensions is straightforward. If our initial configuration is an open triangulated manifold, e.g., with a boundary, then a simple trick is to double cover the simplicial complex while preserving the symmetry. In this way, we can recover the closeness condition and proceed with our algorithm.

\section{Conclusion}
\label{sec: conclusion}
Our work has provided a `dual' approach to~\cite{lloyd2016quantum} and is built upon the (discrete) Hodge theory and de Rham cohomology. There are a few major reasons underlying the advantage of our approach compared to~\cite{lloyd2016quantum}. In~\cite{lloyd2016quantum}, the authors basically quantized the homology approach, associating the chain group to a vector space and finding singular values/ vectors of the boundary map $\partial$. The key step in~\cite{lloyd2016quantum} is the identification of the chain group as a vector space, whereby each simplex is represented by a basis state, which means that the vector space needs to be at least as large as the number of simplexes contained in the given simplicial complex $K$. By doing it this way, the resource scales as a polynomial of $n$. In reality, the high value of $n$ is usually desired (large-scale analysis), which means that poly($n$) could be very high, and hence the algorithm induces a very high computational cost. 

The cohomology approach we have adopted here fits nicely in such a large-scale setting. We recall that in cohomology, a form is a map $C \rightarrow R$ where $C$ is some chain group (space). Equivalently, we can imagine that each simplex is associated with a real number, and, therefore, we can have a more efficient way of storing our data, as we only need a logarithmic number of qubits to store a polynomial-size vector, which further reduces the resources needed for processing. Another major point is that in~\cite{lloyd2016quantum}, it is required to generate the proper simplicial complex state, which contributes substantially to the computational cost and, in certain cases, has a high failing probability (for details, see~\cite{lloyd2016quantum}). Here, in the cohomology approach, the initial form can, in fact, be chosen arbitrarily, which is more convenient to prepare. This randomized form can be deformed to its harmonic part, according to the Hodge decomposition theorem, which is a major building block of (discrete) Hodge theory. The Betti numbers are then revealed by tracking the dimension of this harmonic space. This approach features a distinct one from the homology approach, and as we analyzed above, it potentially yields exponential improvement compared to the homology one in many regimes. It indicates that this cohomology framework is an excellent complementary approach to the homology framework, thus also displaying the power of the cohomology theory, especially with respect to potential real-world applications. Given that TDA is an emerging field and that cohomology theory has shown promise in estimating Betti numbers -- which is a central problem within TDA, it may serve as a useful avenue for future exploration.    \\

\begin{acknowledgements}
The authors thank Trung V. Phan for helpful feedback on the manuscript. Part of the work is done when N.A.N is an intern at QuEra Computing Inc. This work was supported in part by the US Department of Energy, Office of Science, National Quantum Information Science Research Centers, Co-design Center for Quantum Advantage
(C2QA) under contract number DE-SC0012704 (T.-C.W.), and by the
National Science Foundation under Grants  No. CMMI-1762287 (X.D.G.) and No. FAIN-2115095 (X.D.G.), as well as  NIH 3R01LM012434-05S1 (X.D.G.) and 
NIH 1R21EB029733-01A1 (X.D.G.). 
We also acknowledge the support by a Seed Grant from
Stony Brook University’s Office of the Vice President for Research and by the Center for Distributed Quantum Processing.
\end{acknowledgements}

\appendix

\section{Relevant QSVT Techniques}
\label{sec: QSVT}
As mentioned in the main text, in this part, we introduce the main quantum ingredients that are needed to construct our subsequent algorithm. We recapitulate the key results for brevity and leave out the details, which are carefully described in~\cite{gilyen2019quantum}.

\begin{lemma}[\cite{gilyen2019quantum} Product]
\label{lemma: product}
    Given the unitary block encoding of two matrices $A_1$ and $A_2$, then there exists an efficient procedure that constructs a unitary block encoding of $A_1 A_2$ using each block encoding of $A_1$ and $A_2$ one time. 
\end{lemma}

\begin{lemma}[\cite{camps2020approximate} Tensor Product]
\label{lemma: tensorproduct}
    Given the unitary block encoding $\{U_i\}_{i=1}^m$ of multiple operators $\{A_i\}_{i=1}^m$ (assumed to be exact encoding), then, there is a procedure that produces the unitary block encoding operator of $\bigotimes_{i=1}^m A_i$, which requires parallel single uses of 
    $\{U_i\}_{i=1}^m$ and $\mathcal{O}(1)$ SWAP gates. 
\end{lemma}
The above lemma is a result in \cite{camps2020approximate}. 
\begin{lemma}[\cite{gilyen2019quantum} (Lemma 48) Sparse-Access Matrix]
\label{lemma: As}
    Given oracle access to $s$-sparse matrix $A$ (assuming the operator norm $|A| \leq 1$) of dimension $n\times n$, then an $\epsilon$-approximated unitary block encoding of $A/s$ can be prepared with gate/time complexity $\mathcal{O}\Big(\log n + \log^{2.5}(\frac{s^2}{\epsilon})\Big).$
\end{lemma}
This is presented in~\cite{gilyen2019quantum} (see their Lemma 48). We remark further that the scaling factor $s$ in the above lemma can be reduced by the preamplification method with further complexity $\mathcal{O}({s})$~\cite{gilyen2019quantum}.

\begin{lemma}[Linear Combination \cite{gilyen2019quantum}]
    Given the unitary block encoding of multiple operators $\{A_i\}_{i=1}^m$. Then, there is a procedure that produces a unitary block encoding operator of \,$\sum_{i=1}^m \pm A_i/m $ in complexity $\mathcal{O}(m)$, e.g., using the block encoding of each operator $A_i$ a single time. 
    \label{lemma: sumencoding}
\end{lemma}

\begin{lemma}[Scaling Multiplication] 
\label{lemma: scale}
    Given a block encoding of some matrix $A$, as in Definition~(\ref{def: blockencode}), the block encoding of $A/p$ where $p > 1$ can be prepared with an extra $\mathcal{O}(1)$ cost.  
\end{lemma}
To show this, we note that the matrix representation of RY rotational gate is
\begin{align}
   R_Y(\theta) = \begin{pmatrix}
        \cos(\theta/2) & -\sin(\theta/2) \\
        \sin(\theta/2) & \cos(\theta/2) 
    \end{pmatrix}.
\end{align}
If we choose $\theta$ such that $\cos(\theta/2) = 1/p$, then Lemma~\ref{lemma: tensorproduct} allows us to construct block encoding of $R_Y(\theta) \otimes \mathbb{I}_{{\rm dim}(A)}$  (${\rm dim}(A)$ refers to the dimension of the row or column of the square matrix $A$), which contains the diagonal matrix of size ${\rm dim}(A) \times {\rm dim}(A)$ with entries $1/p$. Then Lemma~\ref{lemma: product}  allows us to construct the block encoding of $(1/p) \ \mathbb{I}_{{\rm dim}(A)} \cdot A = A/p$.  \\

The following is called amplification technique:
\begin{lemma}[Amplification, Theorem 30 of~\cite{gilyen2019quantum}]
\label{lemma: amp_amp}
Let $U$, $\Pi$, $\widetilde{\Pi} \in {\rm End}(\mathcal{H}_U)$ be linear operators on $\mathcal{H}_U$ such that $U$ is a unitary, and $\Pi$, $\widetilde{\Pi}$ are orthogonal projectors. 
Let $\gamma>1$ and $\delta,\epsilon \in (0,\frac{1}{2})$. 
Suppose that $\widetilde{\Pi}U\Pi=W \Sigma V^\dagger=\sum_{i}\varsigma_i\ket{w_i}\bra{v_i}$ is a singular value decomposition. 
Then there is an $m= \mathcal{O} \Big(\frac{\gamma}{\delta}
\log \left(\frac{\gamma}{\epsilon} \right)\Big)$ and an efficiently computable $\Phi\in\mathbb{R}^m$ such that
\begin{align}
\left(\bra{+}\otimes\widetilde{\Pi}_{\leq\frac{1-\delta}{\gamma}}\right)U_\Phi \left(\ket{+}\otimes\Pi_{\leq\frac{1-\delta}{\gamma}}\right) \nonumber \\ = \sum_{i\colon\varsigma_i\leq \frac{1- \delta}{\gamma} }\tilde{\varsigma}_i\ket{w_i}\bra{v_i} , \text{ where } \Big|\!\Big|\frac{\tilde{\varsigma}_i}{\gamma\varsigma_i}-1 \Big|\!\Big|\leq \epsilon.
\end{align}
Moreover, $U_\Phi$ can be implemented using a single ancilla qubit with $m$ uses of $U$ and $U^\dagger$, $m$ uses of C$_\Pi$NOT and $m$ uses of C$_{\widetilde{\Pi}}$NOT gates and $m$ single qubit gates.
Here,
\begin{itemize}
\item C$_\Pi$NOT$:=X \otimes \Pi + I \otimes (I - \Pi)$ and a similar definition for C$_{\widetilde{\Pi}}$NOT; see Definition 2 in \cite{gilyen2019quantum},
\item $U_\Phi$: alternating phase modulation sequence; see Definition 15 in \cite{gilyen2019quantum},
\item $\Pi_{\leq \delta}$, $\widetilde{\Pi}_{\leq \delta}$: singular value threshold projectors; see Definition 24 in \cite{gilyen2019quantum}.
\end{itemize}
\end{lemma}

Another important tool is the inversion of a matrix~\cite{harrow2009quantum, childs2017quantum, gilyen2019quantum},  
\begin{lemma}[Matrix Inversion \cite{harrow2009quantum, childs2017quantum, gilyen2019quantum}]
\label{lemma: matrixinversion}
    Given a block encoding of some Hermitian matrix $A$  with operator norm $ \frac{1}{\kappa} \leq ||A|| \leq 1$ and block-encoding complexity $T_A$, then there is a quantum circuit producing an $\epsilon$-approximated block encoding of $ \frac{1}{\kappa} A^{-1}$. The complexity of this quantum circuit is 
    $$ \mathcal{O}\Big( T_A \kappa \log \frac{1}{\epsilon}  \Big). $$
    If the matrix $A$ is not Hermitian, then by a slight abuse of notation, the block encoding of its pseudo-inverse (both left and right) $ \frac{1}{\kappa^2}A^{-1}$ can be implemented with the same complexity as above. 
\end{lemma}

Finally, the most important recipe for our work is the following lemma.
\begin{lemma}[Block Encoding of A Known Matrix]
\label{lemma: entrycomputablematrix}
    Let $A$ be a matrix of sizes $M \times N$ (not necessarily Hermitian) with sparsity $s$ and classical knowledge of its entries. 
    \begin{itemize}
        \item (version 1) Then there is a quantum circuit of complexity $\mathcal{O}\left( \log sN  \right)$ which  block-encodes $\frac{1}{ \sqrt{2}||A||_F}A$ where $||A||_F$ is the Frobenius norm of $A$ using a classical pre-processing of complexity $\mathcal{O}(\log N)$. 
        \item (version 2) In circumstances, for example, all off-diagonal entries of $A$ are the same, all diagonal entries of $A$ are the same, then there is quantum circuit of complexity $\mathcal{O}\left( \log sN  \right)$ which  block-encodes $\frac{1}{ \sqrt{2}||A||_F}A$ where $||A||_F$ is the Frobenius norm of $A$ with $\mathcal{O}(1)$ classical pre-processing time.
    \end{itemize}
\end{lemma}
\noindent
\textbf{Proof.} We first outline the proof of the first version. The proof is based on the recent work \cite{nghiem2025refined} (see their second diagram, or Section III.C), in which the author shows the following:
\begin{lemma}[\cite{nghiem2025refined}]
\label{lemma: nghiem2025refined}
    Let $A$ be a matrix of size $M \times N$, with classically known entries and its Frobenius norm is $||A||_{F}$. Then:
    \begin{itemize}
        \item there is a quantum circuit of depth $\mathcal{O}\big( \log(s\log N)  \big)$ that is an exact block encoding of $ \frac{1}{||A||_F^2} A^\dagger A$
        \item If $A$ is square, Hermitian, then there is a quantum circuit of depth $\mathcal{O}\big( \log(s\log N) \log^2 \frac{1}{\epsilon} \big)$ that is an $\epsilon$-approximated block encoding of $ \frac{1}{||A||_F} A$, 
    \end{itemize}
    using $\mathcal{O}(s)$ ancilla qubits where $s$ is the sparsity (in this case, it is maximum number of nonzero entries in a column) of $A$, and a classical pre-processing with $\mathcal{O}(\log N)$ complexity. 
\end{lemma}
While the original proof of the above lemma can be found in \cite{nghiem2025refined}, here for completeness, we summarize the key steps. Denote $A^i$ as the $i$-th column of $A$. Then the classical knowledge of $A$ yields the classical knowledge of $A^i$ for all $i$. Then as worked out in \cite{zhang2022quantum, mcardle2022quantum, marin2023quantum, nakaji2022approximate}, we have the following efficient state preparation:
\begin{lemma}[State Preparation \cite{zhang2022quantum, mcardle2022quantum, marin2023quantum, nakaji2022approximate}]
\label{lemma: statepreparation}
    Let $\ket{\psi} = \sum_{i=1}^n a_i \ket{i}$ and $\{a_i\}_{i=1}^n$ is classically known. Further, let $s$ denotes the sparsity of $\ket{\psi}$, or the number of non-zero entries of $\ket{\psi}$. Then $\ket{\psi} $ can be deterministically prepared using a quantum circuit of depth $\mathcal{O}\big( \log (s \log n) \big)$. 
\end{lemma}
By using the above lemma, the following state can be prepared with a quantum circuit of complexity $\mathcal{O}(\log (s \log N))$: 
\begin{align}
    \ket{\Phi} = \frac{1}{||A||_F} \sum_{i=1}^N \ket{i} A^i.
\end{align}
If we trace out the second register (the one containing $A^i$), the resulting density state is $\frac{1}{||A||_F^2} \sum_{i=1}^N \ket{j=1}^N \ket{i}\bra{j} (A^i)^\dagger A^j$, which is exactly $\frac{1}{||A||_F^2} A^\dagger A$. This density state can be block-encoded via the following:
\begin{lemma}[\cite{gilyen2019quantum} Block Encoding Density Matrix]
\label{lemma: improveddme}
Let $\rho = \Tr_A \ket{\Phi}\bra{\Phi}$, where $\rho \in \mathbb{H}_B$, $\ket{\Phi} \in  \mathbb{H}_A \otimes \mathbb{H}_B$. Given unitary $U$ that generates $\ket{\Phi}$ from $\ket{\bf 0}_A \otimes \ket{\bf 0}_B$, then there exists a highly efficient procedure that constructs an exact unitary block encoding of $\rho$ using $U$ and $U^\dagger$ a single time, respectively.
\end{lemma}
\revise{Lemma \ref{lemma: improveddme} is then applied} to block-encode $\mathcal{X}^T \mathcal{X}$, with a total circuit of depth $\mathcal{O}( \log mn )$. We can use this block encoding and use Lemma \ref{lemma: scale} with scaling factor $m$, to obtain the block encoding of $ \frac{1}{m}\mathcal{X}^T \mathcal{X}$. 
Thus, the first part of Lemma \ref{lemma: nghiem2025refined} is proved. For the second part, the block encoding of $ \frac{1}{||A||_F^2} A^\dagger A$ can be used with the following lemma:
\begin{lemma}[Positive Power Exponent \cite{gilyen2019quantum}, \cite{chakraborty2018power}]
\label{lemma: positive}
    Given a block encoding of a positive matrix $\mathcal{M}$ such that 
    $$ \frac{\Ibb}{\kappa_M} \leq \mathcal{M} \leq \Ibb. $$
   Let $c \in (0,1)$. Then we can implement an $\epsilon$-approximated block encoding of $\mathcal{M}^c/2$ in time complexity $\mathcal{O}( \kappa_M T_M \log^2 (\frac{\kappa_M}{\epsilon})  )$, where $T_M$ is the complexity to obtain the block encoding of $\mathcal{M}$ 
\end{lemma}
\revise{Lemma \ref{lemma: positive} then can be used} to perform the transformation (of the block-encoded operator) $\frac{1}{||A||_F^2} A^\dagger A \longrightarrow \frac{1}{||A||_F} A $.

Essentially, the key difference between the Lemma \ref{lemma: nghiem2025refined} and Lemma \ref{lemma: entrycomputablematrix} is that, Lemma \ref{lemma: nghiem2025refined} assumes $A$ to be Hermitian. Meanwhile, in our case, as we see earlier (Eqn.~\ref{eqn: coexactdeltaomega}), the matrices of interest are not necessarily Hermitian. In order to apply the above lemma in the context of Lemma \ref{lemma: entrycomputablematrix}, we consider the following matrix:
\begin{align}
    A'  = \begin{pmatrix}
        \textbf{0} & A \\
        A^\dagger & \textbf{0} 
    \end{pmatrix}.
\end{align}
As the entries of $A$ are known, the entries of $A'$ are apparently known. So we can apply Lemma \ref{lemma: nghiem2025refined} to construct the block encoding of $ \frac{1}{||A'||_F} A'$. We note that:
\begin{align}
   \frac{1}{||A'||_F}  A' =\frac{1}{||A'||_F}  \ket{0}\bra{1} \otimes A +\frac{1}{||A'||_F}  \ket{1}\bra{0}\otimes A^\dagger
\end{align}
and the dimension of $A'$ is $(M+N) \times (M+N)$. The gate $X \otimes \Ibb_{ M \times N  }$ is straightforward to construct, as we just need to take a $X$ gate, acting on 1 qubit, and tensor product with remaining $ \log \left( MN \right)$ qubits system without any gate acting on it (effectively an identity gate). So, using Lemma \ref{lemma: product} we can construct the block encoding of: 
\begin{widetext}
    \begin{align}
   \frac{1}{||A'||_F}  A' \cdot \big( X \otimes  \Ibb_{ M \times N  } \big) &=  \left(\frac{1}{||A'||_F}  \ket{0}\bra{1} \otimes A + \frac{1}{||A'||_F} \ket{1}\bra{0}\otimes A^\dagger  \right) \big(  X \otimes  \Ibb_{ M \times N  } \big) \\
    &= \ket{0}\bra{0} \otimes \frac{1}{||A'||_F} A + (...),
\end{align}
\end{widetext}
where $(...)$ means we are ignoring the rest term. According to \ref{def: blockencode}, the above operator is in fact a block encoding of $\frac{1}{||A'||_F} A$. In addition, it can be seen that $||A'||_F = \sqrt{2} ||A||_F$. Thus, we complete the proof of lemma \ref{lemma: entrycomputablematrix}. 

The above procedure can be applied as long as the entries of $A$ are classically provided. However, as stated in the Lemma \ref{lemma: nghiem2025refined}, a classical pre-processing is required (even though it is efficient) as it employs the result of \cite{zhang2022quantum}, e.g., Lemma \ref{lemma: statepreparation}. Here, we point out that, that in certain circumstances, classical pre-processing cost is very modest, and in particular, the matrices appeared in this work, e.g., in Eqn.~\ref{eqn: coexactdeltaomega},~\ref{eqn: coexactdeltaomega}, \ref{eqn: coexactequation}, fall into these circumstances. 

The first particular circumstance is when all the non-zero entries of $A$ having the same value. In this case, the classical pre-processing step required in Lemma \ref{lemma: statepreparation} only needs to perform once, as the output of this pre-processing step can be naturally applied to all entries. From this observation, we extend it more generally as follows. Let $ S= \{A_{ij} \}$ be the set of entries of $A$. Let $S_1,S_2,..., S_K$ be subsets of $S $ and that $ S_1 \cup S_2 \cup \cdots \cup S_K = S$, $S_m \cap S_m = \emptyset$  for any $1\leq m,n \leq K$. For all $1 \leq m \le K$, the entries of $A$ contained in $S_m$ is the same. Then as long as $K = \mathcal{O}(1)$, then the classical pre-processing step is $\mathcal{O}(1)$.

\section{Homology Theory}
\label{sec: crashcoursehomology}
Here, we provide an overview of key concepts from algebraic topology, along with the definition of Betti numbers. 
We quote many results  from~\cite{nakahara2018geometry}, to which we strongly refer the interested readers for more details on the subject. 
\begin{definition}[Simplexes]
Let $p_0,p_1,...,p_r \in \Rbb^m$ be geometrically independent points where $m \geq r$. The $r$-simplex $\sigma_r \equiv [p_0,p_1,...,p_r]$ is defined as 
\begin{align}
    \sigma_r = \big\{  x \in \Rbb^m | x= \sum_{i=0}^r c_i p_i, c_i \geq 0, \sum_{i=0}^r c_i =1 \big\},
\end{align}
where the coefficients $(c_0,c_1,...,c_r)$ are called the barycentric coordinate of $x$. Pictorially, a 0-simplex, $[p_0]$ is a point, a 1-simplex $[p_0,p_1]$ is a line and 2-simplex $[p_0,p_1,p_2]$ is a triangle, as the following figures illustrate. 
\begin{tikzpicture}[scale=2.2]
\filldraw (0,0) circle (0.5pt);
\node[below right] at (0,0) {$p_0$};
\draw (1.2,0) -- (2,0);
\node[below] at (1.2,0) {$p_0$};
\node[below] at (2,0) {$p_1$};
\draw (2.7,0) -- (3.2,0.8) -- (3.7,0) -- cycle;
\node[below left] at (2.7,0) {$p_1$};
\node[below right] at (3.7,0) {$p_2$};
\node[above] at (3.2,0.8) {$p_0$};
\end{tikzpicture}
\begin{center}
\begin{tikzpicture}[scale=2, line join=round]
\coordinate (P0) at (0,1);
\coordinate (P1) at (-0.8,0);
\coordinate (P2) at (0.4,-0.2);
\coordinate (P3) at (0.4,0.4);
\draw[thick] (P0) -- (P1) -- (P2) -- (P0);
\draw[thick] (P0) -- (P3) -- (P2);
\draw[dashed, thick] (P1) -- (P3);
\node[above] at (P0) {$p_0$};
\node[left] at (P1) {$p_1$};
\node[below right] at (P2) {$p_2$};
\node[right] at (P3) {$p_3$};
\end{tikzpicture}
\end{center}
An assignment of orientation can be made to an $r$-simplex. For example, in the above figure, $[p_0,p_1]$ can refer to a 1-simplex with orientation $p_0 \longrightarrow p_1$. In this way, $[p_0,p_1]$ is different from $[p_1,p_0]$. Throughout this work, we make an orientation convention that, for an $r$-simplex $[p_0,p_1,...,p_r]$, the order goes from low to high, indicating the orientation. 
\end{definition}

Polyhedra, e.g., cubes, octahedra, dodecahedra, etc., can be built from simplexes. For any $r$-simplex $[p_0,p_1,...,p_r]$, if we take out $s+1$ points, then these $s+1$ points define a $s$-simplex, called a $s$-face $\sigma_s$. A simplicial complex $K$ is a collection of simplexes with two properties: (i) The Arbitrary face of any simplex is a simplex $\in K$; (ii) The intersections of any two simplexes are either empty or a (joint) face of these two simplexes.  If there is a homeomorphism between a simplicial complex $K$ and some topological space $X$, then $K$ is called a triangulation of $X$. In fact, a central aspect of algebraic topology is to study topological spaces using the algebraic language. By triangulating a topological space, one can assign to it an Abelian group, for which the group structure reveals topological structure.

To be more specific, let $K$ be an $n$-dimensional simplicial complex, where the dimension of a simplicial complex is defined as the maximum dimension of simplexes in $K$. Then we have the following definition:
\begin{definition}[Chain group]
\label{def: chaingroup}
    The $r$-th chain group $C_r^K$ of a simplicial complex $K$ is a free Abelian group generated by the (oriented) $r$-simplexes of $K$. An element of $C_r^K$ is then called an $r$-chain. For $r > \dim (K)$, $C_r^K$ is defined to be $0$. 
    
    More specifically, let $S_r^K$ be the set of $r$-simplexes in $K$, so the total of $r$-simplexes in $K$ is $|S_r^K|$. Then a $r$-chain is $c_r = \sum_{i=1}^{|S_r^K|} c_i \sigma_{r_i} $ where $\sigma_{r_i}$ refers to the $i$-th $r$-simplex and $c_i \in \mathbb{Z}$. Given two different $r$-chains $c_r(1) =\sum_{i=1}^{|S_r^K|} c_i(1)  \sigma_{r_i} $ and $c_r(2) = \sum_{i=1}^{|S_r^K|} c_i(2) \sigma_{r_i}$, we can define an addition operation between them as following:
    \begin{align*}
        c_r(1) + c_r(2) = \sum_{i=1}^{|S_r^K|} \Big( c_i(1)+c_i(2) \Big) \sigma_{r_i}.
    \end{align*}
    We define the zero element of $C_r^K$ to be $\sum_i 0 \sigma_{r_i}$. For an element $c_r$, its inverse element is defined $-c_r = \sum_{i=1}^{|S_r^K|} (-c_i) \sigma_{r_i}$. Thus, under the operation,   elements in $C_r^K$ form a free Abelian group of rank $|S_r^K|$. 
\end{definition}
Another crucial operation on any $r$-chain (for any $r$) is the \textit{boundary operator} $\partial_r$. Previously, for an $r$-simplex $[p_0,p_1,...,p_r]$, we have defined what it means to be a face. A boundary is essentially an $r-1$-face. The boundary operator, once acted on an $r$-simplex $\sigma_r$, will ``break'' $\sigma_r$ into an algebraic summation of its boundary. More specifically, we have:
\begin{align}
    \partial_r [p_0,p_1,...,p_r] = \sum_{i=0}^r (-1)^i [p_0,p_1,...,\hat{p_i}, ..., p_r],
\end{align}
where $\hat{p_i} $ indicates that $p_i$ is omitted and hence, $[p_0,p_1,...,\hat{p_i}, ..., p_r] $ is a $(r-1)$-simplex. Let $\partial_r$ act on a $r$-chain as following $\partial_r c_r = \sum_{i=1}^{|S_r^K|} c_i \partial_r \sigma_{r_i}$. As each $ \partial_r \sigma_{r_i} \in C_{r-1}^K$, it is straightforward to verify that $\sum_{i=1}^{|S_r^K|} c_i \partial_r \sigma_{r_i} \in C_{r-1}^K$ and, thus, $\partial_r$ defines a homomorphism $C_r^K \longrightarrow C_{r-1}^K$. Chain groups and homomorphisms form a sequence:
\begin{align*}
    0 \xrightarrow{\text{inclusion}} C_n^K \xrightarrow{\text{$\partial_n$}} C_{n-1}^K \xrightarrow{\text{$\partial_{n-1}$}} \cdots \xrightarrow{\text{$\partial_1$}} C_0^K \xrightarrow{\text{$\partial_0$}} 0. 
\end{align*}
For an $r$-chain $c_r$, if $\partial_r c_r =0$ then $c_r$ is called an $r$-cycle. A set of $r$-cycle forms a subgroup of $C_r^K$, called an $r$-cycle group and denoted as $Z_r^K$. If there is a $r+1$-chain $d_{r+1}$ such that $\partial_{r+1} d_{r+1} = c_r$ then $c_r$ is called a $r$-boundary. A set of $r$-boundaries forms an $r$-boundary group $B_r^K$. The boundary operator has a property that a boundary of a boundary is $0$, i.e., the composite map $\partial_{r} \partial_{r+1} =0$, which implies that $B_r^K \subset Z_r^K$. The $r$-th homology group is defined as the quotient group $H_r^K \equiv Z_r^K/B_r^K$. The rank of this group is called the $r$-th Betti number, $\beta_r = \rm rank \  H_r^K$. In order to conveniently compute the Betti number of given simplicial complex, one can define the so-called combinatorial Laplacian $\Delta_r = \partial_{r+1} \partial_{r+1}^\dagger + \partial_r^\dagger \partial_r$, and there is a fundamental result, which states that $H_r^K \cong \rm Ker (\Delta_r)$. 

An important result of algebraic topology states that the homology group is, in fact, a topological invariant~\cite{hatcher2005algebraic}. If two topological spaces are homeomorphic to each other, then their triangulations have the same homology groups. Therefore, the Betti numbers can be used to classify topological spaces.

\section{Cohomological Framework for Estimating Betti Numbers}
\label{sec: cohomologicalframework}
This section begins with an introduction to key ingredients from cohomology theory and differential geometry that underlie the classical and subsequent quantum algorithms for estimating Betti numbers. 

\subsection{Basic Notions of Differential Geometry}
\label{sec: crashcourseondifferentialgeometry}
\begin{tikzpicture}[scale = 0.7]
\shade[ball color=yellow!20, opacity=0.8] 
  (0,3) ellipse (3.5 and 1.5);
\filldraw[fill=cyan!30, draw=black, opacity=0.7] (1.0,3.3) ellipse (0.9 and 0.5);
\filldraw[fill=orange!30, draw=black, opacity=0.6, dashed] (-0.6,3.6) ellipse (0.7 and 0.4);
\draw[-{Latex[length=3mm]}, thick] (-0.6,3.6) .. controls (-1.2,2.2) .. (-2.5,0.8);
\draw[-{Latex[length=3mm]}, thick] (1.0,3.3) .. controls (1.2,2.0) .. (2.5,0.8);
\begin{scope}[shift={(-3.5,0)}]
    \foreach \x in {-1.2,-0.6,...,1.2}
        \draw[dashed, black!50] (\x,-1.2) -- (\x,1.2);
    \foreach \y in {-1.2,-0.6,...,1.2}
        \draw[dashed, black!50] (-1.2,\y) -- (1.2,\y);
    \draw[->, thick] (0, -1.4) -- (0,1.4);
    \draw[->, thick] (-1.4,0) -- (1.4, 0);
    \filldraw[fill=orange!30, draw=black, opacity=0.6] (0,0) ellipse (0.7 and 0.4);
\end{scope}
\begin{scope}[shift={(3.5,0)}]
    \foreach \x in {-1.2,-0.6,...,1.2}
        \draw[dashed, black!50] (\x,-1.2) -- (\x,1.2);
    \foreach \y in {-1.2,-0.6,...,1.2}
        \draw[dashed, black!50] (-1.2,\y) -- (1.2,\y);
    \draw[->, thick] (0, -1.4) -- (0,1.4);
    \draw[->, thick] (-1.4,0) -- (1.4, 0);
    \filldraw[fill=cyan!30, draw=black, opacity=0.6] (0,0) ellipse (0.8 and 0.5);
\end{scope}
\node at (0,4.7) {$\mathcal{M}$};
\node at (-1.5, 3) {$U_1$};
\node at (2,3) {$U_2$};
\node at (-1.8, 0.5) {$\phi_1(U_1) \subset \mathbb{R}^2$};
\node at (5.0, 0.5) {$\phi_2(U_2) \subset \mathbb{R}^2$};
\end{tikzpicture}

As illustrated above, a manifold $\mathcal{M}$ is a topological space which is locally Euclidean. The fact that a manifold is a topological space suggests that a manifold can be triangulated by a simplicial complex and that the techniques from homology theory can be used to study manifolds. However, a manifold itself is also a geometric object, and it might contain many more properties that go beyond topology, e.g., (local) curvature, which are not captured by the homological method. More specifically, at a given point $\mathcal{X}$ on the given manifold, tangent vectors are defined, forming the so-called tangent space $\mathcal{T}_{\mathcal{X}} \mathcal{M}$.      

\begin{figure}[H]
    \centering
   \begin{tikzpicture}[scale=0.8]
        \begin{scope}[shift={(-3,0)}]
            \foreach \x in {-2,-1.5,...,2} {
                \draw[gray!60] (\x,-2) to[out=70,in=250] (\x,2);
            }
            \foreach \y in {-2,-1.5,...,2} {
                \draw[gray!60] (-2,\y) to[out=20,in=160] (2,\y);
            }
            
            \draw[purple,thick] (-1.2,-1.2) -- (1.2,-1.2) -- (1.2,1.2) -- (-1.2,1.2) -- cycle;
            \draw[purple, thick] (-1.2,0) -- (1.2,0);
            \draw[purple, thick] (0,-1.2) -- (0,1.2);

            \filldraw[black] (0,0) circle (1pt);
            \node at (0.2,-0.25) {$x$};
            \node[purple] at (-1.6,1.6) {$T_x \mathcal{M}$};
            \node at (1.9,-1.9) {$\mathcal{M}$};
        \end{scope}

        \begin{scope}[shift={(3,0)}]
            \draw[gray!80, thick] (-2,-0.5) to[out=30,in=200] (0,0) to[out=30,in=170] (2,1);
            

            \filldraw[black] (0,0) circle (1pt);
            \draw[->, purple, thick] (0,0) -- (1,0.47);
            \draw[purple,thick] (0,0) -- (1,0.47);
            \node at (0.2,-0.2) {$x$};
            \node at (1.1,0.6) {$\dot{x}$};
            \node at (1.9,1.1) {$\mathcal{M}$};
        \end{scope}
\end{tikzpicture}
    \caption{A tangent vector $\mathcal{X'}$ is defined at a point $\mathcal{X} \in \mathcal{M}$. The collection of these tangent vectors forms a linear vector space at $\mathcal{X}$. }
    \label{fig: tangent}
\end{figure}
It is known that the tangent space is a linear space having a dimension similar to the dimension of the given manifold. There exists a dual to this linear space, which is defined as the space of maps $f:\mathcal{T}_{\mathcal{X}} \mathcal{M} \longrightarrow \mathbb{R} $. Such a map $f$, which takes a tangent vector and outputs a real number is called a 1-form, and there is a generalization of this map, namely, the $r$-form, which is multi-linear map that takes $r$ tangent vectors to a real number, i.e., $ \mathcal{T}_{\mathcal{X}} \mathcal{M} \times \mathcal{T}_{\mathcal{X}} \mathcal{M} \times \cdots \times \mathcal{T}_{\mathcal{X}} \mathcal{M} \longrightarrow \mathbb{R}$. The collection of such $r$-forms form a vector space $\Omega^r(\mathcal{M})$. We recall from homology theory (described in Section \ref{sec: crashcoursehomology}) that the $r$-chain group $C_r^K$ is formed by linear combinations of $r$-simplexes, and there is a boundary map  $\partial_r: C_r^K \longrightarrow C_{r-1}^K$. In the context of a manifold, there is an analogous map, so-called differential map $d^r:\Omega^r(\mathcal{M}) \longrightarrow \Omega^{r+1}(\mathcal{M}) $, which is between the space of $r$-forms and $(r+1)$-forms.

The above definitions and terminologies are considered in a general, continuous setting. In a discrete setting, for example, when the manifold is triangulated, one needs a discrete theory, which is a fascinating research area at the intersection of modern geometry and topology, computer science, and mathematics. A more detailed treatment can be found in~\cite{gu2020computational}. To keep it neat and sufficient for our purpose, we mention that in the context of a manifold $\mathcal{M}$ admitting a triangulation represented by a complex $K$, an $r$-form is a linear functional taking $r$-chains $\{ c_{r_i} \}_{i=1}^{|S_r^K|}$ to a real number, i.e., $\omega^r: C_r^K \longrightarrow \mathbb{R}$. As all $r$-simplexes $\{ \sigma_{r_i}\}$ form the basis of a vector space $C_r$, one can define any $r$-form by its action on the basis: for all $i=1,2,...,|S_r^K|$, let $\omega^r( c_{r_i}) \in \mathbb{R}$. As we will see in Sections~\ref{sec: derhamcohomologyhodgetheory} \&~\ref{sec: sketchbettinumbers}, by analyzing the space of these forms, one can reveal Betti numbers of the underlying triangulated manifold. In the next section, we proceed to describe essential concepts from de Rham cohomology theory and Hodge theory, which underlie our algorithms for estimating Betti numbers.

\subsection{de Rham Cohomology and Hodge Theory }
\label{sec: derhamcohomologyhodgetheory}
De Rham cohomology is a very important tool that encompasses both algebraic and differential topology and can be used to probe the topological structure of smooth manifolds. While in homology, the main objects are spaces (or groups) of chains, in de Rham cohomology, the main objects are spaces of \textit{forms}. The homology group is formed via the equivalence classes of closed chains, whereas in de Rham cohomology,  the cohomology group is formed via the equivalence classes of closed forms. Hodge's theory is built on an important observation that each cohomology class has a canonical representative, the so-called harmonic form. A standard result  is the Hodge decomposition theorem: 
\begin{theorem}[Hodge Decomposition]
Suppose $\mathcal{M}$ is an $n$-dimensional closed Riemannian manifold, then 
\begin{align}
    \Omega^r = {\rm Img}(d^{r -1}) \oplus {\rm Img}(\delta^{r +1}) \oplus H^r _{\Delta}(M),
    \label{theorem: hodgedecompo}
\end{align}
where $\Omega^r $ is the space of $r $-forms, {\rm Img} denotes the image of a map, $d^{r -1}$ is the exterior derivative map: $\Omega^{r -1}(M) \rightarrow \Omega^{r }(M)$, $\delta^{r +1}$ is the codifferential operator: $\Omega^{r +1} \rightarrow \Omega^{r }$,  and $H^r _{\Delta}$ is the space of harmonic $r $-forms. 
\end{theorem}
In other words, if $\omega \in \Omega^r $, it can be expressed and decomposed as 
\begin{align}
    \omega =  \delta \Omega+ d\eta  + h,
    \label{eqn: hodge}
\end{align}
where $\eta \in \Omega_{r -1}$, $\Omega \in \Omega_{r +1}$, and a special property that \revise{the harmonic form $h$} vanishes under the action of $d$ and $\delta$, i.e., $dh = 0$ and $\delta h = 0$. Most importantly, $h$ is unique for each cohomology class, i.e., if $\omega$ and $\omega'$ lie in the same cohomology class, then they have the same $h$. Throughout this work, we may abuse the notation by writing $\delta$ and $d$ without the superscript.

Generally, de Rham cohomology and Hodge theory work for the smooth setting. But the extension to the discrete setting can be achieved by simply replacing de Rham cohomology with simplicial cohomology by, e.g., identifying $k$-forms with $k$-cochains, and so on. The discrete version has been developed and applied extensively in real applications; see, e.g., Refs~\cite{gu2008computational, gu2003genus}, as $d$ and $\delta$ operators can be represented as a linear transformation on a set of corresponding simplexes.  We use the same notations for both cases, and we treat groups and vector spaces in the same manner, as a vector space is also an Abelian group, and we only work in the Abelian (commutative) setting.

Next, we state two important results that provide a useful insight into our algorithm, to be described below. As mentioned above, each cohomology class has a unique representative, and, therefore, it directly implies the correspondence between the two spaces (or groups), as stated in the two theorems below. 
\begin{theorem}[\cite{bott1982differential, hatcher2005algebraic}]
Given an $n$-dimensional closed Riemannian manifold. The $r$-th de Rham cohomology group is isomorphic to the harmonic $r$-form group 
\begin{align}
    H^r _{dR}(M) \approx H^r _{\Delta}(M).
\end{align}
\label{theorem: dualharmonicderham}
\end{theorem}
Another standard and important result that we will employ is the following.
\begin{theorem}[\cite{bott1982differential, hatcher2005algebraic}]
The de Rham cohomology group is isomorphic to the simplicial cohomology group
\begin{align}
    H^r _{dR}(M) \approx H^r  (M).
\end{align}
\label{theorem: dualsimderham}
\end{theorem}

We remark that the above two theorems are standard in the areas of differential geometry and algebraic topology, which are explained in standard textbooks, e.g., see~\cite{bott1982differential, hatcher2005algebraic}.

\subsection{Calculating Betti Numbers}
\label{sec: sketchbettinumbers}
Theorems \ref{theorem: dualharmonicderham} and \ref{theorem: dualsimderham}, plus the duality between the simplicial homology and cohomology, show that all these groups are isomorphic, which reveals a potential approach to calculate Betti numbers of a given simplicial complex $\sum$. Given some $r $, the $r $-th Betti number is the rank of the $r $-th homology group, which is also the rank of the $r $-th cohomology group. If we regard them as vector spaces, the rank becomes the dimension. We remark that the harmonic forms also form a vector space; therefore, the dimension of such space can be inferred if we know the maximum number of linearly independent vectors. The Hodge decomposition theorem~\ref{theorem: dualharmonicderham} allows us to find the harmonic form given an initial $r $-form $\omega$,  as $h = \omega - d\eta - \delta \Omega$. 

\section{Further elaboration on linear equation}
\label{sec: elaboration}
In this section, we discuss in detail how to solve for $\Omega$ and  $\eta$ that appear in the Hodge decomposition~(\ref{eqn: hodge}), which constitutes the linear equation appeared to Eqn.~\ref{eqn: coexactequation}, Eqn.~\ref{eqn: coexactdeltaomega}. The Hodge decomposition holds for arbitrary dimensions, similar to the main text, we begin with a description on 2-d triangulated manifold first, for simplicity. Further generalization to higher dimensions will be left for the next section.

Let $\omega$ be some 1-form. Recall that the Hodge decomposition allows $\omega$ to be written as
\begin{align}
    \omega = \delta \Omega + d\eta + h,
    \label{eqn: hodge1}
\end{align}
where $\Omega$ is a 2-form, $\eta$ is a 0-form and $h$ is called the harmonic form, $d$ is the exterior derivative, which maps from the space of 0-form to 1-form, and $\delta$ is the codifferential operator, which maps a 2-form to a 1-form. We note that $h$ vanishes under $d$ and $\delta$: $dh = 0$ and $\delta h = 0$. These two operators  have a special property: 
$$ d^2 \equiv d \circ d = 0, $$
and 
$$ \delta^2 \equiv \delta \circ \delta = 0.  $$

\noindent
\textbf{Computing 2-form $\Omega$. } Applying the  derivative map 
$d$ to both sides of Eqn.~\ref{eqn: hodge1}, and using the above properties of both the derivative map and  the harmonic form, we obtain:
\begin{align}
    d\omega = d\delta \Omega. 
\end{align}
To proceed, we need the following recipe, which is the discrete form for the boundary/coboundary operator: 
\begin{lemma}[\cite{gu2008computational, gu2023classical}]
\label{lemma: discreteform}
   In the setting of 2-dimensional manifold, let $\{ \sigma_{1_i}\}$ and $\{ \sigma_{2_j}\}$ be the set of 1-simplexes and 2-simplexes (as in the main text). Let $\eta$ be some 0-form, $\omega$ be some 1-form and $\Omega$ be some 2-form. Then for arbitrary 2-simplex $\sigma_{2_j}$, it holds that:
    \begin{align}
        d \omega( \sigma_{2_j} ) = \omega_{\sigma_{2_j}} \big(\partial \sigma_{2_j}\big).
    \end{align}
    In addition, it holds that:
    \begin{align}
        \delta \Omega (\sigma_{1_i}) = \frac{1}{w_{\sigma_{1_i}}} \left( \sum_{j: \sigma_{1_i} \subseteq \sigma_{2_j} } (-1)^{j-1} \frac{\Omega (\sigma_{2_j})}{|\sigma_{2_j}|} \right),
    \end{align}
     where $  w_{\sigma_{2_j}} $ is called cotangent edge weight of $\sigma_{1_i} $ and $|f_j|$ is the area of 2-simplex $\sigma_{2_j}$. Further, we have that:
     \begin{align}
         \delta \omega (\sigma_{0_i}) = - \frac{1}{|\overline{\sigma_{0_i}} |} \sum_{ j, \sigma_{0_i} \subseteq \sigma_{1_j}  } \omega_{j} \omega( \sigma_{1_j}  )
     \end{align}
     where $|\overline{\sigma_{0_i}} | $ is the area of the dual cell of $\sigma_{0_i}$ (e.g., see Appendix \ref{sec: poincaredualcell} for definition and construction of dual cell).
\end{lemma}
More details and explanation of the above derivation can be found in Chapter 16 of Ref.~\cite{gu2023classical}. We comment that the last two formulas are special cases of a more general version, e.g., see Lemma \ref{lemma: generalcodiffoperator} of the next section where we consider a triangulated manifold of higher dimension. 

\begin{figure}[htbp]
    \centering
    \includegraphics[width=0.3\textwidth]{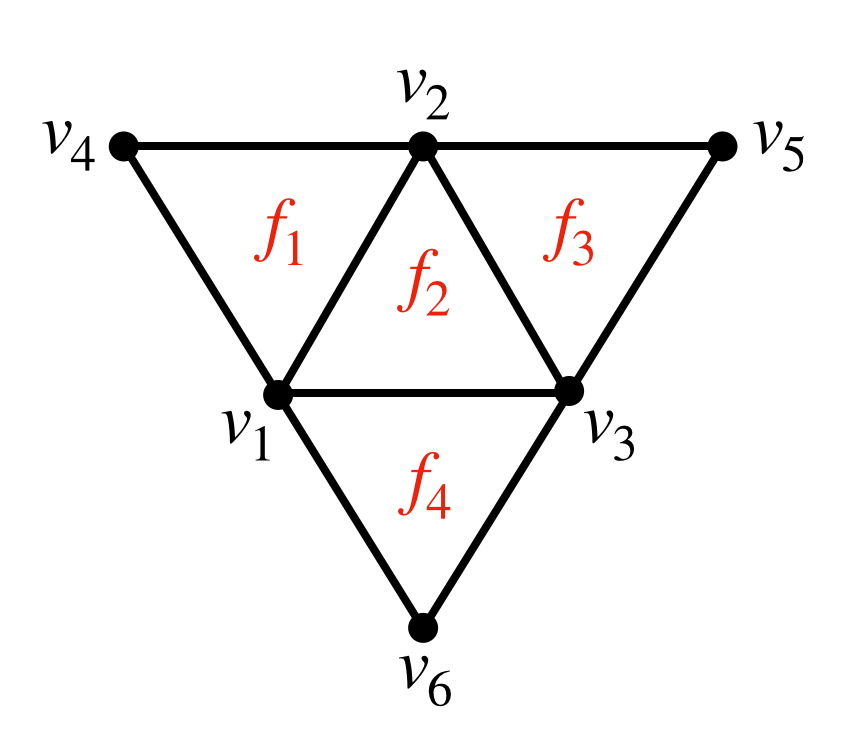}
    \caption{ A part of a triangulated 2-manifold. Each face (2-simplex) is labeled with the red color.}
    \label{fig: hodge}
\end{figure}

Let $[v_1,v_2,v_3]$ be a 2-simplex (see Fig.~\ref{fig: hodge}), then we have:
\begin{align}
    d\omega[v_1,v_2,v_3] = d\delta \Omega [v_1,v_2,v_3].
    \label{eqn: solveforexact}
\end{align}
From Lemma \ref{lemma: discreteform}, we have
\begin{align}
    d\omega[v_1,v_2,v_3] = \omega[v_1,v_2] + \omega[v_2,v_3] + \omega[v_3,v_1]. 
\end{align}
If $\omega$ is known, then we can compute $d\omega[v_1,v_2,v_3]$. Likewise:
\begin{align}
   d\delta \Omega [v_1,v_2,v_3] = \delta \Omega [v_1,v_2] + \delta \Omega [v_2,v_3] + \delta \Omega [v_3,v_1].  
\end{align}
From Lemma \ref{lemma: discreteform}, we then have the following formula for $\delta\Omega [v_1,v_2]$: 
\begin{align}
    \delta \Omega [v_1,v_2] = \frac{1}{ w_{12}} \Big(\frac{\Omega(f_1)}{|f_1|} - \frac{\Omega(f_2)}{|f_2| } \Big),
    \label{eqn: deltaformula}
\end{align}
where $|f_{1,2}|$ is the area of the corresponding faces, or 2-simplexes, and $w_{12}$ is the cotangent edge weight (see in  Fig.~\ref{fig: hodge} for the illustration),
$$ w_{12} = \frac{1}{2}\big( \cot(\angle v_1v_3v_2 ) + \cot(\angle v_1v_4v_2) \big) $$


Recall that we have:
\begin{align}
    d\omega[v_1,v_2,v_3] = d\delta \Omega [v_1,v_2,v_3].
\end{align}
which leads to:
\begin{align}
    &\omega[v_1,v_2] + \omega[v_2,v_3] + \omega[v_3,v_1] \\ &= \frac{1}{ w_{12}} \Big(\frac{\Omega(f_1)}{|f_1|} - \frac{\Omega(f_2)}{|f_2| } \Big) + \frac{1}{ w_{23}} \Big(\frac{\Omega(f_3)}{|f_3|} - \frac{\Omega(f_2)}{|f_2| } \Big) \nonumber
     \\ & \quad+ \frac{1}{ w_{13}} \Big(\frac{\Omega(f_4)}{|f_4|} - \frac{\Omega(f_2)}{|f_2| } \Big)  
\end{align}
where $w_{ij}$ is a cotangent edge weight (of the edge $v_i,v_j$ in the Fig.~\ref{fig: hodge}), and $|f_1|, |f_2|,|f_3|,|f_4|$ is the area of 2-simplexes/triangles in Fig.~\ref{fig: hodge}. In general, these quantities can be computed provided the pairwise distance between the vertices that form the corresponding simplexes. If the input is a uniformly, or nearly uniformly triangulated manifold, which means that all the 2-simplexes or triangles are (almost) equilateral. Therefore, the cotangent edge weight is simply:
$$ w = \frac{1}{2} ( \frac{1}{\sqrt{3}} + \frac{1}{\sqrt{3}} ) = \frac{1}{\sqrt{3}} $$
for all cotangent edge weights (including the internal ones) and the value of $|f|$ is
$$ |f| = \frac{\sqrt{3}a^2}{2}, $$
where $a$ is the distance between arbitrary two connected points, or 0-simplexes. The value $a$, in principle, can be arbitrary, as the length does not affect the topology of a given configuration. Thus, for simplicity, we set $a=1$, which means that:
\begin{align}
    \frac{1}{\omega |f| }= 2
\end{align}
and thus we achieve a simplified equation:
\begin{align}
    &\omega[v_1,v_2] + \omega[v_2,v_3] + \omega[v_3,v_1] \\ &= \frac{1}{\omega |f|} \Big( \Omega(f_1)+ \Omega(f_3) + \Omega(f_4) - 3 \Omega(f_2) \Big) \\
    &= 2 \Big( \Omega(f_1)+ \Omega(f_3) + \Omega(f_4) - 3 \Omega(f_2) \Big)
\end{align}
The above equation, which features a linear combination, could be written in the following matrix-vector form:
\begin{align}
\label{eqn: a9}
    \begin{pmatrix}
        1 & 1 & 1 \\
    \end{pmatrix}
    \cdot \begin{pmatrix}
        \omega[v_1,v_2]\\
        \omega[v_2,v_3]\\
         \omega[v_3,v_1]
    \end{pmatrix}
    = 2 \begin{pmatrix}
        1 & -3 & 1 & 1 \\
    \end{pmatrix}
    \cdot \begin{pmatrix}
        \Omega(f_1)\\
        \Omega(f_2) \\
        \Omega(f_3) \\
        \Omega(f_4) 
    \end{pmatrix}.
\end{align}
Similarly, if we proceed with the action of $d\delta \Omega$ to different faces, we obtain a list of dot product of two vectors similarly to the right-hand side of the above equation, which constitutes a linear action of a matrix on a vector $\Vec{\Omega} = \big( \Omega(f_1), \Omega(f_2), ..., \Omega(f_{c_2})  \big)^T$. To see this, we note that from the above equation, we can embed it in a higher dimension one as follows: 
\begin{align}
\label{eqn: a10}
    & 2\begin{pmatrix}
        1 & -3 & 1 & 1 \\
    \end{pmatrix}
    \cdot \begin{pmatrix}
        \Omega(f_1)\\
        \Omega(f_2) \\
        \Omega(f_3) \\
        \Omega(f_4) 
    \end{pmatrix}\nonumber \\
    &\rightarrow 
    2\begin{pmatrix}
        1 & -3 & 1 & 1 & 0 & ... & 0 \\
    \end{pmatrix}
    \cdot \begin{pmatrix}
        \Omega(f_1)\\
        \Omega(f_2) \\
        \Omega(f_3) \\
        \Omega(f_4)  \\
        \vdots  \\
        \Omega(f_i) \\
        \vdots \\
        \Omega(f_{|S_2^K|}) 
    \end{pmatrix}.
\end{align}
By defining $A^i$ to be a $1 \times |S_2^K|$ vector as the left term on the above product, then the above equation is essentially a dot product $A^i \Vec{\Omega}$. By packing all the action of $d\Omega$ to all faces, or equivalently, organizing $\{A^i\}$ as rows of a matrix $A$, we obtain exactly $A \Vec{\Omega}$. The non-zero entries of matrix $A$ in Eqn. \ref{eqn: coexactdeltaomega} is $1/\omega |f|$ for off-diagonal elements and $-3/\omega |f|$ for on-diagonal elements, as derived from above. The location of these non-zero entries depends completely on the mutual relation between a pair of faces, as from the above construction, we see that all faces $f_1,f_3,f_4$ are adjacent to face $f_2$. Therefore, if we treat each face as a vertex, and put an edge connecting two vertices if their corresponding faces (in the given triangulated manifold) share an edge, then it results in the so-called ``dual'' graph. Then matrix $A$ is essentially the adjacency matrix of such a graph, with the diagonal element being $-3$ instead (the regular adjacency matrix has diagonals being 0). We summarize the property of $A$ as follows. 
\begin{center}
    \textit{ The matrix $A$ of Eqn.~\ref{eqn: coexactdeltaomega} has size $|S_2^K| \times |S_2^K|$ with $-6$ on the diagonals. For an $i$-th row, the off-diagonal elements of $A$ are 2 and their column indexes are those faces, or 2-simplexes that are adjacent to $\sigma_{2_i}$. The Frobenius norm of $A$ is $||A||_F = \sqrt{40}|S_2^K| $}
\end{center}
Thus, the first part of Lemma \ref{lemma: ACP} is proved. 

The above matrix has diagonal elements to be the same, and off diagonal elements to be the same, thus being appropriate to the second version of Lemma \ref{lemma: entrycomputablematrix}, which allows such a matrix to be block-encoded (up to a factor of Frobenius norm) with very modest classical pre-processing time. We note that if the input triangulation is not uniform, then the algorithm still works, because
\begin{align}
     &\frac{1}{ w_{12}} \Big(\frac{\Omega(f_1)}{|f_1|} - \frac{\Omega(f_2)}{|f_2| } \Big) + \frac{1}{ w_{23}} \Big(\frac{\Omega(f_3)}{|f_3|} - \frac{\Omega(f_2)}{|f_2| } \Big) \\
      & \quad+ \frac{1}{ w_{13}} \Big(\frac{\Omega(f_4)}{|f_4|} - \frac{\Omega(f_2)}{|f_2| } \Big)  
\end{align}
is still a linear combination of entries of $\Vec{\Omega}$. Therefore, via the same construction as above, we can still obtain $A \Vec{\Omega}$. In fact, the matrix $A$ has the same structure: for a given row, the column indexes of those non-zero entries are the same as above, but the value of these entries are changed (according to $\frac{1}{w_{ij} |f_k|} $). 

Now we consider the left-hand side of Eqn. \ref{eqn: a9} (again in the uniform/nearly uniform setting), which is obtained by the action of $d\omega$ on all faces, e.g., $[v_1,v_2,v_3]$. Recall that $d\omega([v_1,v_2,v_3] = \omega([v_1,v_2]) + \omega([v_2,v_3]) + \omega([v_3,v_1])$, which suggests that the summation is done over all edges on the face $[v_1,v_2,v_3]$. Extension on different faces is straightforward to form the linear matrix action $C \Vec{\omega}$. More specifically, we perform a similar embedding to a higher dimension as above:
\begin{align}
    \begin{pmatrix}
        1 & 1 & 1 \\
    \end{pmatrix}
    \cdot \begin{pmatrix}
        \omega[v_1,v_2]\\
        \omega[v_2,v_3]\\
         \omega[v_3,v_1]
    \end{pmatrix} \nonumber \\
    \longrightarrow \begin{pmatrix}
        1 & 1 & 1 & 0 & 0 & ... & 0 
    \end{pmatrix} \begin{pmatrix}
        \omega (e_2) \\
        \omega( e_2) \\
        \vdots \\
        \omega (e_{|S_1^K|})
    \end{pmatrix},
\end{align}
where we have used $e_i$ to denote the $i$-th edge, or 1-simplex, for brevity. By taking the action of $d\omega$ on all faces, we obtain a system of inner products as above, which can be packed into matrix form  $ C \overrightarrow{\omega}$. Therefore, the location of these non-zero entries is characterized completely by which three edges enclose a face. The structure of $C$ is summarized in the following:
\begin{center}
    \textit{The matrix $C$ in Eqn.~\ref{eqn: coexactdeltaomega} has size $|S_2^K|\times |S_1^K|$. For a given row $i$-th, the $C$ has 3 non-zero elements (which are 1), and their column positions are those edges, or 1-simplexes that are boundary of $\sigma_{2_i}$. The Frobenius norm is $ \sqrt{3}| S_2^K|$.}
\end{center}
thus proving the second part of Lemma \ref{lemma: ACP}.

Again, the above matrix has all non-zero entries to be 1, thus the second version of Lemma \ref{lemma: entrycomputablematrix} is applicable. The above details reveal that the crucial input to our algorithm is the classical knowledge which encodes the relation between different building cells, e.g., edges (1-simplexes), triangles (2-simplex), etc. 

In the above, we have provided the structure of $A$ and $C$. Now we complete the remaining matrix appeared in Eqn.~\ref{eqn: coexactdeltaomega}, which is $P$. We recall that $P$ encodes the action of $\delta$ on the 2-form $\Omega$ as $\overrightarrow{\delta \Omega} = P \cdot \Vec{\Omega}$. From Lemma \ref{lemma: discreteform}, we have that, for a 1-simplex $\sigma_{1_i}$, 
\begin{align}
        \delta \Omega (\sigma_{1_i}) = \frac{1}{w_{\sigma_{1_i}}} \left( \sum_{j: \sigma_{1_i} \subseteq \sigma_{2_j} } (-1)^{j-1} \frac{\Omega (\sigma_{2_j})}{|f_j|} \right) 
\end{align}
Note that the right-hand side of the above equation can be expressed as $P^i \cdot \Vec{\Omega}$ where $ \Vec{\Omega}$ was defined earlier, and $P^i$ is a vector of dimension $1 \times |S_2^k|$. The non-zero entries of this vector are $\pm \frac{1}{\omega |f|}$ (assuming uniform/nearly uniform triangulation) which is 2 as we pointed out before, and their column indexes are those triangles or 2-simplexes that includes $\sigma_{1_i}$ as a joint faces. The matrix $P$ is formed by organizing all the vectors $\{ P^i\}$ as rows of $A$. We thus have the following:
\begin{center}
    \textit{The matrix P of Eqn.~\ref{eqn: coexactdeltaomega} has size $|S_1^K| \times |S_2^K|$. For a given row $i$-th, $P$ has 2 non-zero entries element, which are $2$ and $-2$. Their column indexes are those 2-simplexes, or triangles that include $\sigma_{1_i}$ as a face/ or a part of the boundary. The Frobenius norm $||P||_F$ is $ \sqrt{2} |S_1^K|$.}
\end{center}
which completes the last part of Lemma \ref{lemma: ACP}.

\noindent
\textbf{Computing 0-form $\eta$. }  In an analogous manner, we elaborate on the computation of the 0-form $\eta$ in Eqn.~\ref{eqn: hodge1}. By applying $\delta$ to both sides of Eqn.~\ref{eqn: hodge1}, $\omega = \delta \Omega + d\eta + h$, we have:
\begin{align}
    \delta \omega = \delta d\eta
\end{align}
\begin{figure}[htbp]
    \centering
    \includegraphics[width = 0.3 \textwidth ]{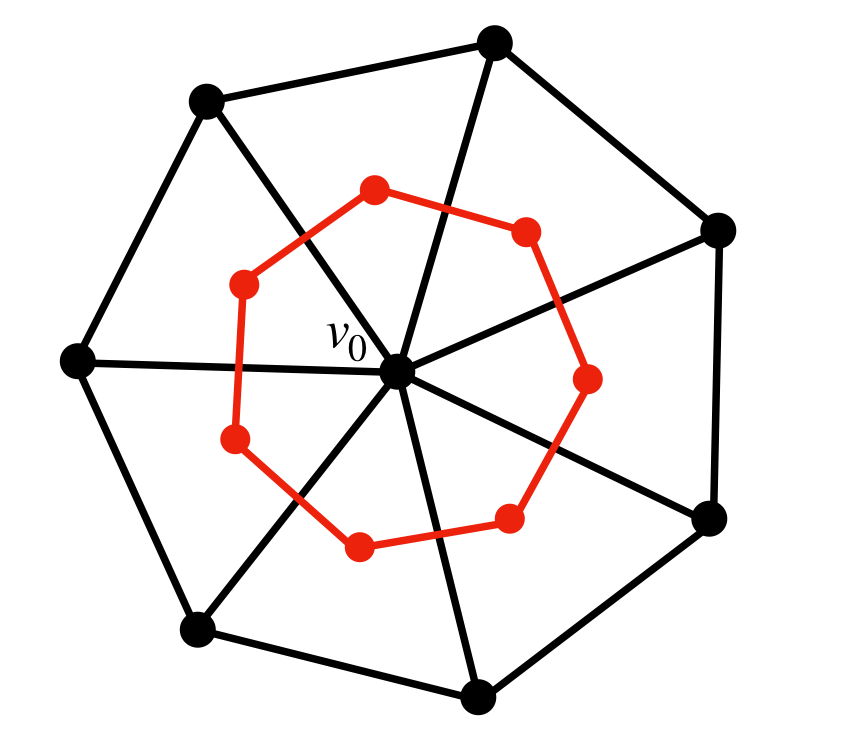}
    \caption{Illustration of dual cell of 0-simplex $v_0$. The whole region inside the red-colored boundaries is the dual cell of $v_0$, with area denoted as $\Tilde{|v_0|}$. This area could be computed given the pairwise distance between $v_0$ and its adjacent vertices. In our case where the given triangulation is nearly uniform and the distance between two vertices is $a$, it is easy to calculate the area $|\Tilde{v_0}|$, which is $p a^2/ 4\sqrt{3}$, where $p$ is the number of adjacent points to $v_0$. }
    \label{fig: 0form}
\end{figure}
As $\delta\omega$ is a 0-form, according to Lemma \ref{lemma: discreteform}, the formula for $\delta \omega$ acting on some 0-simplex or a point $v_0$ is: 
\begin{align}
    \delta\omega [v_0] = -\frac{1}{\overline{|v_0|}} \sum_{v_i \sim v_0} w_{0j} \cdot \omega[v_0,v_j],  
    \label{eqn: deltaomega}
\end{align}
where the symbol $\sim$ denotes the adjacency (i.e., vertex $v_j$ is connected to $v_0$ via an edge), $w_{0,j}$ is the cotangent edge weight (defined above), and $|\overline{v_0}|$ is the area of the dual cell complex of $v_0$ (the area enclosed by the red curves in the Fig.~\ref{fig: 0form}). We note that we will elaborate further the dual cell complex in the subsequent section. 

The formulation of $\delta d\eta$ is the same, as $d\eta$ is also a 1-form. More explicitly, we have:
\begin{align}
    \delta d\eta[v_0] &= -\frac{1}{\overline{|v_0|}} \sum_{v_i \sim v_0} w_{0j} \cdot d\eta[v_0,v_j]\\& = -\frac{1}{\overline{|v_0|}} \sum_{v_i \sim v_0} w_{0j} \cdot ( \eta[v_j] - \eta[v_0] ).
\end{align}
Since $\delta \omega = \delta d\eta$, the area term $\Tilde{|v_0|}$ cancels out, fortunately. In a very similar fashion to the previous case with 2-form $\Omega$, by applying the above formulas for all vertices, eventually, we obtain a linear equation for $\eta$ with known coefficients $w$'s, i.e.,  from the cotangent edge weights.

\section{General Version of Discrete Hodge Theory}
\label{sec: generalhodge}
As we have emphasized, going to higher dimensional manifolds incurs more complication, as the rule for $d\delta$ and $\delta d$ that we described in the previous section will no longer hold exactly in the same form. The main reason is due to the use of the Poincar\'e duality. The dual cell to each simplex is generally different for different dimensions. Hence, it affects how we formulate the codifferential operator $\delta$ (and its composition with $d$). We remark that the underlying theory of our algorithm relies on results from non-trivial geometry and topology, which means that we do not expect to cover everything here, and we refer the readers to the literature, such as~\cite{nakahara2018geometry}, and particularly, more relevant to our perspective from computational algorithms, Ref.~\cite{gu2008computational}. In this section, we aim to exemplify the subtle points regarding the generalization of our algorithm to higher dimensions in more detail than what we have described in Sec.~\ref{sec: generalization}. 

\subsection{Poincar\'e Dual Cell Complex}
\label{sec: poincaredualcell}
We first review the concept of Poincar\'e dual cell complex, which is generally related to the concept of CW complex. The key point is that, given a triangulated manifold of some dimension $n$, the dual cell of some $k$-simplex is not necessarily a proper simplex. A standard procedure to construct the Poincar\'e dual cell can be found in Section 3.3 of \cite{hatcher2005algebraic}, Section V.2 of \cite{edelsbrunner2010computational} or chapter 9 of \cite{munkres2018elements}. For convenience, we quote the procedure in the following lemma. 
\begin{lemma}[Section 3.3 of \cite{hatcher2005algebraic}, Chapter 9 of \cite{munkres2018elements}]
\label{lemma: poincaredualityconstruction}
    Let $M$ be a triangulation of a $n$-dimensional manifold. Let $\sigma_{r}$ be some $r$-simplex in the triangulation. \\
    \noindent
    \textbf{Step 1. Barycentric Subdivision} 
    \begin{itemize}
        \item Subdivide $M$ by inserting the barycenter of each simplex $\sigma \in M$.
        \item For each simplex $\sigma_r$, its barycenter is denoted as $B(\sigma_r)$.
    \end{itemize}
    \noindent
    \textbf{Step 2. Dual Cell as Union of Barycenter Simplices} 
    \begin{itemize}
        \item The dual cell $\overline{\sigma}$ is the union of all simplices in the barycentric subdivision whose sequence of barycenters 
        $$ B(\sigma_r)  \leq  B (\sigma_{r+1})  \leq \cdots \leq B(\sigma_n) $$
        start at $\sigma_r$ and ends at an $n$-simplex $\sigma_r \subset \sigma_n$ 
        \item These chains define a cell of dimension $n-r$ that intersects $\sigma^r$ at the barycenter. 
    \end{itemize}
\end{lemma}

\begin{figure*}[htbp]
    \centering
    \includegraphics[width = 0.6\textwidth]{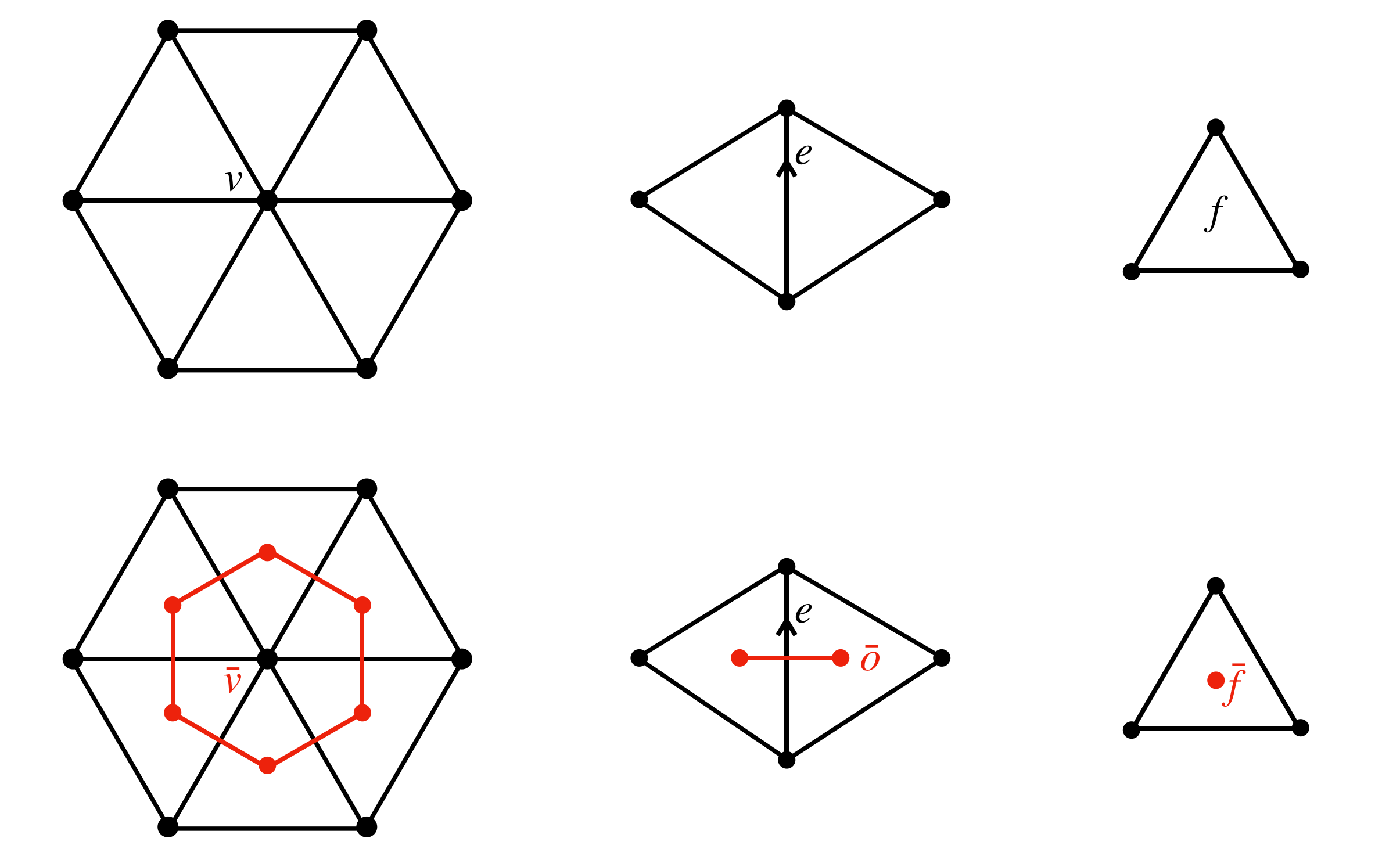}
    \includegraphics[width = 0.6\textwidth]{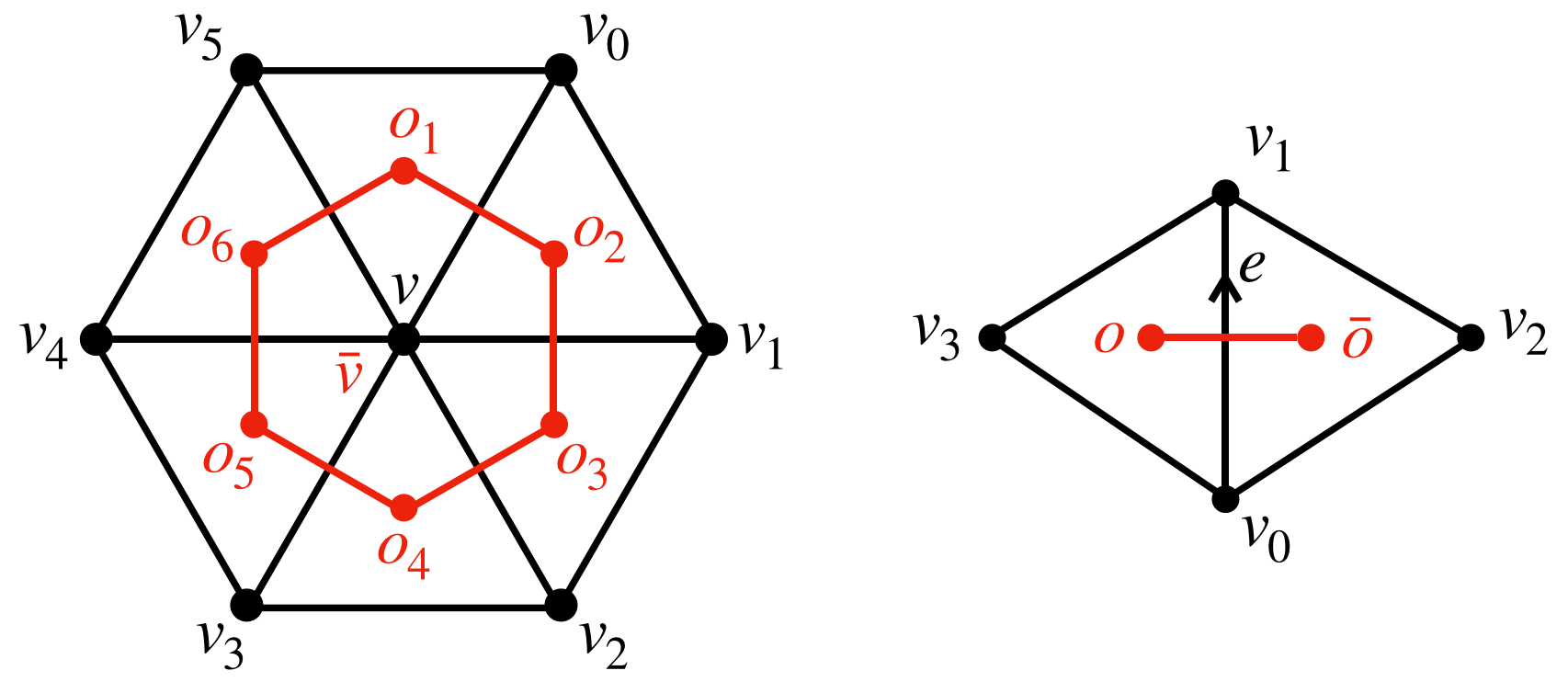}
    \caption{Dual cell complex in 2-d case. }
    \label{fig: duallcell2d}
\end{figure*}

As we can see from Fig.~\ref{fig: duallcell2d}, the order of the given simplex (e.g., 0, 1, or 2-simplex) and of its dual cell match the dimension of a given manifold, which is 2. In the 2d case, the dual cell to a 2-simplex, denoted by $f$ (rightmost ones), is a point $\Bar{f}$ (of zero order), which is the center of the circle that passes all three points of the original 2-simplex $f$. In the middle two figures, the 1-simplex $e$ has the dual cell $\Bar{e}$. The way $\Bar{e}$ is constructed is as follows. First, we construct the dual cell of the two triangles, or 2-simplexes, that are adjacent to $e$, which are two points. Then we connect them, hence forming a line $\Bar{e}$. Likewise, for the leftmost figures, the dual cell of a given point or 0-simplex $v$   is formed by connecting the dual cells of all triangles (2-simplex) that $v$ belongs to, including the interior. Hence, according to the bottom left figure, the dual cell of $v$, $\Bar{v}$ is the region inside the red lines, including the boundary. A similar construction holds for a triangulated manifold of arbitrary dimension. 

\subsection{Codifferential Operator}
\label{sec: generalcodifferentialoperator}
In the previous section, we have provided a formal definition as well as the construction of Poincar\'e dual cell complex. The formal definition of $\delta $ involves cap product and Poincar\'e duality, which can be found in Section 3.3 of \cite{hatcher2005algebraic}. A discrete form can be found in Chapter 16 of \cite{gu2023classical} or \cite{gu2008computational}, or \cite{edelsbrunner2010computational}. For completeness and convenience, we summarize the procedure in the following lemma.
\begin{lemma}[\cite{edelsbrunner2010computational, hatcher2005algebraic}, Chapter 16 of \cite{gu2023classical}]
\label{lemma: generalcodiffoperator}
    Let $M$ be a triangulation of an $n$-dimensional manifold. The following operators are linear operators defined on $M$, or more precisely, on the tangent bundle of $M$.
    \begin{itemize}
        \item The codifferential operator $\delta: \Omega^{r} \rightarrow \Omega^{r-1}$, where $\Omega^r$ is the space of $r$-form $(r \leq n)$. 
        \item The Hodge star operator $\star: \Omega^r (M) \longrightarrow \Omega^{n-r}(M)$ with its inverse is defined as $\star^{-1} = (-1)^{r (n-r)} \star$
    \end{itemize}
    Let $\sigma_r$ be some $(r-1)$ simplex, and $\overline{\sigma}$ be its corresponding Poincar\'e dual-cell complex. We have the following relation~\cite{gu2023classical, hatcher2005algebraic, munkres2018elements}: 
    \begin{align}
        \frac{\omega(\sigma_r) }{|\sigma_r|} = \frac{\star \omega( \overline{\sigma_r} )}{| \overline{\sigma_r} |},
    \end{align}
    where $|.|$ refers to the Hausdorff measure, e.g., length, surface area, volume, etc. The codifferential operator $\delta$ is defined as:
    \begin{align}
        \delta = (-1)^r \star^{-1} d \star,
    \end{align}
    where $d$ is the usual coboundary operator: let $\sigma_r$ be some $r$-simplex, $\omega^{r-1}$ be some $(r-1)$-form then
    \begin{align}
     (d\omega^{r-1}) (\sigma_{r-1}) = \omega^{r-1} \partial \sigma_{r-1}.
    \end{align}
    Let $\sigma$ be some $(r-1)$-simplex and $\overline{\sigma}$ be its dual cell. Let $\partial \Bar{\sigma} = \sum_i \gamma_i$ be the boundary of $\Bar{\sigma}$. Let $\Delta$ be some $r$-form. The following formula features an action of $\delta$ in an arbitrary dimension: 
\begin{align}
    \delta \Delta (\sigma) = (-1)^{nr+n+1} (-1)^{(n-r+1)(r-1)} \frac{|\sigma|}{ |\overline{\sigma} |} \sum_i \frac{ |\gamma_i | }{|\overline{\gamma_i}|} \Delta ( \overline{\gamma_i}).
    \label{eqn: generaldelta}
\end{align}
In particular, $\overline{\gamma_i}$ is an $r$-simplex in $M$ and $\sigma \subseteq \overline{\gamma_i}$.
\end{lemma}
We recall that earlier, we encountered a similar formula, Lemma \ref{lemma: discreteform}, which is in fact the above lemma applied to the 2-dimensional setting. Generally, given an $n$-dim triangulated manifold, we can, in principle, construct the corresponding dual-cell complex and can then carry out the formula for $\delta$ algorithmically. The above formula is somewhat complicated, but it is straightforward since, in the previous section, we have provided a systematic way to construct a dual-cell complex. The most important aspect is that, the right-hand side of the above equation contains a linear combination $\sum_i \Delta ( \bar{\gamma_i})$, and hence eventually it reduces to a linear equation Eqn.~\ref{eqn: generallinearequation}.

\subsection{Proof of Lemma \ref{lemma: generallinearequation}}
\label{sec: proof}
The proof of this lemma is a straightforward corollary of the above lemma, e.g., Lemma~\ref{lemma: generalcodiffoperator}. We recall the linear equation in Eqn.~\ref{eqn: generallinearequation}:
\begin{align}
    A^r \overrightarrow{\Omega} = C^r \overrightarrow{ \omega^r}.
\end{align}
The left-hand side of the above equation is obtained by considering the action $d \delta \ \Omega(\sigma_{ (r+1)_i})$ on all $(r+1)$-simplexes. For a simplex $\sigma_{(r+1)_i}$, we have:
\begin{align}
    d \delta \ \Omega(\sigma_{ (r+1)_i}) &= \delta \Omega ( \partial \sigma_{ (r+1)_i} ) \\
    &= \delta \Omega ( \sum_{j, \sigma_{r_j} \subseteq \sigma_{(r+1)_i} } \sigma_{r_j}  ).
\end{align}
Because $ \sigma_{ (r+1)_i}$ is a $(r+1)$-simplex, its boundary contains $(r+2)$ r-simplexes. Thus, the summation $\sum_{j, \sigma_{r_j} \subseteq \sigma_{(r+1)_i} } \sigma_{r_j}  $ contains $r+2$ terms. Now we consider the action of $\delta \Omega $ on a single $ \sigma_{r_j} $. According to Lemma~\ref{lemma: generalcodiffoperator}, or more precisely, Eqn.~\ref{eqn: generaldelta}, as $ \sigma_{r_j} $ is a $r$-simplex, we have:
\begin{widetext}
    \begin{align}
    \delta \Omega (\sigma_{r_j}) = (-1)^{nr+n+1} (-1)^{(n-r+1)(r-1)} \frac{|\sigma_{r_j}|}{ |\overline{\sigma_{(r+1)_k}} |} \sum_{k, \sigma_{r_j} \subseteq \sigma_{(r+1)_k}  } \frac{ \overline{|\sigma_{(r+1)_k |} }}{|\sigma_{(r+1)_k}|} \Omega (  \sigma_{(r+1)_k} ).
    \label{eqn: e8}
\end{align}
Thus, we have: 
    \begin{align}
    d \delta \ \Omega(\sigma_{ (r+1)_i}) &= \delta \Omega ( \partial \sigma_{ (r+1)_i} ) = \delta \Omega ( \sum_{j, \sigma_{r_j} \subseteq \sigma_{(r+1)_i} } \sigma_{r_j}  ) =  \sum_{j, \sigma_{r_j} \subseteq \sigma_{(r+1)_i} } \delta \Omega \sigma_{r_j}  \\
    &= \sum_{j, \sigma_{r_j} \subseteq \sigma_{(r+1)_i} } (-1)^{nr+n+1} (-1)^{(n-r+1)(r-1)} \frac{|\sigma_{r_j}|}{ \overline{|\sigma_{(r+1)_k} |}} \sum_{k, \sigma_{r_j} \subseteq \sigma_{(r+1)_k}  } \frac{ \overline{|\sigma_{(r+1)_k |} }}{|\sigma_{(r+1)_k}|} \Omega (  \sigma_{(r+1)_k} ),
\end{align}
\end{widetext}
which is a linear combination of entries of $\Vec{\Omega} =\Big(  \Omega ( \sigma_{(r+1)_i}  )   \Big)^T$. We note that the involvement of two summations $ \sum_{j, \sigma_{r_j} \subseteq \sigma_{(r+1)_i} }  \sum_{k, \sigma_{r_j} \subseteq \sigma_{(r+1)_k}  } $ result in a summation over those $(r+1)$-simplexes that are adjacent to $\sigma_{(r+1)_i}$. Further, we note that the above can be rewritten as $  d \delta \ \Omega(\sigma_{ (r+1)_i})  = (A^r)^i \Vec{\Omega}$, where $ (A^r)^i$ is a $1 \times |S_{r+1}^K|$ vector. Thus, by organizing $ \{ (A^r)^i \}$ as rows, we obtain the matrix $A^r$ in Eqn.~\ref{eqn: generallinearequation}. For a given row $(A^r)^i$, the value and position of non-zero entries of $(A^r)^i$ is determined via the above equation. The locations of non-zero entries corresponds to those $ (r+1)$-simplexes that are adjacent, i.e., sharing a common $r$-simplex with $\sigma_{(r+1)_i}$. As a $(r+1)$-simplex has $(r+2)$ faces, the number of non-zero entries are $r+2$, which is $\mathcal{O}(r)$. Thus, the sparsity of $A^r$ is $\mathcal{O}(r)$, which was stated in Lemma~\ref{lemma: generallinearequation}. 

For the remaining part of Lemma \ref{lemma: generallinearequation}, we note that, given a $r$-form $\omega^r$ and $(r+1)$-simplex $\sigma_{(r+1)_i}$, then from Lemma~\ref{lemma: generalcodiffoperator}, we have: 
\begin{align}
    (d \omega^r) (\sigma_{(r+1)_i}) &= \omega^r \partial \sigma_{(r+1)_i} \\
    &= \omega^r \sum_{j, \sigma_{r_j} \subseteq \sigma_{(r+1)_i}} \sigma_{r_j} \\
    &= \sum_{j, \sigma_{r_j} \subseteq \sigma_{(r+1)_i}} \omega^r (\sigma_{r_j}),
\end{align}
which is a linear combination of entries of $\Vec{\omega^r}$ (with coefficients 1), thus it can be rewritten as $(C^r)^i \cdot \Vec{\omega^r}$ where $(C^r)^i  $ is a $1 \times |S_r^K|$ vector. By considering the action of $d \omega^r $ on all $(r+1)$-simplexes, we obtain a matrix-vector product $C^r  \Vec{\omega^r}$, where $C^r$ is a matrix of size $|S_{r+1}^K| \times |S_r^K|$. For a given row $i$-th, the locations of non-zero entries of this vector correspond to those $ r$-simplexes along the boundary of $\sigma_{(r+1)_i}$ and apparently, the value of these non-zero entries are 1. It can be seen that an $(r+1)$-simplex has $(r+2)$-simplexes on its boundary. Thus, it implies that the sparsity of $C^r$ is $\mathcal{O}(r)$. To prove the final part involving matrix $P^r$ in Lemma \ref{lemma: generallinearequation}, we simply remind Eqn.~\ref{eqn: e8}, which was derived from Lemma \ref{lemma: generalcodiffoperator}. Essentially, Eqn.~\ref{eqn: e8} features an action of $\delta \Omega$ on $\sigma_{r_j}$, which results in a linear combination of entries of $\Vec{\Omega}$, thus can be rewritten as $(P^r)^i \Vec{\Omega}$. By combining all $r$-simplexes, we can obtain $P^r \Vec{\Omega}$. For a given row $i$-th, the locations of non-zero entries corresponding to those $(r+1)$-simplexes that include $\sigma_{r_j}$ as faces. Thus, the sparsity of $P^r$ is $\mathcal{O}(r)$, and Lemma \ref{lemma: generallinearequation} is completed.

Given distances between points on the original manifold $M$, in principle, we can compute the corresponding length, area, and volume for elements of simplexes of $M$, as well as for their dual cell complex. In the case of uniform/nearly uniform triangulation, the calculation is much easier, as the Hausdorff measure, e.g., length, area, volume, etc., of a uniform simplex can be trivially calculated. Therefore, all the linear systems (such as in Eqn.~(\ref{eqn: coexactequation})) are entry-wise computable.  

\subsection{More examples}
We will exemplify the formula~\ref{eqn: generaldelta} further for both 2-dim and 3-dim cases by direct calculation. In the previous section~\ref{sec: elaboration}, we already encounter the formulation. We will try to reproduce the same formulas for $\delta \Omega$ and $\delta w$ in 2-d case and 3-d case.

\subsubsection{2-dimensional Triangulated Manifold}
Let $\Omega$ be 2-form (or 2-cochain) and $\omega$ be 1-form (or 1-cochain). First, we try to apply~(\ref{eqn: generaldelta}) to compute $\delta \Omega (e)$ on the right of Fig.~\ref{fig: duallcell2d}. According to the figure, the dual cell of the triangle $f_1 = [v_0,v_1,v_2]$ is denoted as $\Tilde{o}$, and the dual cell of the triangle $f_2 = [v_0,v_1,v_3]$ is $o$. The dual cell of $e$, or 1-simplex $[v_0,v_1]$, is $\bar{e} = o\bar{o}$. Applying Eqn.~\ref{eqn: generaldelta}, we have
$$ \delta \Omega (e) = \frac{|e|}{|\bar{e}|} \Big( \frac{|o|}{|\bar{ o }|} \Omega( \bar{o} ) - \frac{ |\Tilde{o}| }{|\bar{\Tilde{o}} | } \Omega( \bar{\Tilde{o}} ) \Big).    $$
Since both $o$ and $\Tilde{o}$ are points, $|o|$ and $| \Tilde{o}|$ are equal to 1. Furthermore, the dual of $o$ and $\Tilde{o}$ are $[v_0,v_1,v_3], [v_0,v_1,v_2] \equiv f_1,f_2$, respectively, as we defined in the previous section.  A result from elementary geometry~\cite{gu2008computational} shows that 
$$ \frac{|\bar{e}|}{|e|} = \frac{1}{2} \big( \cot( \angle v_1v_3v_0) + \cot( \angle v_1v_2v_0 )\big) = w_{12}, $$
which is defined as the cotangent edge weight. Combining everything, we obtain:
\begin{align}
    \delta \Omega(e) = \frac{1}{w_{12}}\Big( \frac{1}{f_1} \Omega(f_1) - \frac{1}{f_2} \Omega(f_2)    \Big), 
\end{align}
which is exactly what we got in Eqn.~\ref{eqn: deltaformula}. A similar calculation can be done for $\delta \omega$, which leads to a similar form as in Eqn.~\ref{eqn: deltaomega}.

\subsubsection{3-dimensional Triangulated Manifold}
\begin{figure}[t]
    \centering
    \includegraphics[width = 0.25\textwidth]{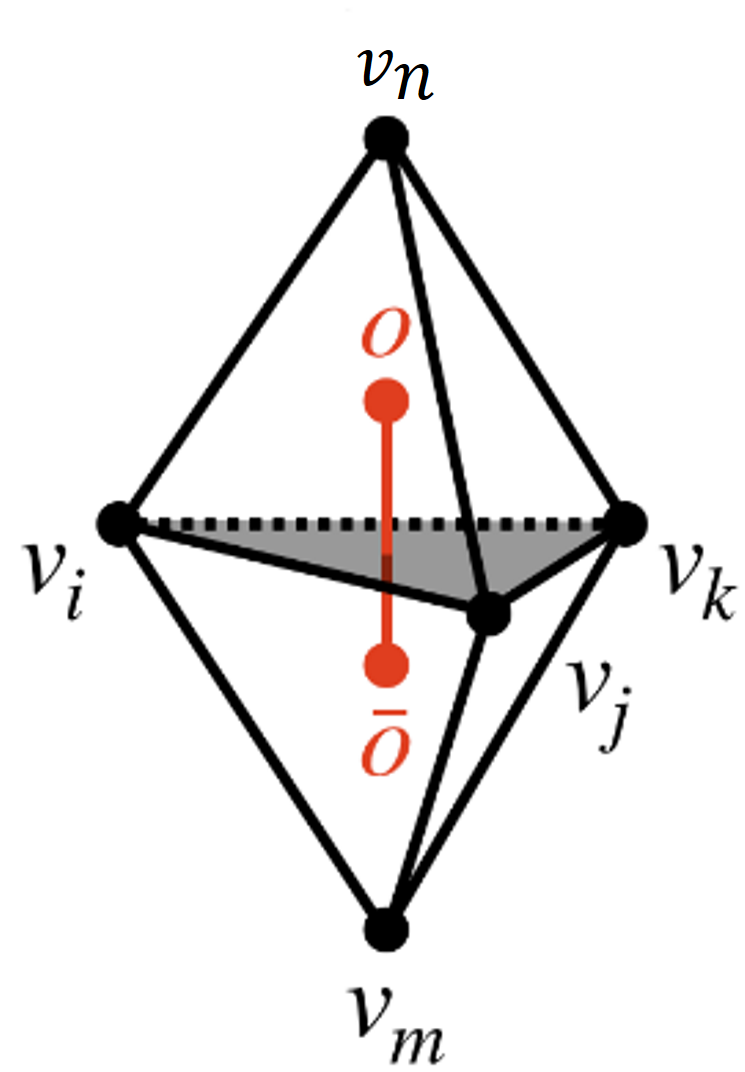}
    \caption{Example in 3-D. }
    \label{fig: 6}
\end{figure}

In the 3d case, the situation is a bit more complicated because the dual of each simplex is no longer the same as in the 2d case. However, the formulation is still straightforward, as we can systematically construct the dual cell and use the same formula ~\ref{eqn: generaldelta}. We work out the following example; see figure~\ref{fig: 6}.

We denote the 2-simplex $\sigma =  [v_i, v_j, v_k]$ and its dual cell $\Bar{\sigma} = o\Bar{o}$; we further denote 3-simplex $[v_n,v_i,v_j,v_k],[v_m,v_i,v_j,v_k]$ as $\alpha, \beta $ . Let $\Omega$ be some 3-form. The boundary of $\Bar{\sigma}$ is:
$$ \partial \Bar{\sigma} = \Bar{o}- o. $$
The action of $\delta \Omega (\sigma)$ is:
\begin{align}
    \delta \Omega(\sigma) =  \frac{ |\sigma| } { |o\Bar{o}| } \Big( \frac{1}{ |\alpha |} \Omega( \alpha   ) - \frac{1}{|\beta|} \Omega(\beta)  \Big).
\end{align}
Since the 3-simplex is a tetrahedron, its volume is $V = |\alpha| = |\beta| = a^3/6\sqrt{2}$, and the length of $o\Bar{o}$ is $a/\sqrt{6} $; the area of $\sigma$ is $|\sigma| = \sqrt{3}a^2 /4$.

\section{Inverting $\frac{A}{||A||_F}$ to obtain $\frac{A^{-1}}{\kappa_A ||A||_F}$}
\label{sec: invertinglarge}
We remind that a direct application of Lemma \ref{lemma: matrixinversion} to invert $  \frac{A}{||A||_F}$ and obtain (the block-encoding of) $ \frac{A^{-1}}{\kappa_A} $ achieves complexity $\mathcal{O}(||A||_F)$ (we are ignoring other factors),  due to the fact that the minimum eigenvalue of  $\frac{A}{||A||_F}$ is $\mathcal{O}(\frac{1}{||A||_F})$. In the following, we show that we can slightly modify the technique behind Lemma \ref{lemma: matrixinversion} to obtain the block-encoding of $ \frac{A^{-1}}{||A||_F \kappa_A}  $ instead.  

The key recipe underlying the Lemma \ref{lemma: matrixinversion} is the polynomial approximation to $\frac{1}{x}$ for $ |x| \in (\frac{1}{\kappa}, 1)$ for some $\kappa \geq 1$. According to \cite{gilyen2019quantum, childs2017quantum}, there exists a polynomial $P(x) = \sum_{i=1}^p a_i x^i$ of degree $p = \mathcal{O}\left(  \frac{1}{\kappa} \log \frac{1}{\epsilon} \right)$ such that:
\begin{align}
   \Big\vert  P(x) - \frac{1}{x} \Big\vert_1 \leq \epsilon,
\end{align}
where $|.|_1$ refers to $l_1$ norm. The above equations also imply that:
\begin{align}
       \Big\vert  P(x) - \frac{1}{x\kappa } \Big\vert_1 \leq \frac{\epsilon}{\kappa} \leq \epsilon
\end{align}
Suppose that the matrix $A$ has eigenvalues in the range $ (\frac{1}{\kappa_A},1)$, e.g., $\frac{1}{\kappa_A} \leq |A| \leq 1$ (where $|.|$ refers to the operator norm), and $\kappa_A > 1$. Using this polynomial, one can take advantage of the quantum singular value transformation technique \cite{gilyen2019quantum} or the linear combination of unitaries plus the Hamiltonian simulation technique \cite{childs2017quantum} to transform the singular values (or eigenvalues if $A$ is Hermitian) of $A$ into their respective reciprocal. In other words, one can construct the (block encoding of) operator $P(x)= \sum_{i=1}^p a_i x^i$ such that:
\begin{align}
    \Big\vert P(A) - \frac{A^{-1}}{\kappa_A} \Big\vert \leq \epsilon.
\end{align}
The reason for the division of $\kappa$ is to guarantee that the operator $ \frac{A^{-1}}{\kappa_A} $ has norm bounded by 1. 

Now suppose that instead of $A$, we want to obtain $\frac{A^{-1}}{\kappa_A ||A||_F}$ from $A/||A||_F$. We consider the following polynomial:
\begin{align}
   P'(x) = \sum_{i=1}^p \frac{a_i}{||A||_F^{p-i}} \frac{1}{||A||_F^i}x^i = \frac{1}{||A||_F^{p}}  P(x)
\end{align}
Then we have that:
\begin{align}
    \Big\vert P'(x) - \frac{1}{||A||_F^p } \frac{1}{x \kappa_A} \Big \vert \leq \frac{1}{||A||_F^p} | P(x) - \frac{1}{x\kappa_A} |  \leq \epsilon
\end{align}
Thus if we choose $P'(x)$ to transform the block-encoded operator $A/||A||_F$ via QSVT, then we obtain the block-encoding of:
\begin{align}
    P'(A) = \frac{1}{||A||_F^p} P(A) \approx \frac{1}{||A||_F^p} \frac{A^{-1}}{\kappa_A}
\end{align}

\section{Stochastic rank estimation and another simple rank estimation}
\label{sec: stochasticrankestimation}
In this section, we first elaborate more on the stochastic rank estimation method proposed in \cite{ubaru2016fast,ubaru2017fast,ubaru2021quantum}. Then we describe another method for rank estimation, building upon one recipe from stochastic rank estimation. Let $A \in \mathbb{C}^N$ be a matrix of interest with operator norm $ \frac{1}{\kappa_A}\leq |A| \leq 1$. The goal is to estimate its rank. 
\subsection{Stochastic rank estimation}
The underlying strategy of the stochastic rank estimation method consists of two parts. First, the trace of a matrix $A$ is estimated by generating random vector states $\ket{v_l}$ with random i.i.d. entries for $l=1,2,...,L$, followed by an average over the moments:
\begin{align}
    \Tr (A) \approx \frac{1}{L} \sum_{l=1}^L \bra{v_l} A \ket{v_l}.
\end{align}
According to \cite{ubaru2016fast,ubaru2017fast}, any vector with zero mean and uncorrelated coordinates can be used. In addition, choosing $L = \mathcal{O}\left( \frac{1}{\epsilon^2} \right) $ guarantee the estimation above has precision $\epsilon$. 

Next, we perform the step function $h(.)$ on $A$, to obtain $h(A)$. Then the rank estimation is approximated by:
\begin{align}
    \rm rank (A) \approx \Tr h(A)
\end{align}
According to \cite{gilyen2019quantum}, for $x$ with $|x| \in ( \frac{1}{\delta},1)$, there is a polynomial of degree $p = \mathcal{O}(\frac{1}{\delta}\log \frac{1}{\epsilon})$ that approximates $h(x)$ to a precision $\epsilon$. Denote this polynomial by $P(x) = \sum_{i=1}^p a_i x^i $. So, assuming that the eigenvalues of $A$ is falling wihtin $(\frac{1}{\delta},1)$ (in magnitude), the step function $h(A)$ can be approximated as:
\begin{align}
    h(A) \approx \sum_{i=1}^p a_i A^i
\end{align}
So the trace of $h(A)$ is:
\begin{align}
    \Tr  h(A) \approx \sum_{i=1}^p a_i \Tr( A^i ) \approx \sum_{i=1}^p a_i \sum_{l=1}^L \frac{1}{L} \bra{v_l} A^i \ket{v_l}.
\end{align}

\subsection{Another approach for rank estimation (also Proof of Lemma \ref{lemma: traceestimation})}
The above method, relies on approximating the trace by computing its moments. Another straightforward solution to the estimation of $\Tr (A)$ is to calculate:
\begin{align}
    \sum_{i=1}^N \bra{i} A\ket{i}
\end{align}

In a quantum context, suppose that we have the block-encoding of $A$. Then we can use Lemma \ref{lemma: tensorproduct} to construct the block-encoding of $\Ibb_N \otimes A$, denoted by $U_A$. To estimate the trace, we first prepare the state $\ket{\bf 0} \sum_{i=1}^N \frac{1}{\sqrt{N}}\ket{i-1}\ket{i-1}$. Then we use $U_A$ to apply to this state, then according to Definition \ref{def: blockencode}, it yields the state:
\begin{align}
    \ket{\phi_1}   = \ket{\bf 0} \frac{1}{\sqrt{N}} \ket{i-1}\sum_{i=1}^N A \ket{i-1} + \ket{\rm Garbage}
\end{align}
The overlaps between the above state and $\ket{\bf 0} \sum_{i=1}^N \frac{1}{\sqrt{N}}\ket{i-1} $ is $\frac{1}{N} \sum_{i=1}^N \bra{i-1}\bra{i-1} \big(\ket{i-1}A \ket{i-1} \big)  = \frac{1}{N} \Tr A$. Via Hadamard test, or SWAP test, this overlaps can be estimated to an additive accuracy $\epsilon$,  with success probability $1- \xi$, using a quantum circuit of complexity $\mathcal{O}\left(  \frac{1}{\sqrt{\epsilon}}  \log \frac{1}{\xi}\right)$. 

For completeness, we describe the procedure behind Hadamard test as follows. Supose that we desire to estimate the overlaps between $\ket{\phi_1},\ket{\phi_2} $ with known preparation unitaries $U_1, U_2$. Then we first generate the state $\frac{1}{\sqrt{2}}\left( \ket{0 } + \ket{1}  \right)\ket{0}$, followed by a controlled-$U_1,U_2$ to generates the state $\frac{1}{\sqrt{2}} \left(  \ket{0} \ket{\phi_1} + \ket{1}\ket{\phi_2}   \right) $. Then applying the Hadamard gate on the first qubit, resulting in the state:
\begin{align}
    \frac{1}{2} \left( \ket{0} +\ket{1}\right)  \ket{\phi_1} + \frac{1}{2}\left( \ket{0} - \ket{1}\right) \ket{\phi_2}
\end{align}
which is:
\begin{align}
    \frac{1}{2} \ket{0}( \ket{\phi_1}+ \ket{\phi_2}) + \frac{1}{2}\ket{1} ( \ket{\phi_1}-\ket{\phi_2})
\end{align}
The amplitude of the state $  \frac{1}{2} \ket{0}( \ket{\phi_1}+ \ket{\phi_2}) $ is:
\begin{align}
    a &= \frac{1}{2} \sqrt{ (\bra{\phi_1}+ \bra{\phi_2}) ( \ket{\phi_1}+ \ket{\phi_2}) } \\
    &= \frac{1}{2} \sqrt{ 1 + 2 \Re \braket{\phi_1,\phi_2}}
\end{align}
Therefore, an estimation of $a$ via amplitude estimation primitive allows us to infer the value of $ \Re \braket{\phi_1,\phi_2} $. In our case above, the desired overlaps is real so a real estimation suffices. 

Now we point out the following error propagation property. Assume that $0 \leq x \leq 1$, if we can estimate $\sqrt{x}$ to an additive accuracy $\sqrt{\epsilon}$, then we can also estimate $x$ to an additive accuracy $ \epsilon$. To see this, let $\widetilde{\sqrt{x}}$ denote the estimation of $\sqrt{x}$. Then we have:
\begin{align}
  | \widetilde{\sqrt{x}}^2 - \sqrt{x}^2 | & \leq | ( \widetilde{\sqrt{x}} -x ) ( \widetilde{\sqrt{x}}+ x) \\ 
  &\leq (   \widetilde{\sqrt{x}}-x)^2  \\
  &=  |  \widetilde{\sqrt{x}} -x|^2 \leq \epsilon
\end{align}
where in the last line we use $ |  \widetilde{\sqrt{x}}- \sqrt{x}| \leq \sqrt{\epsilon}$. 

To apply in our context, we simply replace $x \leftrightarrow  1 + 2 \Re \braket{\phi_1,\phi_2} $. As we can estimate $2a = \sqrt{x}$ to an additive accuracy, say $ \sqrt{\epsilon}$, via amplitude estimation, so we can guarantee that we can have the estimation of $  1 + 2 \Re \braket{\phi_1,\phi_2}$ to an accuracy $\epsilon$, which directly implies the estimation of $  \Re \braket{\phi_1,\phi_2}$ to an accuracy $\epsilon$. The circuit complexity of amplitude estimation is $\mathcal{O}((\log \frac{1}{\xi})/\sqrt{\epsilon} )$.

In order to estimate rank (A), we first transform the block-encoding of $A$ to $h(A)$, using QSVT technique. According to \cite{gilyen2019quantum}, the polynomial approximation to the step function $h(.)$ on the interval $[\frac{1}{\kappa_A},1]$ has degree $\mathcal{O}\left( \kappa_A \log \frac{1}{\epsilon}\right)$. Then we use the above method to compute its trace, which also yields the value of $ \frac{1}{N}\rm rank (A)$. 

\bibliography{ref.bib}{}
\bibliographystyle{unsrt}

\clearpage
\newpage
\onecolumngrid

\end{document}